arXiv:1502.00814v2 [cs.LO] 10 Aug 2015

# Grafting Hypersequents onto Nested Sequents

Roman Kuznets* and Björn Lellmann†

Institut für Computersprachen, Technische Universität Wien

*Dedicated to the memory of Grigori Mints who, not surprisingly, anticipated many ideas discussed in this paper.*

**Abstract**

We introduce a new Gentzen-style framework of grafted hypersequents that combines the formalism of nested sequents with that of hypersequents. To illustrate the potential of the framework, we present novel calculi for the modal logics K5 and KD5, as well as for extensions of the modal logics K and KD with the axiom for shift reflexivity. The latter of these extensions is also known as $\mathsf{SDL}^+$ in the context of deontic logic. All our calculi enjoy syntactic cut elimination and can be used in backwards proof search procedures of optimal complexity. The tableaufication of the calculi for K5 and KD5 yields simplified prefixed tableau calculi for these logic reminiscent of the simplified tableau system for S5, which might be of independent interest.

## 1 Introduction

The framework of *sequent calculi* has proven quite successful in providing analytic calculi for a number of normal modal logics such as K, KT, or S4 [Wan02]. Unfortunately, there are also a number of reasonably simple modal logics for which no acceptable cut-free sequent calculus seems to exist. Perhaps, the easiest way of demonstrating this limitation of the sequent framework is by considering various extensions of the standard modal logic K that validate the Euclideanness axiom

$$(5) \quad \Diamond\Box p \rightarrow \Box p \ .$$

In particular, the logics K5, KD5, and S5 have so far resisted all efforts to provide them with a cut-free sequent formulation, and for some formats of rules it can even be shown that no such calculus can exist [LP13]. To overcome this difficulty, several extensions of the sequent framework have been suggested, including

- the framework of *hypersequent calculi*, which was introduced independently in [Min74, Pot83, Avr96] and which provided numerous cut-free formulations for the logic S5, and
- the framework of *nested sequents* [Brü09, Pog10], which supplies cut-free calculi for all the logics of the modal cube, including S5, K5, and KD5.

The latter framework is, in fact, more general than the former in the sense that every hypersequent can be viewed as a nested sequent. As a consequence, nested rules are more complex in that they are allowed to operate deep inside a given structure. Under a translation of these structures to modal formulae, this corresponds to rules operating under an unbounded number of nested boxes, in the spirit of *deep*

*Electronic address: `roman@logic.at`. Funded by the Austrian Science Fund (FWF) project P 25417-G15 and Lise Meitner project M 1770-N25.

†Electronic address: `lellmann@logic.at`. Funded by the Austrian Science Fund (FWF) START project Y 544-N23, and by the European Union's Horizon 2020 research and innovation programme under the Marie Skłodowska-Curie grant agreement No 660047.

*inference*. In contrast, hypersequent rules can be called *shallow* because, under corresponding formula translations, they operate either directly on formulae or, at worst, under one layer of modalities. In the usual trade-off between simplicity and expressivity, while the shallow structures of uniform depth from the hypersequent framework succeed in capturing the logic S5, they do not seem to suffice for the logics K5 or KD5.

In this article, we propose to combine the benefits of both frameworks by suggesting a novel framework of *grafted hypersequents*, which is, at the same time, sufficiently expressive to capture the logics K5 and KD5 and sufficiently simple to retain most of the nice properties of the hypersequent framework. In particular, grafted hypersequents employ only shallow inferences.

Our intuitions in constructing this framework have been guided by two main considerations. From a syntactical point of view, the structures we consider arise from the general study of the connections between Hilbert-style axioms and rules in sequent-style calculi, including hypersequent calculi [Lel14]. On the one hand, axioms formed by a disjunction of boxed formulae exactly match the shape of the formula translation of a hypersequent and, thus, lend themselves to a conversion into rules in this framework. On the other hand, since the standard formula translation of a single-component hypersequent is a boxed formula, the soundness of the resulting calculus under this translation seems to require that the rule $\Box A/A$ be admissible in the logic under scrutiny. To lift this limitation and be able to capture logics that lack this property, such as K5 or KD5, it is natural to extend the sequent-like structure by a new part to be interpreted as an unboxed formula, thus motivating the move to the nested sequent style setting with minimal nesting. Apart from these purely syntactic considerations, our intuitions have been guided by the connections with the semantics for the logic in question. Fitting in [Fit12] demonstrated the correspondence between prefixed tableaux, whose prefixes encode tree-like Kripke frames, and purely internal nested sequent calculi, which can be viewed as trees of sequents. A similar correspondence between simplified tableaux for S5 with integer prefixes and hypersequent calculi for S5 has already been known. Thus, it is natural also from this perspective to reflect the structure of Euclidean Kripke frames, i.e., totally connected components partially accessible from a single "observer" world, in the proof structures used.

Informally, the idea is to consider a *trunk* in the form of a nested sequent of bounded depth[1] and to glue or *graft* a hypersequent onto its ends. Grafted hypersequent systems are obtained by combining suitable systems of nested sequent rules applied to the trunk and of hypersequent rules applied to the grafts. This leads to bounded-depth calculi for the logics K5 and KD5, as well as for the extensions of K and KD with the axiom

$$(\mathsf{T}_\Box) \quad \Box(\Box p \to p)$$

of *shift reflexivity*. Apart from syntactic cut elimination, we show how these calculi can be used in decision procedures of (optimal) complexity for these logics:

- coNP for K5 and KD5;
- PSPACE for the logics of shift reflexivity.

Moreover, extending the well-known correspondence between simplified tableaux and hypersequent calculi for S5 to our setting, we obtain simplified tableaux for K5 and KD5 with integer prefixes of three different types that closely mirror the semantics for these logics. Using these tableau systems, we obtain alternative semantic proofs of cut-free completeness for our grafted hypersequent calculi.

**Related Work.** Finding cut-free internal sequent-style calculi for the logics K5 and KD5 was an open problem for a rather long time. While such calculi for other modal logics had been around for more than 50 years [OM57], it was not until recently that they were developed for these two logics as well.

Among the earliest approaches are the purported analytic calculi for these logics given by Massacci in [Mas94] using the framework of *prefixed tableaux* introduced by Fitting in [Fit72]. In this framework formulae are prefixed with names representing a world in which the formula holds in the Kripke semantics for the logics, and the modal rules make essential use of the accessibility structure of the represented worlds. It was later discovered that in the calculi for K5 and KD5 as given in [Mas94] one crucial rule was missing, but this was fixed by Goré and Massacci in [Gor99, Mas00]. In the original [Mas94], Massacci also presented prefixed tableau calculi for the logics $\mathsf{KT}_\Box$ and $\mathsf{SDL}^+$ of shift reflexivity, called the logics OM and OM$^+$ of almost reflexivity there. The prefixed tableaux calculi for all these logics give rise to decision procedures for the respective logics, and in contrast to our calculi also can be used to show derivability from a set of global assumptions. However, since prefixed tableaux are essentially nested sequents upside-down, they do not provide a natural medium for obtaining optimal complexity bounds either, unlike our grafted hypersequents.[2]

[1] In this article, this depth is always zero, i.e., the trunk consists of only one sequent that corresponds to the observer world.

Perhaps closer than the framework of prefixed tableaux to Gentzen's original sequent calculus is Belnap's framework of *display logic* [Bel82]. Cut-free calculi for the logics K5 and KD5 in this framework were introduced by Wansing in [Wan94]. Alternatively, the calculi follow from a more general result by rewriting the axioms (5) and (D) as axioms of tense logic and then applying Kracht's algorithm [Kra96] to convert the resulting formulae into structural rules. However, since the resulting calculi are based on modal tense logic instead of non-temporal modal logic, they contain a structural connective for the backwards directed tense modalities as well as for the standard (forward directed) modalities. Thus again there is no formula translation of the structures occurring in the derivations in the language of standard (non-temporal) modal logic, and hence the calculi cannot be considered fully internal (see also [Wan02]). Furthermore, since these calculi do not satisfy the substructure property, it is not clear whether they can be used in decision procedures for the logics considered here.

The very general results about conversion of first-order frame properties for modal logics into structural rules of a *labelled sequent* calculus established by Negri and von Plato in [NP01, Neg05] also can be used to obtain cut-free calculi for all the logics considered in this article. In this framework formulae are labelled with worlds, and the sequents also contain a relational part describing the accessibility relation on these worlds. Thus again the sequents considered do not admit a direct formula translation and the calculi should not be counted as fully internal. It is also not clear whether the resulting calculi would yield decision procedures of optimal complexity.

One way to avoid the use of labels which are not part of the object language is to explicitly include them as first class citizens in the object language, as done in *hybrid modal logic* (see, e.g., [AtC07]). This leads to very natural and elegant formulations of terminating tableaux calculi for many extensions of the hybrid version of modal logic K, including the hybrid versions of the modal logics considered in this paper [BB09, Bra11]. However, the move to hybrid modal logic results in additional expressivity, since now the formulae can explicitly refer to names for worlds in a Kripke model. So while tableaux calculi for the hybrid versions of K5 or KD5 can be used to decide theoremhood in the non-hybrid logics, the calculi again can not be considered fully internal with respect to the latter.

As a first step towards such fully internal cut-free calculi, standard sequent calculi with a form of the *analytic cut rule* for these logics were introduced by Takano in [Tak01]. These calculi contain the obvious sequent rule for the axiom (5) which adds a boxed context on the right hand side of the standard rule for modal logic K. It is then shown that the cut rule can be limited to sub-formulae of the conclusion and formulae of the form $\Box\neg\Box B$ or $\neg\Box B$ where $\Box B$ occurs under a $\Box$ in the conclusion of the cut. At about the same time, tableau calculi for a number of modal logics including K5 and KD5 with an analytic superformula property similar to the limited sub-formula property of the above calculi were introduced by Nguyen in [Ngu01]. These calculi make use of additional connectives denoting a blocked version of $\Box$ and the existence of a predecessor world. Of course, since a restricted form of the cut rule is necessary in both of these approaches, they cannot be considered cut-free.

The problem of finding calculi for K5 and KD5 which are at the same time fully internal and cut-free was finally solved by Brünnler's framework of *nested sequents* in [Brü06] for K5 and [Brü09] for KD5. This framework allows for an arbitrary nesting of a structural connective for $\Box$ which essentially changes the underlying structures from sequents to trees of sequents. This additional structure is then used to provide fully internal cut-free calculi for all logics in the modal cube, including the logics K5 and KD5. Cut-free completeness is shown both by syntactical cut elimination and by a counter-model construction from a failed proof search. While the calculi thus can be used to decide memberships for these logics, the provided decision procedure runs in exponential time and thus is of suboptimal complexity. The nested sequent framework was also independently introduced by Poggiolesi under the name of *tree-hypersequents* and using a different notation [Pog09], but the logics we are concerned with here were not covered.

Mints states in [Min97, p. 690] that he used a structure similar to our grafted hypersequents to establish a calculus for the logic S5 in [Min71]. In particular, the formula translation of the structures considered in [Min97] is exactly our formula translation of the grafted hypersequents. However, while the credited work [Min71] does present essentially a hypersequent calculus for S5 (which by the way predates the hypersequent calculi for this logic given by Pottinger [Pot83] or Avron [Avr96] by more than ten years), we were unable to verify the interpretation of the structures there as analogous to the interpretation of our grafted hypersequents. In his seminal [Avr96], Avron also credits Mints with introducing a hypersequential calculus for S5 in [Min74, Min92] where one of the components is designated, but we could only find the hypersequent calculus (in tableau form) both in the Russian original [Min74] and in the English publication [Min92]. In any case, while it seems clear that Mints considered structures similar to our grafted hypersequents at some point, none of the calculi considered in all these works deal with the modal logics K5 or KD5.

[2]Simplified prefixed tableaux with structure of prefixes simpler than a generic tree have been used successfully to prove optimal complexity bounds, see, e.g., [Mas00]. Grafted hypersequents can be seen as upside-down versions of simplified tableaux suitable for K5, which we present in Section 6.

The optimal coNP bound on the complexity of the theoremhood problem for the logics K5 and KD5 was established by Halpern and Rêgo in [HR07]. There the authors in fact established a much more general result to the effect that deciding theoremhood is coNP complete for *every* logic containing K5 using a small model construction.

**Preliminaries and Notation.** As usual, the *language of modal logic* contains a set $\mathcal{V}$ of countably many *propositional variables* $p, q, \ldots$, the binary *Boolean operators* $\wedge, \vee, \to$, the Boolean constant $\bot$, and the *modal operator* $\Box$. *Modal formulae* are constructed from these operators in the usual way, and are usually denoted by $A, B, C, \ldots$. We introduce the abbreviations $\neg A$ for $A \to \bot$ and $\Diamond A$ for $\neg\Box\neg A$ and adopt the standard conventions about omitting brackets: The modal connective $\Box$ binds stronger than $\wedge$ and $\vee$ which in turn bind stronger than $\to$. The *size* of a formula $A$ is the number $|A|$ of symbols occurring in it. We write $\mathbb{N}$ for the set $\{0, 1, 2, 3, \ldots\}$ of natural numbers.

Our calculi are based on finite *multisets*, i.e., on sets counting multiplicities of elements. Formally, given a set $F$, a finite *multiset over* $F$ is given by a function $\Gamma : F \to \mathbb{N}$ with finite support. We usually write $\Gamma, \Delta, \ldots$ for finite multisets over the set of modal formulae and use the standard notation $A \in \Gamma$ for $\Gamma(A) > 0$. For $A \in \Gamma$ we also say that $A$ is *contained in* $\Gamma$. If $\Gamma$ and $\Delta$ are multisets over a set $F$, we write $\Gamma, \Delta$ for the *union* of $\Gamma$ and $\Delta$, defined by $(\Gamma, \Delta)(A) = \Gamma(A) + \Delta(A)$ for $A \in F$. Finally, for $A \in F$ we also write $A$ for the multiset containing exactly one occurrence of $A$ and nothing else. Using this notation we standardly denote a finite multiset by the comma-separated list of objects contained in it, respecting their multiplicities. E.g., we write $A, B, A$ for the multiset $\Gamma$ with $\Gamma(A) = 2$, $\Gamma(B) = 1$ and $\Gamma(C) = 0$ for $C \notin \{A, B\}$.

The semantics of the logics we investigate are given as usual in terms of Kripke frames and models. A *Kripke frame* is a tuple $(W, R)$ consisting of a set $W$ of *possible worlds* and an *accessibility relation* $R \subseteq W \times W$ on this set. If in a Kripke frame for two worlds $w, v \in W$ we have $wRv$, then we also say that $v$ is a *successor of* $w$. A *Kripke model* or simply *model* then is a Kripke frame together with a *valuation* $\sigma : \mathcal{V} \to \mathcal{P}(W)$ assigning to each propositional variable in $\mathcal{V}$ a subset of the set $W \neq \varnothing$ of possible worlds. Truth of a formula $A$ in a world $w$ of a Kripke model $(W, R, \sigma)$ is then written as $(W, R, \sigma), w \Vdash A$ and is defined recursively, as usual, by the clauses

$$\begin{array}{rcl} (W,R,\sigma), w \Vdash p & \text{iff} & w \in \sigma(p) \\ (W,R,\sigma), w \Vdash \Box B & \text{iff} & \text{for all } v \text{ with } wRv \text{ we have } (W,R,\sigma), v \Vdash B \end{array}$$

together with the standard clauses for the propositional connectives. If $(W, R, \sigma), w \Vdash A$, then we also say that the formula $A$ *holds at world* $w$ *in the model* $(W, R, \sigma)$. If a formula $A$ holds in every world of a model, then it *holds in the model*, and if it holds in every model $(W, R, \sigma)$ based on a particular Kripke frame $(W, R)$, then it is *valid* in the Kripke frame $(W, R)$. Dually, a formula $A$ is *satisfiable* in a Kripke frame $(W, R)$ if there is a valuation $\sigma : \mathcal{V} \to \mathcal{P}(W)$ and a world $w \in W$ such that $A$ holds at $w$ in $(W, R, \sigma)$. More details can be found, e.g., in [BdRV01].

## 2 Grafted Hypersequents

The notion of a grafted hypersequent can be seen both as a generalisation of the notion of a hypersequent and as a restriction of the notion of a nested sequent. Here, as usual, a *hypersequent* [Avr96] is a structure $\Gamma_1 \Rightarrow \Delta_1 \mid \cdots \mid \Gamma_n \Rightarrow \Delta_n$, where each *component* $\Gamma_i \Rightarrow \Delta_i$ of the hypersequent is an ordinary *sequent*, i.e., a pair of multisets of formulae. The standard formula interpretation of such a structure is the formula

$$\Box\left(\bigwedge\Gamma_1 \to \bigvee\Delta_1\right) \vee \cdots \vee \Box\left(\bigwedge\Gamma_n \to \bigvee\Delta_n\right) \quad . \tag{1}$$

Rules in a hypersequent calculus either modify a single component or transfer formulas between different components. Importantly, all components have equal status, and the interpretation of each component is prefixed with the modality $\Box$. The latter fact means that the hypersequent framework with the standard interpretation (1) is suitable mainly for logics in which the rule $\Box A/A$ is admissible, e.g., for *reflexive* modal logics, which validate formulae $\Box A \to A$ (such as $\mathsf{S5}$).

In contrast, *nested sequents* [Brü09, Pog10] exchange the homogenuous list structure of hypersequents for a hierarchical tree structure: a nested sequent is a structure $\Gamma \Rightarrow \Delta, [\mathcal{N}_1], \ldots, [\mathcal{N}_n]$, where $\Gamma \Rightarrow \Delta$ is an ordinary sequent and each $\mathcal{N}_i$ is again a nested sequent. The standard formula interpretation of nested sequents is given recursively as

$$\left(\Gamma \Rightarrow \Delta, [\mathcal{N}_1], \ldots, [\mathcal{N}_n]\right)^{\mathrm{tr}} := \bigwedge\Gamma \to \bigvee\Delta \vee \Box(\mathcal{N}_1)^{\mathrm{tr}} \vee \cdots \vee \Box(\mathcal{N}_n)^{\mathrm{tr}} \quad ,$$

where $(\mathcal{N}_i)^{\mathrm{tr}}$ is the standard formula interpretation of the nested sequent $\mathcal{N}_i$. Thus, in particular, the standard formula interpretation of a hypersequent $\Gamma_1 \Rightarrow \Delta_1 \mid \cdots \mid \Gamma_n \Rightarrow \Delta_n$ is the same modulo trivial propositional reasoning as that of the nested sequent $\Rightarrow [\Gamma_1 \Rightarrow \Delta_1], \ldots, [\Gamma_n \Rightarrow \Delta_n]$. While the full tree structure of nested sequents greatly extends the framework's expressivity, there are also drawbacks. The additional structure makes proofs of cut elimination (both in the context-sharing [Brü09] and the context-splitting [Pog10] formulation) very complex and typically prevents the extraction of optimal upper bounds on the complexity of decision procedures from cut-free calculi. Moreover, some calculi, such as the one for the logic $\mathsf{K5}$ given in [Brü09], use rules linking two parts of the structure that might be arbitrarily far apart in terms of nestings of the structural box [ ].

The main idea for constructing *grafted hypersequents* is to take an ordinary sequent $\Gamma \Rightarrow \Delta$ as a *root* or *trunk* and add or *graft* a hypersequent $\mathcal{H}$ to it. Depending on which framework one is most comfortable with, these structures can also be viewed as rooted hypersequents, i.e., hypersequents with a designated *root* component, or as truncated nested sequents, i.e., nested sequents of bounded depth. However, since we use hypersequent-style rules to reason in the grafted part of the structure and nested-sequent-style rules to govern the interaction between the grafted part and the trunk, we choose to use this terminology. Formal definitions are as follows.

**Definition 2.1.** A *sequent* is a pair of multisets $\Gamma$ and $\Delta$, written as $\Gamma \Rightarrow \Delta$. A *hypersequent* is a multiset of sequents, written $\Gamma_1 \Rightarrow \Delta_1 \mid \cdots \mid \Gamma_n \Rightarrow \Delta_n$, where each $\Gamma_i \Rightarrow \Delta_i$ is called a *component*. A *grafted hypersequent* is given by a sequent $\Gamma \Rightarrow \Delta$, called its *trunk*, together with a hypersequent $\mathcal{H}$, called its *crown*, and is written as $\Gamma \Rightarrow \Delta \mathbin{\|} \mathcal{H}$. If the crown is the empty hypersequent, the double-line separator can be omitted: a grafted hypersequent $\Gamma \Rightarrow \Delta$ is understood as $\Gamma \Rightarrow \Delta \mathbin{\|} \varnothing$. Formulae occurring on the left hand side of the sequent arrow in the trunk or a component of the crown are called *antecedent formulae*, those occurring on the right hand side *consequent formulae*.

Thus, a grafted hypersequent $\Gamma \Rightarrow \Delta \mathbin{\|} \Sigma_1 \Rightarrow \Pi_1 \mid \cdots \mid \Sigma_n \Rightarrow \Pi_n$ is essentially the same as the nested sequent $\Gamma \Rightarrow \Delta, [\Sigma_1 \Rightarrow \Pi_1], \ldots, [\Sigma_n \Rightarrow \Pi_n]$. The interpretation of grafted hypersequents is adapted from the nested sequent setting as well:

**Definition 2.2.** Let $\mathcal{G}$ be a grafted hypersequent $\Gamma \Rightarrow \Delta \mathbin{\|} \Sigma_1 \Rightarrow \Pi_1 \mid \cdots \mid \Sigma_n \Rightarrow \Pi_n$. Its *formula interpretation* is the formula $\iota(\mathcal{G}) := \bigwedge\Gamma \to \bigvee\Delta \vee \bigvee_{i=1}^{n} \Box\left(\bigwedge\Sigma_i \to \bigvee\Pi_i\right)$.

Since the structure of a grafted hypersequent only encodes a bounded nesting depth of structural boxes, grafted hypersequent calculi can still be considered *shallow inference* calculi, in contrast to e.g. nested sequent calculi which allow *deep inference* (with respect to the nesting depth of structural boxes).

# 3 A Grafted Hypersequent Calculus for K5

The logic we are mainly interested in is the modal logic $\mathsf{K5}$. This logic has a Hilbert-style presentation given by a complete set of axioms for classical propositional logic in our language of modal logic, by the axioms

$$(\mathsf{K})\ \ \Box(A \to B) \to (\Box A \to \Box B) \qquad \text{and} \qquad (\mathsf{5})\ \ \Diamond\Box A \to \Box A \quad ,$$

and by the inference rules *modus ponens* $\mathsf{MP}$ and *necessitation* $\mathsf{nec}$, shown below, where $\vdash$ denotes membership in the logic:

$$\frac{\vdash A \qquad \vdash A \to B}{\vdash B}\ \mathsf{MP} \qquad \text{and} \qquad \frac{\vdash A}{\vdash \Box A}\ \mathsf{nec}\ .$$

In semantical terms, it is the logic of *Euclidean Kripke frames*, i.e., it is the set of all modal formulae valid in every Kripke-frame $(W, R)$ with the accessibility relation $R$ satisfying $\forall x, y, z\ (xRy \wedge xRz \to yRz)$. For more details, see, e.g., [BdRV01].

According to the construction of a grafted hypersequent as a trunk with an additional hypersequent grafted onto it, the rules of the grafted hypersequent calculi are split into two groups: the trunk rules and the crown rules. The *trunk rules* consist of the structural and logical rules that govern inferences at the *trunk level* and are given in Figure 1. The structural rules and the logical rules for the propositional connectives at this level are standard. The *transfer rules* that introduce the connective $\Box$ on the left or on the right of the sequent arrow $\Rightarrow$ follow the treatment of the modality in nested sequent calculi [Pog10].

*Crown rules* govern inferences in the crown of the grafted hypersequent. While the propositional crown rules are analogous to the propositional trunk rules, the modal crown rules are modelled after the hypersequent calculus for the modal logic $\mathsf{S5}$ from [Res07]. This group of the rules is given in Figure 2. The semantic intuition for why we use the rules for $\mathsf{S5}$ at the crown level is that the class of Euclidean Kripke frames is the class of frames where the successors of every node form a totally connected component, i.e., an $\mathsf{S5}$ subframe. Thus, if the trunk sequent is evaluated in a given world, the crown is evaluated in its successors and as such should follow the inference rules of $\mathsf{S5}$. The syntactic intuition is that, converting the axioms of $\mathsf{K5}$ into rules using a method similar to the one described in [Lel14] for the hypersequent framework, we obtain exactly the rules $\mathsf{5}$ and $\mathsf{K}$ from Figure 2. Furthermore, the resulting system permits the adaption of the general cut elimination proof in [Lel14] to the crown layer, and thus serves as a blueprint for the construction of further calculi: in principle the modal crown rules could be exchanged for rules modelled after an arbitrary hypersequent calculus satisfying the sufficient conditions for cut elimination in *op. cit.*, and the cut elimination proof of Section 4 would still go through. Of course certain details such as derivability of the axioms of the logic at the trunk level or soundness of the rules still would need to be checked. Notwithstanding, we do not doubt that it would be possible to use one of the many other hypersequent calculi for $\mathsf{S5}$ proposed in the literature as basis for the crown rules, if the cut elimination proof is adapted suitably.

*Remark* 3.1. One technical peculiarity of the crown rules is that in order to be able to show soundness of the rule $\mathsf{K}$, we need to stipulate the trunk sequent to be empty, making it necessary to extend this restriction to all the crown rules. It might seem unexpected that the rule $\mathsf{K}$, which closely resembles in shape the standard modal $\mathsf{nec}$ rule, is not generally sound without this restriction and, moreover, relies on the Euclideanness of the frame for the soundness proof even in the case of the empty trunk sequent. The mystery is easy to clarify. The meaning of the $\mathsf{K}$ rule under the formula interpretation, in the simplest case, is that $\Box H \vee \Box\Box A$ is inferred from $\Box H \vee \Box A$, an inference clearly invalid for the class of all Kripke frames.

Finally, the *cut rule* has two versions: one for the the trunk level and the other for the crown level. The rules are given in Figure 3. Similar to the crown logical rules, the trunk in the rule $\mathsf{Cut_c}$ needs to be empty.

**Definition 3.2.** The rule set $\mathcal{R}_{\mathsf{K5}}$ consists of the rules given in Figure 1 together with the rules given in Figure 2. We write $\mathcal{R}_{\mathsf{K5}}\mathsf{Cut_c}$ for the system with the rule $\mathsf{Cut_c}$ from Figure 3 added to $\mathcal{R}_{\mathsf{K5}}$ and $\mathcal{R}_{\mathsf{K5}}\mathsf{Cut}$ for the system with the rule $\mathsf{Cut_t}$ from Figure 3 added to $\mathcal{R}_{\mathsf{K5}}\mathsf{Cut_c}$.

The notions of a *derivation*, *derivability*, *derivable* and *admissible rules*, etc., are defined in the standard way. For a grafted hypersequent $\mathcal{G}$ we write $\mathcal{R}_{\mathsf{K5}} \vdash \mathcal{G}$ if $\mathcal{G}$ is derivable in the system $\mathcal{R}_{\mathsf{K5}}$, and analogously for the systems $\mathcal{R}_{\mathsf{K5}}\mathsf{Cut_c}$ and $\mathcal{R}_{\mathsf{K5}}\mathsf{Cut}$. As usual, we sometimes write a double line to abbreviate multiple applications of the same rule. Following [TS00], in the rules of $\mathcal{R}_{\mathsf{K5}}\mathsf{Cut}$ as given in Figures 1-3 we call the formulae in the $\Gamma, \Delta, \Sigma, \Pi, \Gamma_i, \Delta_i, \Sigma_i, \Pi_i$ or the components of $\mathcal{H}, \mathcal{H}'$ the *context formulae* or *contextual*. All non-contextual formulae occurring in the conclusion are called *principal*, and all non-contextual formulae in the premisses are called *active formulae*. Thus in particular the contracted formulae in the contraction rules and the weakened formulae in the weakening rules are principal. These notions extend naturally to the level of crown components: all the crown components in the $\mathcal{H}$ are *context components*, while the non-contextual crown components in the conclusion are *principal components*, those in the premisses *active components*.

Figure 1: The trunk rules of the calculus for K5.

The trunk propositional rules:

$$\frac{\Gamma, A, B \Rightarrow \Delta \mid\mid \mathcal{H}}{\Gamma, A \wedge B \Rightarrow \Delta \mid\mid \mathcal{H}} \wedge_L \qquad \frac{\Gamma, A \Rightarrow \Delta \mid\mid \mathcal{H} \quad \Gamma, B \Rightarrow \Delta \mid\mid \mathcal{H}}{\Gamma, A \vee B \Rightarrow \Delta \mid\mid \mathcal{H}} \vee_L \qquad \frac{\Gamma \Rightarrow \Delta, A \mid\mid \mathcal{H} \qquad \Gamma, B \Rightarrow \Delta \mid\mid \mathcal{H}}{\Gamma, A \to B \Rightarrow \Delta \mid\mid \mathcal{H}} \to_L$$

$$\frac{\Gamma \Rightarrow \Delta, A \mid\mid \mathcal{H} \qquad \Gamma \Rightarrow \Delta, B \mid\mid \mathcal{H}}{\Gamma \Rightarrow \Delta, A \wedge B \mid\mid \mathcal{H}} \wedge_R \qquad \frac{\Gamma \Rightarrow \Delta, A, B \mid\mid \mathcal{H}}{\Gamma \Rightarrow \Delta, A \vee B \mid\mid \mathcal{H}} \vee_R \qquad \frac{\Gamma, A \Rightarrow \Delta, B \mid\mid \mathcal{H}}{\Gamma \Rightarrow \Delta, A \to B \mid\mid \mathcal{H}} \to_R$$

The trunk initial structures and the transfer rules:

$$\frac{}{\Gamma, p \Rightarrow \Delta, p \mid\mid \mathcal{H}} \mathsf{Init} \qquad \frac{}{\Gamma, \bot \Rightarrow \Delta \mid\mid \mathcal{H}} \bot_L \qquad \frac{\Gamma \Rightarrow \Delta \mid\mid \mathcal{H} \mid \Rightarrow A}{\Gamma \Rightarrow \Delta, \Box A \mid\mid \mathcal{H}} \Box_R \qquad \frac{\Gamma \Rightarrow \Delta \mid\mid \mathcal{H} \mid \Sigma, A \Rightarrow \Pi}{\Gamma, \Box A \Rightarrow \Delta \mid\mid \mathcal{H} \mid \Sigma \Rightarrow \Pi} \Box_L$$

The trunk structural rules:

$$\frac{\Gamma, A, A \Rightarrow \Delta \mid\mid \mathcal{H}}{\Gamma, A \Rightarrow \Delta \mid\mid \mathcal{H}} \mathsf{C}_L \qquad \frac{\Gamma \Rightarrow \Delta, A, A \mid\mid \mathcal{H}}{\Gamma \Rightarrow \Delta, A \mid\mid \mathcal{H}} \mathsf{C}_R \qquad \frac{\Gamma \Rightarrow \Delta \mid\mid \mathcal{H}}{\Gamma, \Omega \Rightarrow \Delta, \Xi \mid\mid \mathcal{H}} \mathsf{W}$$

Figure 2: The crown rules of the calculus for K5

The crown propositional rules:

$$\frac{\Rightarrow \mid\mid \mathcal{H} \mid \Gamma, A, B \Rightarrow \Delta}{\Rightarrow \mid\mid \mathcal{H} \mid \Gamma, A \wedge B \Rightarrow \Delta} \wedge_L \qquad \frac{\Rightarrow \mid\mid \mathcal{H} \mid \Gamma \Rightarrow \Delta, A \quad \Rightarrow \mid\mid \mathcal{H} \mid \Gamma \Rightarrow \Delta, B}{\Rightarrow \mid\mid \mathcal{H} \mid \Gamma \Rightarrow \Delta, A \wedge B} \wedge_R$$

$$\frac{\Rightarrow \mid\mid \mathcal{H} \mid \Gamma, A \Rightarrow \Delta \quad \Rightarrow \mid\mid \mathcal{H} \mid \Gamma, B \Rightarrow \Delta}{\Rightarrow \mid\mid \mathcal{H} \mid \Gamma, A \vee B \Rightarrow \Delta} \vee_L \qquad \frac{\Rightarrow \mid\mid \mathcal{H} \mid \Gamma \Rightarrow \Delta, A, B}{\Rightarrow \mid\mid \mathcal{H} \mid \Gamma \Rightarrow \Delta, A \vee B} \vee_R$$

$$\frac{\Rightarrow \mid\mid \mathcal{H} \mid \Gamma \Rightarrow \Delta, A \quad \Rightarrow \mid\mid \mathcal{H} \mid \Gamma, B \Rightarrow \Delta}{\Rightarrow \mid\mid \mathcal{H} \mid \Gamma, A \to B \Rightarrow \Delta} \to_L \qquad \frac{\Rightarrow \mid\mid \mathcal{H} \mid \Gamma, A \Rightarrow \Delta, B}{\Rightarrow \mid\mid \mathcal{H} \mid \Gamma \Rightarrow \Delta, A \to B} \to_R$$

The crown initial structures and the crown modal rules:

$$\frac{}{\Rightarrow \mid\mid \mathcal{H} \mid \Gamma, p \Rightarrow \Delta, p} \mathsf{Init} \qquad \frac{}{\Rightarrow \mid\mid \mathcal{H} \mid \Gamma, \bot \Rightarrow \Delta} \bot_L \qquad \frac{\Rightarrow \mid\mid \mathcal{H} \mid \Sigma, A \Rightarrow \Pi}{\Rightarrow \mid\mid \mathcal{H} \mid \Box A \Rightarrow \mid \Sigma \Rightarrow \Pi} \mathsf{5} \qquad \frac{\Rightarrow \mid\mid \mathcal{H} \mid \Rightarrow A}{\Rightarrow \mid\mid \mathcal{H} \mid \Rightarrow \Box A} \mathsf{K}$$

The crown structural rules:

$$\frac{\Rightarrow \mid\mid \mathcal{H} \mid \Omega \Rightarrow \Xi \mid \Omega \Rightarrow \Xi}{\Rightarrow \mid\mid \mathcal{H} \mid \Omega \Rightarrow \Xi} \mathsf{EC} \qquad \frac{\Rightarrow \mid\mid \mathcal{H}}{\Rightarrow \mid\mid \mathcal{H} \mid \Omega \Rightarrow \Xi} \mathsf{EW}$$

$$\frac{\Rightarrow \mid\mid \mathcal{H} \mid \Sigma, A, A \Rightarrow \Pi}{\Rightarrow \mid\mid \mathcal{H} \mid \Sigma, A \Rightarrow \Pi} \mathsf{IC}_L \qquad \frac{\Rightarrow \mid\mid \mathcal{H} \mid \Sigma \Rightarrow \Pi, A, A}{\Rightarrow \mid\mid \mathcal{H} \mid \Sigma \Rightarrow \Pi, A} \mathsf{IC}_R \qquad \frac{\Rightarrow \mid\mid \mathcal{H} \mid \Sigma \Rightarrow \Pi}{\Rightarrow \mid\mid \mathcal{H} \mid \Sigma, \Omega \Rightarrow \Pi, \Xi} \mathsf{IW}$$

Figure 3: The trunk and crown cut rules

$$\frac{\Gamma_1 \Rightarrow \Delta_1, A \mid\mid \mathcal{H} \qquad \Gamma_2, A \Rightarrow \Delta_2 \mid\mid \mathcal{H}'}{\Gamma_1, \Gamma_2 \Rightarrow \Delta_1, \Delta_2 \mid\mid \mathcal{H} \mid \mathcal{H}'} \mathsf{Cut_t} \qquad \frac{\Rightarrow \mid\mid \mathcal{H} \mid \Sigma_1 \Rightarrow \Pi_1, A \qquad \Rightarrow \mid\mid \mathcal{H}' \mid \Sigma_2, A \Rightarrow \Pi_2}{\Rightarrow \mid\mid \mathcal{H} \mid \mathcal{H}' \mid \Sigma_1, \Sigma_2 \Rightarrow \Pi_1, \Pi_2} \mathsf{Cut_c}$$

**Proposition 3.3** (Soundness)**.** *All the rules of* $\mathcal{R}_{\mathsf{K5}}\mathsf{Cut}$ *preserve* $\mathsf{K5}$*-validity under the formula interpretation, i.e., for each (instance of a) rule of* $\mathcal{R}_{\mathsf{K5}}\mathsf{Cut}$*, if* $\iota(\mathcal{P}) \in \mathsf{K5}$ *for each premiss* $\mathcal{P}$ *of this rule, then* $\iota(\mathcal{C}) \in \mathsf{K5}$ *for the conclusion* $\mathcal{C}$ *of this rule.*

*Proof.* By inspection of all the different cases it is shown that whenever $\iota(\mathcal{C})$ is not $\mathsf{K5}$-valid, i.e., whenever $\neg\iota(\mathcal{C})$ is satisfiable in a Euclidean frame, then also one of the $\iota(\mathcal{P})$ is not $\mathsf{K5}$-valid, i.e., one of the $\neg\iota(\mathcal{P})$ is satisfiable. We show this for the modal rules; the cases for the propositional and structural rules, trunk or crown alike, are standard. Throughout the proof, we use the letter $\mathcal{H}$ to denote the side hypersequent of the rule and write $H$ for the formula $\iota(\Rightarrow \mid\mid \mathcal{H})$. We also write $A \equiv_{\mathsf{K5}} B$ to mean that $(A \to B) \wedge (B \to A) \in \mathsf{K5}$.

For the rule $\mathsf{5}$, assume that the formula interpretation of the conclusion of an instance of $\mathsf{5}$ is not $\mathsf{K5}$-valid. Given that

$$\neg\iota(\Rightarrow \mid\mid \mathcal{H} \mid \Box A \Rightarrow \mid \Sigma \Rightarrow \Pi) \quad \equiv_{\mathsf{K5}} \quad \neg H \wedge \Diamond\Box A \wedge \Diamond\left(\bigwedge\Sigma \wedge \neg\bigvee\Pi\right) \ ,$$

this means that the latter formula holds in a Euclidean model $(W, R, \sigma)$ at a world $w$. Then, in particular, there are worlds $v_1$ and $v_2$ in $W$ with $wRv_1$ and $wRv_2$ such that $\Box A$ holds at $v_1$ and $\bigwedge\Sigma \wedge \neg\bigvee\Pi$ holds at $v_2$. Given that $v_1Rv_2$ by Euclideanness of $R$, the formula $A$ holds at $v_2$ and, thus, $\bigwedge\Sigma \wedge A \wedge \neg\bigvee\Pi$ holds at $v_2$. Given that

$$\neg\iota(\Rightarrow \mid\mid \mathcal{H} \mid \Sigma, A \Rightarrow \Pi) \quad \equiv_{\mathsf{K5}} \quad \neg H \wedge \Diamond\left(\bigwedge\Sigma \wedge A \wedge \neg\bigvee\Pi\right) \ ,$$

it follows that the formula interpretation of the premise of this rule does not hold at the world $w$, meaning that it is not $\mathsf{K5}$-valid either.

For the rule $\mathsf{K}$, assume that $\iota(\Rightarrow \mid\mid \mathcal{H} \mid \Rightarrow \Box A)$ is not $\mathsf{K5}$-valid, i.e., that $\neg H \wedge \Diamond\Diamond\neg A$ holds in a Euclidean model $(W, R, \sigma)$ at a world $w$. Let $v$ be a world in $W$ such that $wRv$ and $\Diamond\neg A$ holds at $v$. Note that $\neg H \equiv_{\mathsf{K5}} \bigwedge\Diamond\Upsilon$ for some finite (possibly empty) set $\Upsilon$ of formulas, where $\Diamond\Upsilon := \{\Diamond B \mid B \in \Upsilon\}$. Since $\Diamond B$ holds at $w$ for every $B \in \Upsilon$, for every such $B$ there is a $u \in W$ such that $wRu$ and $B$ holds at $u$. By Euclideanness of $R$, we also have $vRu$, meaning that $\Diamond B$ holds at $v$ too. Thus, $\neg H \wedge \Diamond\neg A$ holds at $v$, invalidating $\iota(\Rightarrow \mid\mid \mathcal{H} \mid \Rightarrow A)$.

For the rule $\Box_L$, assume that $\iota\,(\Gamma, \Box A \Rightarrow \Delta \mid\mid \mathcal{H} \mid \Sigma \Rightarrow \Pi)$ is not $\mathsf{K5}$-valid, i.e., that

$$\bigwedge\Gamma \wedge \Box A \wedge \neg\bigvee\Delta \wedge \neg H \wedge \Diamond\left(\bigwedge\Sigma \wedge \neg\bigvee\Pi\right)$$

holds at a world $w$ of a Kripke model $(W, R, \sigma)$. Then, using the standard Kripke semantics for $\mathsf{K}$,

$$\bigwedge\Gamma \wedge \neg\bigvee\Delta \wedge \neg H \wedge \Diamond\left(\bigwedge\Sigma \wedge A \wedge \neg\bigvee\Pi\right)$$

also holds at the world $w$, invalidating $\iota\,(\Gamma \Rightarrow \Delta \mid\mid \mathcal{H} \mid \Sigma, A \Rightarrow \Pi)$.

The case of the rule $\Box_R$ is trivial because the formula interpretations of the premiss and the conclusion of the rule are clearly logically equivalent. □

To enhance the readability of the derivations we introduce the following rule as an abbreviation, which allows us to merge two components in the crown:

**Lemma 3.4.** *The rule*

$$\frac{\Rightarrow \mid\mid \mathcal{H} \mid \Gamma \Rightarrow \Delta \mid \Sigma \Rightarrow \Pi}{\Rightarrow \mid\mid \mathcal{H} \mid \Gamma, \Sigma \Rightarrow \Delta, \Pi}\ \mathsf{merge}$$

*is derivable in* $\mathcal{R}_{\mathsf{K5}}$.

*Proof.* Every application of the rule $\mathsf{merge}$ can be replaced by the following derivation:

$$\dfrac{\dfrac{\Rightarrow \mid\mid \mathcal{H} \mid \Gamma \Rightarrow \Delta \mid \Sigma \Rightarrow \Pi}{\Rightarrow \mid\mid \mathcal{H} \mid \Gamma, \Sigma \Rightarrow \Delta, \Pi \mid \Gamma, \Sigma \Rightarrow \Delta, \Pi}\ \mathsf{IW}}{\Rightarrow \mid\mid \mathcal{H} \mid \Gamma, \Sigma \Rightarrow \Delta, \Pi}\ \mathsf{EC}$$

□

Spelling out the abbreviation of $\neg A$ as $A \to \bot$ it is also easy to see that the trunk- and crown-level rules for negation

$$\frac{\Gamma \Rightarrow \Delta, A \,||\, \mathcal{H}}{\Gamma, \neg A \Rightarrow \Delta \,||\, \mathcal{H}}\,\neg_L \qquad \frac{\Gamma, A \Rightarrow \Delta \,||\, \mathcal{H}}{\Gamma \Rightarrow \Delta, \neg A \,||\, \mathcal{H}}\,\neg_R \qquad \frac{\Rightarrow \,||\, \mathcal{H} \mid \Sigma \Rightarrow \Pi, A}{\Rightarrow \,||\, \mathcal{H} \mid \Sigma, \neg A \Rightarrow \Pi}\,\neg_L \qquad \frac{\Rightarrow \,||\, \mathcal{H} \mid \Sigma, A \Rightarrow \Pi}{\Rightarrow \,||\, \mathcal{H} \mid \Sigma \Rightarrow \Pi, \neg A}\,\neg_R$$

are derivable in $\mathcal{R}_{\mathsf{K5}}$. Thus, from now on we use these rules to abbreviate the corresponding derivations. In order to get a feel for the derivations in this calculus it is instructive to show completeness. In order to do so, we first show that the generalised axioms with arbitrary formulae instead of propositional variables are derivable. This enables us to formulate derivations of the axioms of $\mathsf{K5}$ without having to construct them fully up to the atomic initial sequents.

**Lemma 3.5** (Generalised axioms)**.** *For every formula $A$ the grafted hypersequents $\Gamma, A \Rightarrow \Delta, A \,||\, \mathcal{H}$ and $\Rightarrow \,||\, \mathcal{H} \mid \Gamma, A \Rightarrow \Delta, A$ are derivable in $\mathcal{R}_{\mathsf{K5}}$.*

*Proof.* We first show the claim at the crown level by induction on the complexity of $A$. If the main connective is propositional, the proof is standard. In the modal case we have a derivation

$$\cfrac{\cfrac{\cfrac{\cfrac{\Rightarrow \,||\, \mathcal{H} \mid B \Rightarrow B}{\Rightarrow \,||\, \mathcal{H} \mid \Box B \Rightarrow \,\mid\, \Rightarrow B}\,\mathsf{5}}{\Rightarrow \,||\, \mathcal{H} \mid \Box B \Rightarrow \,\mid\, \Rightarrow \Box B}\,\mathsf{K}}{\Rightarrow \,||\, \mathcal{H} \mid \Box B \Rightarrow \Box B}\,\mathsf{merge}}{\Rightarrow \,||\, \mathcal{H} \mid \Gamma, \Box B \Rightarrow \Delta, \Box B}\,\mathsf{IW} \tag{2}$$

of $\Rightarrow \,||\, \mathcal{H} \mid \Gamma, \Box B \Rightarrow \Delta, \Box B$ from $\Rightarrow \,||\, \mathcal{H} \mid B \Rightarrow B$, which is derivable by the induction hypothesis.

To apply the same reasoning on the trunk level we only need to replace the above derivation by

$$\cfrac{\cfrac{\cfrac{\Rightarrow \,||\, \mathcal{H} \mid B \Rightarrow B}{\Box B \Rightarrow \,||\, \mathcal{H} \mid \Rightarrow B}\,\Box_L}{\Box B \Rightarrow \Box B \,||\, \mathcal{H}}\,\Box_R}{\Gamma, \Box B \Rightarrow \Delta, \Box B \,||\, \mathcal{H}}\,\mathsf{W}$$

where $\Rightarrow \,||\, \mathcal{H} \mid B \Rightarrow B$ is derived using the claim for the crown level. $\Box$

The completeness proof for $\mathcal{R}_{\mathsf{K5}}\mathsf{Cut}$ now proceeds by deriving all the axioms of $\mathsf{K5}$ both at the trunk level and at the crown level, showing that the necessitation rule is admissible, and simulating modus ponens using cuts. Deriving each axiom twice, at the crown level in addition to the trunk level, may seem redundant. However, this is necessary to show admissibility of the necessitation rule.

**Theorem 3.6** (Completeness with Cut)**.** *Every $\mathsf{K5}$ theorem is derivable in $\mathcal{R}_{\mathsf{K5}}\mathsf{Cut}$, i.e., $\mathcal{R}_{\mathsf{K5}}\mathsf{Cut} \vdash \Rightarrow A$ for every $A \in \mathsf{K5}$.*

*Proof.* We show a stronger statement that $\mathcal{R}_{\mathsf{K5}}\mathsf{Cut} \vdash \Rightarrow A$ and $\mathcal{R}_{\mathsf{K5}}\mathsf{Cut} \vdash \Rightarrow \,||\, \Rightarrow A$ by a simultaneous induction on the Hilbert-style derivation of $A$ in $\mathsf{K5}$. We omit the standard derivations of propositional axioms. An instance $\Box(B \to C) \to (\Box B \to \Box C)$ of $(\mathsf{K})$ at the trunk level is derived by

$$\cfrac{\cfrac{\cfrac{\cfrac{\cfrac{}{\Rightarrow \,||\, B \Rightarrow C, B}\,\text{Lem. 3.5} \qquad \cfrac{}{\Rightarrow \,||\, B, C \Rightarrow C}\,\text{Lem. 3.5}}{\Rightarrow \,||\, B, B \to C \Rightarrow C}\,\to_L}{\Box(B \to C), \Box B \Rightarrow \,||\, \Rightarrow C}\,\Box_L}{\Box(B \to C), \Box B \Rightarrow \Box C}\,\Box_R}{\Rightarrow \Box(B \to C) \to (\Box B \to \Box C)}\,\to_R$$

An instance $\Diamond \Box B \to \Box B = \neg \Box \neg \Box B \to \Box B$ of $(\mathsf{5})$ at the trunk level is derived by

$$\cfrac{\cfrac{\cfrac{\cfrac{\cfrac{\cfrac{}{\Rightarrow \,||\, B \Rightarrow B}\,\text{Lem. 3.5}}{\Rightarrow \,||\, \Box B \Rightarrow \,\mid\, \Rightarrow B}\,\mathsf{5}}{\Rightarrow \,||\, \Rightarrow \neg \Box B \mid \Rightarrow B}\,\neg_R}{\Rightarrow \Box B, \Box \neg \Box B}\,\Box_R}{\neg \Box \neg \Box B \Rightarrow \Box B}\,\neg_L}{\Rightarrow \neg \Box \neg \Box B \to \Box B}\,\to_R$$

An instance $\Box(B \to C) \to (\Box B \to \Box C)$ of $(\mathsf{K})$ at the crown level is derived by

$$
\begin{prooftree}
\AxiomC{}
\RightLabel{Lem. 3.5}
\UnaryInfC{$\Rightarrow \,||\, B \Rightarrow C, B$}
\AxiomC{}
\RightLabel{Lem. 3.5}
\UnaryInfC{$\Rightarrow \,||\, B, C \Rightarrow C$}
\RightLabel{$\to_L$}
\BinaryInfC{$\Rightarrow \,||\, B, B \to C \Rightarrow C$}
\doubleLine
\RightLabel{$\mathsf{5}$}
\UnaryInfC{$\Rightarrow \,||\, \Box(B \to C) \Rightarrow \,|\, \Box B \Rightarrow \,|\, \Rightarrow C$}
\RightLabel{$\mathsf{K}$}
\UnaryInfC{$\Rightarrow \,||\, \Box(B \to C) \Rightarrow \,|\, \Box B \Rightarrow \,|\, \Rightarrow \Box C$}
\doubleLine
\RightLabel{$\mathsf{merge}$}
\UnaryInfC{$\Rightarrow \,||\, \Box(B \to C), \Box B \Rightarrow \Box C$}
\doubleLine
\RightLabel{$\to_R$}
\UnaryInfC{$\Rightarrow \,||\, \Rightarrow \Box(B \to C) \to (\Box B \to \Box C)$}
\end{prooftree}
\tag{3}
$$

An instance $\neg\Box\neg\Box B \to \Box B$ of $(\mathsf{5})$ at the crown level is derived by

$$
\begin{prooftree}
\AxiomC{}
\RightLabel{Lem. 3.5}
\UnaryInfC{$\Rightarrow \,||\, B \Rightarrow B$}
\RightLabel{$\mathsf{5}$}
\UnaryInfC{$\Rightarrow \,||\, \Box B \Rightarrow \,|\, \Rightarrow B$}
\RightLabel{$\neg_R$}
\UnaryInfC{$\Rightarrow \,||\, \Rightarrow \neg\Box B \,|\, \Rightarrow B$}
\doubleLine
\RightLabel{$\mathsf{K}$}
\UnaryInfC{$\Rightarrow \,||\, \Rightarrow \Box\neg\Box B \,|\, \Rightarrow \Box B$}
\RightLabel{$\mathsf{merge}$}
\UnaryInfC{$\Rightarrow \,||\, \Rightarrow \Box\neg\Box B, \Box B$}
\RightLabel{$\neg_L$}
\UnaryInfC{$\Rightarrow \,||\, \neg\Box\neg\Box B \Rightarrow \Box B$}
\RightLabel{$\to_R$}
\UnaryInfC{$\Rightarrow \,||\, \Rightarrow \neg\Box\neg\Box B \to \Box B$}
\end{prooftree}
$$

If $\Box B$ is inferred from $B$ by $\mathsf{nec}$, then $\mathcal{R}_{\mathsf{K5}}\mathsf{Cut} \vdash\, \Rightarrow \,||\, \Rightarrow B$ by the induction hypothesis for the crown. Hence, $\mathcal{R}_{\mathsf{K5}}\mathsf{Cut} \vdash\, \Rightarrow \Box B$ and $\mathcal{R}_{\mathsf{K5}}\mathsf{Cut} \vdash\, \Rightarrow \,||\, \Rightarrow \Box B$ by the rules $\Box_R$ and $\mathsf{K}$ respectively.

The rule $\mathsf{MP}$ is simulated in the crown by means of the grafted hypersequent $\Rightarrow \,||\, B, B \to C \Rightarrow C$ derived above and the rule $\mathsf{Cut_c}$, whereas in the trunk we use the rule $\mathsf{Cut_t}$ and the analogous derivation of $B, B \to C \Rightarrow C$. $\Box$

While this establishes completeness of the calculus with the cut rule, we are mainly interested in cut-free completeness of the system. As usual there are two ways of showing this. The first is purely syntactical and relies on proving a cut elimination theorem, stating that derivations using the cut rule can be transformed into derivations not using this rule. While proving a cut elimination theorem is often quite tedious, it provides deep insights into the calculus since it gives a constructive step-by-step method to eliminate applications of the cut rule. We will follow this path in the next section, culminating in Thm. 4.13 at the end of the next section.

The second method is to provide a semantic proof of cut-free completeness, usually achieved by showing how to construct a counter-model from a failed proof search. This method provides insights of its own, since it connects the calculus to the semantics more directly. However, a more natural setting for this kind of proofs is that of *prefixed tableaux* [FM98]. We will explore this direction by making use of an analogue of the correspondence between nested sequent calculi and prefixed tableaux established in [Fit12]. Since this correspondence is seen clearest for grafted hypersequent calculi in which the structural rules are admissible, we postpone this exploration to Sec. 6, until after the modification of the calculus for $\mathsf{K5}$ to this effect in Sec. 5. The reader who is not interested in the intricacies of cut elimination but only in cut-free completeness of the calculus is therefore advised to skip the description of cut elimination starting with Definition 4.5 of the next section and continue with Sec. 5 instead.

# 4 Cut Elimination

While the fact that grafted hypersequents can be used to give a calculus which is sound and in the presence of the cut rules complete for the logic $\mathsf{K5}$ is perhaps not so surprising, it might be more remarkable that it is possible to show admissibility of the cut rules for this calculus. Of course this result is highly desirable, since it entails the subformula property for the calculus and thus provides the basis for a decision procedure via backwards proof search. The proof of cut elimination itself has several ingredients. At its core lies the fact that the formulation of the crown rules with the empty root sequent entails a layering of the derivations into the *crown layer* modifying only the crown part at the top of the proof tree, followed by a layer involving the trunk rules only. This is further strengthened by the following lemma stating that the bottom layer of the proof tree can be assumed to be divided into the *transfer layer* in which formulae are transferred from the crown to the trunk using the transfer rules and the *trunk layer* in which only non-modal trunk rules are applied. The first step to seeing this is the following observation.

**Lemma 4.1.** *In any derivation in $\mathcal{R}_{\mathsf{K5}}\mathsf{Cut_c}$, no trunk rule from Figure 1 occurs above any crown rule from Figure 2 nor above $\mathsf{Cut_c}$.*

*Proof.* Follows directly from the fact that the trunk is empty in all the crown rules and the fact that no rule moves formulae from the trunk to the crown. □

Moreover, the transfer rules can be permuted upwards over all the other trunk rules.

**Lemma 4.2.** *In any derivation in $\mathcal{R}_{\mathsf{K5}}\mathsf{Cut_c}$ or in $\mathcal{R}_{\mathsf{K5}}$, the rule $\Box_L$ can be permuted upwards over every trunk rule and the rule $\Box_R$ can be permuted upwards over every trunk rule other than $\Box_L$. All the permutations are depth-preserving.*

*Proof.* Both rules $\Box_L$ and $\Box_R$ replace a formula in the crown with another formula in the trunk. The reason they can be permuted upwards over all the other trunk rules is that the latter operate exclusively on formulae in the trunk (all principal and active formulae are in the trunk), have no context restrictions, and do not modify the context. It should also be noted that neither $\Box_L$ nor $\Box_R$ restricts the trunk context. (Strictly speaking, the permutation of a transfer rule over a trunk initial structure means that the transfer rule disappears, leaving another instance of the same initial structure.)

To permute $\Box_L$ upwards over $\Box_R$, use the transformation

$$\dfrac{\dfrac{\begin{array}{c}\vdots\\ \Gamma \Rightarrow \Delta \mid\mid \mathcal{H} \mid \Sigma, A \Rightarrow \Pi \mid \Rightarrow B\end{array}}{\Gamma \Rightarrow \Delta, \Box B \mid\mid \mathcal{H} \mid \Sigma, A \Rightarrow \Pi}\ \Box_R}{\Gamma, \Box A \Rightarrow \Delta, \Box B \mid\mid \mathcal{H} \mid \Sigma \Rightarrow \Pi}\ \Box_L \quad \rightsquigarrow \quad \dfrac{\dfrac{\begin{array}{c}\vdots\\ \Gamma \Rightarrow \Delta \mid\mid \mathcal{H} \mid \Sigma, A \Rightarrow \Pi \mid \Rightarrow B\end{array}}{\Gamma, \Box A \Rightarrow \Delta \mid\mid \mathcal{H} \mid \Sigma \Rightarrow \Pi \mid \Rightarrow B}\ \Box_L}{\Gamma, \Box A \Rightarrow \Delta, \Box B \mid\mid \mathcal{H} \mid \Sigma \Rightarrow \Pi}\ \Box_R$$

It is also easy to see that any two applications of $\Box_L$ are permutable and any two applications of $\Box_R$ are permutable. It is, however, impossible to permute $\Box_R$ upwards over $\Box_L$ if their active formulae belong to the same sequent in the crown. Since the rule permutations are local, the depth of the derivations is not increased. □

The previous two lemmata allow us to reorder derivations so that they are layered in a specific way.

**Definition 4.3** (Normal derivations)**.** A derivation in $\mathcal{R}_{\mathsf{K5}}$ or $\mathcal{R}_{\mathsf{K5}}\mathsf{Cut_c}$ is called *normal* if

1. no crown rule occurs below a trunk rule,
2. $\Box_L$ does not occur below any trunk rules other than $\Box_L$, and
3. $\Box_R$ does not occur below any trunk rules other than $\Box_L$ and $\Box_R$.

**Proposition 4.4** (Layering of derivations)**.** *If a grafted hypersequent $\mathcal{G}$ is derivable in $\mathcal{R}_{\mathsf{K5}}$ (resp. in $\mathcal{R}_{\mathsf{K5}}\mathsf{Cut_c}$) with a derivation of depth $n$, then it is derivable in $\mathcal{R}_{\mathsf{K5}}$ (resp. in $\mathcal{R}_{\mathsf{K5}}\mathsf{Cut_c}$) with a normal derivation of depth at most $n$.*

*Proof.* By permuting upwards topmost instances of rules violating the proposition using Lemma 4.1 and Lemma 4.2. □

Thus w.l.o.g. we may assume that all derivations are normal, i.e., layered in such a way that in every branch we have from top to bottom:

the crown layer (possibly with applications of $\mathsf{Cut_c}$),

applications of $\Box_L$,

applications of $\Box_R$,

the trunk layer.

Moreover, if a branch has no crown layer, it has only the trunk layer. This layering of the derivations enables us to eliminate the two cut rules $\mathsf{Cut}_\mathsf{t}$ and $\mathsf{Cut}_\mathsf{c}$ in two stages. First we show how to eliminate $\mathsf{Cut}_\mathsf{c}$ in the crown using techniques from the hypersequent framework. Then we essentially run a cut elimination proof from the nested sequent (or standard sequent) framework to eliminate $\mathsf{Cut}_\mathsf{t}$ in the trunk, reducing principal $\mathsf{Cut}_\mathsf{t}$-cuts on boxed formulae to $\mathsf{Cut}_\mathsf{c}$-cuts in the crown which are eliminated using the results of the first stage.

We first give a description of the procedure to eliminate $\mathsf{Cut}_\mathsf{c}$ in the crown. The method is based on the cut elimination proof for extensions of the fuzzy logic MTL with truth stresser modalities given in [CMM10] and generalised in [Lel14]. The method uses the following notion.

**Definition 4.5** (Cut rank)**.** Let $\mathcal{D}$ be a derivation in $\mathcal{R}_{\mathsf{K5}}\mathsf{Cut}$ or in $\mathcal{R}_{\mathsf{K5}}\mathsf{Cut}_\mathsf{c}$. The *cut rank* of $\mathcal{D}$ is the maximum over sizes of all cut formulae in $\mathcal{D}$ and is denoted by $\rho(\mathcal{D})$. If $\mathcal{D}$ is cut-free, we set $\rho(\mathcal{D}) := 0$.

The proof proceeds by first shifting a topmost $\mathsf{Cut}_\mathsf{c}$ on a cut formula with the largest size upwards into the *left* premiss using a generalised induction hypothesis that, similar to a one-sided version of multicut, one occurrence of the cut formula in the right premiss can be cut against several occurrences of the cut formula in the left premiss. Once this $\mathsf{Cut}_\mathsf{c}$ reaches the place where the cut formula is introduced in the left premiss, we start shifting the $\mathsf{Cut}_\mathsf{c}$ upwards into the *right* premiss, using another generalised induction hypothesis that one occurrence of the cut formula in the left premiss can be cut against several occurrences of the cut formula in the right premiss. This last step is captured in the following lemma, where, for a formula $A$ and a natural number $n > 0$, we write $A^n$ for $\underbrace{A, \dots, A}_{n \text{ times}}$.

**Lemma 4.6** (Shift Right)**.** *For positive natural numbers $n, m_1, \dots, m_n$, let $\mathcal{D}_L$ and $\mathcal{D}_R$ be derivations in $\mathcal{R}_{\mathsf{K5}}\mathsf{Cut}_\mathsf{c}$ of grafted hypersequents*

$$\Rightarrow || \mathcal{H}_L \mid \Gamma \Rightarrow \Delta, A \qquad \textit{and} \qquad \Rightarrow || \mathcal{H}_R \mid \Sigma_1, A^{m_1} \Rightarrow \Pi_1 \mid \cdots \mid \Sigma_n, A^{m_n} \Rightarrow \Pi_n$$

*respectively such that the last applied rule in $\mathcal{D}_L$ is not structural, the displayed occurrence of $A$ is principal in it, $\rho(\mathcal{D}_L) < |A|$, and $\rho(\mathcal{D}_R) < |A|$. Then there is a derivation $\mathcal{D}$ in $\mathcal{R}_{\mathsf{K5}}\mathsf{Cut}_\mathsf{c}$ of the grafted hypersequent*

$$\Rightarrow || \mathcal{H}_L \mid \mathcal{H}_R \mid \Gamma, \Sigma_1 \Rightarrow \Delta, \Pi_1 \mid \cdots \mid \Gamma, \Sigma_n \Rightarrow \Delta, \Pi_n \tag{4}$$

*such that $\rho(\mathcal{D}) < |A|$.*

*Proof.* The proof is by induction on the depth of the derivation $\mathcal{D}_R$, distinguishing cases based on the main connective of $A$ and on the last applied rule in $\mathcal{D}_R$. By by the same reasoning as used in Lemma 4.1 this rule must have been a crown rule.

Since the proof is very similar to the one given in [CMM10] and, apart from the (empty) trunk, is an instance of the general proof contained in [Lel14], here we only provide details for the cases where the formula $A$ is of the form $\Box B$. In particular, in these cases the last applied rule in $\mathcal{D}_L$ must have been $\mathsf{K}$ and, hence, both $\Gamma$ and $\Delta$ are empty. The remaining cases are handled similarly.

If the last applied rule in $\mathcal{D}_R$ was $\mathsf{Init}$ or $\bot_L$, then none of the occurrences of $\Box B$ is principal in it and (4) is another instance of the same rule.

If the last applied rule $\mathsf{r}$ in $\mathcal{D}_R$ was the rule $\mathsf{K}$, the rule $\mathsf{Cut}_\mathsf{c}$ with the cut formula simpler than $A$, or a propositional rule, then, since the rule $\mathsf{5}$ is the only non-structural rule introducing a boxed formula in an antecedent, again none of the occurrences of $\Box B$ are principal in this application. Thus, we first apply the induction hypothesis to the premiss(es) of $\mathsf{r}$ and then apply the same rule, followed by structural rules if necessary, to obtain (4). For instance, if the last applied rule in $\mathcal{D}_R$ was $\mathsf{K}$:

$$\dfrac{\begin{array}{c}\mathcal{D}'_R \\ \vdots \\ \Rightarrow || \mathcal{H}'_R \mid \Sigma_1, \Box B^{m_1} \Rightarrow \Pi_1 \mid \cdots \mid \Sigma_n, \Box B^{m_n} \Rightarrow \Pi_n \mid \Rightarrow C\end{array}}{\Rightarrow || \mathcal{H}'_R \mid \Sigma_1, \Box B^{m_1} \Rightarrow \Pi_1 \mid \cdots \mid \Sigma_n, \Box B^{m_n} \Rightarrow \Pi_n \mid \Rightarrow \Box C}\ \mathsf{K}$$

we would obtain

$$\dfrac{\dfrac{\begin{array}{c}\mathcal{D}_L\\ \vdots\\ \Rightarrow \,||\, \mathcal{H}_L \mid \,\Rightarrow \Box B\end{array} \qquad \begin{array}{c}\mathcal{D}'_R\\ \vdots\\ \Rightarrow \,||\, \mathcal{H}'_R \mid \Sigma_1, \Box B^{m_1} \Rightarrow \Pi_1 \mid \cdots \mid \Sigma_n, \Box B^{m_n} \Rightarrow \Pi_n \mid \,\Rightarrow C\end{array}}{\Rightarrow \,||\, \mathcal{H}_L \mid \mathcal{H}'_R \mid \Sigma_1 \Rightarrow \Pi_1 \mid \cdots \mid \Sigma_n \Rightarrow \Pi_n \mid \,\Rightarrow C}\ \mathsf{IH}}{\Rightarrow \,||\, \mathcal{H}_L \mid \mathcal{H}'_R \mid \Sigma_1 \Rightarrow \Pi_1 \mid \cdots \mid \Sigma_n \Rightarrow \Pi_n \mid \,\Rightarrow \Box C}\ \mathsf{K}$$

where IH signifies the application of the induction hypothesis, $\mathcal{D}'_R$ is $\mathcal{D}_R$ without the last application of the rule K, and $\mathcal{H}'_R$ is $\mathcal{H}_R$ without the component $\Rightarrow \Box C$.

The case where the last applied rule in $\mathcal{D}_R$ was the rule 5 or a structural rule with none of the displayed occurrences of $\Box B$ principal is similar. The case where the last applied rule was EC or $\mathsf{IC}_L$ with at least one of the displayed occurrences of $\Box B$ principal is taken care of by the induction hypothesis. The argument for the remaining structural rules is standard and left to the reader.

The case somewhat peculiar to our system is the one where the last applied rule in $\mathcal{D}_R$ was the rule 5 introducing one of the displayed $\Box B$. Then the derivation $\mathcal{D}_R$ ends with

$$\dfrac{\begin{array}{c}\mathcal{D}'_R\\ \vdots\\ \Rightarrow \,||\, \mathcal{H}_R \mid \Sigma_1, \Box B^{m_1} \Rightarrow \Pi_1 \mid \cdots \mid \Sigma_{n-2}, \Box B^{m_{n-2}} \Rightarrow \Pi_{n-2} \mid \Sigma_n, \Box B^{m_n}, B \Rightarrow \Pi_n\end{array}}{\Rightarrow \,||\, \mathcal{H}_R \mid \Sigma_1, \Box B^{m_1} \Rightarrow \Pi_1 \mid \cdots \mid \Sigma_{n-2}, \Box B^{m_{n-2}} \Rightarrow \Pi_{n-2} \mid \Box B \Rightarrow \mid \Sigma_n, \Box B^{m_n} \Rightarrow \Pi_n}\ \mathsf{5}$$

(Here $m_n$ can also be zero, making $\Sigma_n \Rightarrow \Pi_n$ a part of $\mathcal{H}_R$.) In the other premiss, since $\Box B$ was principal in the last rule application in $\mathcal{D}_L$, the latter ends with

$$\dfrac{\begin{array}{c}\mathcal{D}'_L\\ \vdots\\ \Rightarrow \,||\, \mathcal{H}_L \mid \,\Rightarrow B\end{array}}{\Rightarrow \,||\, \mathcal{H}_L \mid \,\Rightarrow \Box B}\ \mathsf{K}$$

In order to derive (4), which has the form

$$\Rightarrow \,||\, \mathcal{H}_L \mid \mathcal{H}_R \mid \Sigma_1 \Rightarrow \Pi_1 \mid \Sigma_2 \Rightarrow \Pi_2 \mid \cdots \mid \Sigma_{n-2} \Rightarrow \Pi_{n-2} \mid \,\Rightarrow\, \mid \Sigma_n \Rightarrow \Pi_n \ ,$$

we first eliminate those displayed instances of $\Box B$ that are contextual in the application of the rule 5 (if any) by performing a *cross cut*, i.e., by applying the induction hypothesis to the premiss of this application and the conclusion of $\mathcal{D}_L$. This yields a derivation of the grafted hypersequent

$$\Rightarrow \,||\, \mathcal{H}_L \mid \mathcal{H}_R \mid \Sigma_1 \Rightarrow \Pi_1 \mid \Sigma_2 \Rightarrow \Pi_2 \mid \cdots \mid \Sigma_{n-2} \Rightarrow \Pi_{n-2} \mid \Sigma_n, B \Rightarrow \Pi_n \ ,$$

To remove the extra $B$, we cut it against $\Rightarrow \,||\, \mathcal{H}_L \mid \,\Rightarrow B$ and remove possible duplicates of sequents in $\mathcal{H}_L$ caused by the cross cut by means of EC:

$$\Rightarrow \,||\, \mathcal{H}_L \mid \mathcal{H}_R \mid \Sigma_1 \Rightarrow \Pi_1 \mid \Sigma_2 \Rightarrow \Pi_2 \mid \cdots \mid \Sigma_{n-2} \Rightarrow \Pi_{n-2} \mid \Sigma_n \Rightarrow \Pi_n \ .$$

It remains to use EW to add the sequent $\Rightarrow$. Since the size of the new cut formula $B$ is smaller than the size of the original cut formula $\Box B$, this yields the desired derivation. □

The idea of shifting cuts up on the left is similarly captured in the proof of the following lemma, which also makes use of the Shift Right Lemma.

**Lemma 4.7** (Shift Left). *For positive natural numbers $n, m_1, \ldots, m_n$, let $\mathcal{D}_L$ and $\mathcal{D}_R$ be derivations in $\mathcal{R}_{\mathsf{K5}}\mathsf{Cut_c}$ of grafted hypersequents*

$$\Rightarrow \,||\, \mathcal{H}_L \mid \Gamma_1 \Rightarrow \Delta_1, A^{m_1} \mid \cdots \mid \Gamma_n \Rightarrow \Delta_n, A^{m_n} \qquad \text{and} \qquad \Rightarrow \,||\, \mathcal{H}_R \mid \Sigma, A \Rightarrow \Pi$$

*respectively such that $\rho(\mathcal{D}_L) < |A|$ and $\rho(\mathcal{D}_R) < |A|$. Then there is a derivation $\mathcal{D}$ in $\mathcal{R}_{\mathsf{K5}}\mathsf{Cut_c}$ of the grafted hypersequent*

$$\Rightarrow \,||\, \mathcal{H}_L \mid \mathcal{H}_R \mid \Gamma_1, \Sigma \Rightarrow \Delta_1, \Pi \mid \cdots \mid \Gamma_n, \Sigma \Rightarrow \Delta_n, \Pi$$

*such that $\rho(\mathcal{D}) < |A|$.*

*Proof.* The proof is by induction on the the depth of the derivation $\mathcal{D}_L$. Once again, all the rules applied in $\mathcal{D}_L$ must have been crown rules. Since the proof again is essentially the one from [CMM10, Lel14] and since all the cases where no displayed occurrence of the formula $A$ is principal in the last rule application in $\mathcal{D}_L$ are analogous to the proof of the Shift Right Lemma, we only give the case where $A$ is the formula $\Box B$ and one occurrence of it is principal in the last applied rule in $\mathcal{D}_L$. In this case the derivation $\mathcal{D}_L$ ends with

$$\dfrac{\begin{array}{c}\mathcal{D}'_L\\ \vdots \\ \Rightarrow \mid\mid \mathcal{H}_L \mid \Gamma_1 \Rightarrow \Delta_1, \Box B^{m_1} \mid \cdots \mid \Gamma_{n-1} \Rightarrow \Delta_{n-1}, \Box B^{m_{n-1}} \mid \Rightarrow B\end{array}}{\Rightarrow \mid\mid \mathcal{H}_L \mid \Gamma_1 \Rightarrow \Delta_1, \Box B^{m_1} \mid \cdots \mid \Gamma_{n-1} \Rightarrow \Delta_{n-1}, \Box B^{m_{n-1}} \mid \Rightarrow \Box B}\ \mathsf{K}$$

If $n = 1$, then we can directly apply the Shift Right Lemma 4.6. Otherwise, we first perform a cross cut by applying the induction hypothesis to the premiss of this rule $\mathsf{K}$, and then apply $\mathsf{K}$ to the result:

$$\dfrac{\dfrac{\begin{array}{c}\mathcal{D}'_L\\ \vdots \\ \Rightarrow \mid\mid \mathcal{H}_L \mid \Gamma_1 \Rightarrow \Delta_1, \Box B^{m_1} \mid \cdots \mid \Gamma_{n-1} \Rightarrow \Delta_{n-1}, \Box B^{m_{n-1}} \mid \Rightarrow B\end{array} \qquad \begin{array}{c}\mathcal{D}_R\\ \vdots \\ \Rightarrow \mid\mid \mathcal{H}_R \mid \Sigma, \Box B \Rightarrow \Pi\end{array}}{\Rightarrow \mid\mid \mathcal{H}_L \mid \mathcal{H}_R \mid \Gamma_1, \Sigma \Rightarrow \Delta_1, \Pi \mid \cdots \mid \Gamma_{n-1}, \Sigma \Rightarrow \Delta_{n-1}, \Pi \mid \Rightarrow B}\ \mathsf{IH}}{\Rightarrow \mid\mid \mathcal{H}_L \mid \mathcal{H}_R \mid \Gamma_1, \Sigma \Rightarrow \Delta_1, \Pi \mid \cdots \mid \Gamma_{n-1}, \Sigma \Rightarrow \Delta_{n-1}, \Pi \mid \Rightarrow \Box B}\ \mathsf{K}$$

Since now there is only one displayed occurrence of the formula $\Box B$ which is moreover principal in the last applied rule, we can use the Shift Right Lemma 4.6 for the cut formula $\Box B$, applied to the conclusions of this derivation and of $\mathcal{D}_R$, obtaining a derivation of

$$\Rightarrow \mid\mid \mathcal{H}_L \mid \mathcal{H}_R \mid \Gamma_1, \Sigma \Rightarrow \Delta_1, \Pi \mid \cdots \mid \Gamma_{n-1}, \Sigma \Rightarrow \Delta_{n-1}, \Pi \mid \mathcal{H}_R \mid \Sigma \Rightarrow \Pi\ .$$

It now only remains to remove duplicate sequents using $\mathsf{EC}$. $\Box$

As an immediate consequence we obtain the procedure to eliminate applications of the crown cut rule $\mathsf{Cut_c}$ by repeated applications of the Shift Left Lemma.

**Theorem 4.8** (Crown Cut Elimination). *The system $\mathcal{R}_{\mathsf{K5}}\mathsf{Cut_c}$ enjoys $\mathsf{Cut_c}$-elimination.*

*Proof.* A derivation $\mathcal{D}$ is turned into a cut-free derivation using a double induction on the cut rank $\rho(\mathcal{D})$ of $\mathcal{D}$ and on the number of applications of $\mathsf{Cut_c}$ with cut formulae of size $\rho(\mathcal{D})$. Each topmost application of $\mathsf{Cut_c}$ with a cut formula of size $\rho(\mathcal{D})$ can be reduced to cuts with cut formulae of smaller size by using the Shift Left Lemma 4.7, thereby either preserving the cut rank and decreasing the number of applications of $\mathsf{Cut_c}$ with cut formulae of size $\rho(\mathcal{D})$ or decreasing the cut rank. $\Box$

This gives us consistency of the crown rules of our calculus.

**Corollary 4.9.** $\mathcal{R}_{\mathsf{K5}}\mathsf{Cut_c} \nvdash \Rightarrow \mid\mid \Rightarrow$ *and* $\mathcal{R}_{\mathsf{K5}}\mathsf{Cut_c} \nvdash \Rightarrow$ .

*Proof.* Assume towards a contradiction that one of the above grafted hypersequents were derivable in $\mathcal{R}_{\mathsf{K5}}\mathsf{Cut_c}$. By the Crown Cut Elimination Theorem, it would then also be derivable in $\mathcal{R}_{\mathsf{K5}}$. It is easy to see, however, that $\Rightarrow$ cannot be a conclusion of any non-trivial rule of $\mathcal{R}_{\mathsf{K5}}$. Similarly, writing $[\Rightarrow]^n$ for $\overbrace{\Rightarrow \mid \cdots \mid \Rightarrow}^{n\text{-times}}$, the grafted hypersequent $\Rightarrow \mid\mid [\Rightarrow]^n$ with $n > 0$ can be obtained in $\mathcal{R}_{\mathsf{K5}}$ either by $\mathsf{EW}$ from $\Rightarrow \mid\mid [\Rightarrow]^{n-1}$ or by $\mathsf{EC}$ from $\Rightarrow \mid\mid [\Rightarrow]^{n+1}$. Since none of $\Rightarrow \mid\mid [\Rightarrow]^n$ for $n > 0$ are initial grafted hypersequents, $\Rightarrow \mid\mid \Rightarrow$ is not derivable either. $\Box$

In preparation for the trunk cut elimination proof we formulate a slightly more general version of crown cut elimination.

**Lemma 4.10.** *For positive natural numbers $n, m_1, \ldots, m_n$, if*

$$\mathcal{R}_{\mathsf{K5}} \vdash \Rightarrow \mid\mid \mathcal{H}_L \mid \Gamma_1 \Rightarrow \Delta_1, A^{m_1} \mid \cdots \mid \Gamma_n \Rightarrow \Delta_n, A^{m_n} \qquad \textit{and} \qquad \mathcal{R}_{\mathsf{K5}} \vdash \Rightarrow \mid\mid \mathcal{H}_R \mid \Sigma, A \Rightarrow \Pi\ ,$$

*then*

$$\mathcal{R}_{\mathsf{K5}} \vdash \Rightarrow \mid\mid \mathcal{H}_L \mid \mathcal{H}_R \mid \Gamma_1, \Sigma \Rightarrow \Delta_1, \Pi \mid \cdots \mid \Gamma_n, \Sigma \Rightarrow \Delta_n, \Pi\ .$$

*Proof.* Since the derivations of the two given grafted hypersequents are cut free, their cut rank is $0 < |A|$. Thus, applying the Shift Left Lemma 4.7 gives a derivation $\mathcal{D}$ in $\mathcal{R}_{\mathsf{K5}}\mathsf{Cut}_{\mathsf{c}}$ of the desired grafted hypersequent with $\rho(\mathcal{D}) < |A|$, and applying Theorem 4.8 yields a cut-free derivation. ☐

*Remark* 4.11. Since the rules of the system $\mathcal{R}_{\mathsf{K5}}$ considered here do not include any restrictions on the context, in this particular case it would also be possible to reverse the order of the Shift Lemmata in the proof of crown cut elimination, i.e., to first shift cuts upwards into the derivation of the right premiss, and then into that of the left premiss. However, to emphasise the connection to the hypersequent cut elimination proofs contained in [CMM10, Lel14] we chose to keep this order.

To eliminate cuts at the root level we now essentially run Gentzen's original reductive cut elimination proof for the sequent calculus [Gen34]. The proof eliminates applications of the *trunk multicut rule*

$$\frac{\Gamma \Rightarrow \Delta, A^n \mid\mid \mathcal{H}_L \qquad \Sigma, A^m \Rightarrow \Pi \mid\mid \mathcal{H}_R}{\Gamma, \Sigma \Rightarrow \Delta, \Pi \mid\mid \mathcal{H}_L \mid \mathcal{H}_R}\ \mathsf{MCut}_{\mathsf{t}}\ ,$$

which allows to cut several instances of the cut formula at the same time. It is clear that in presence of contraction the trunk multicut rule is derivable using a normal cut $\mathsf{Cut}_{\mathsf{t}}$. Moreover, since the rule $\mathsf{Cut}_{\mathsf{t}}$ is just an instance of the multicut rule $\mathsf{MCut}_{\mathsf{t}}$, it is clear that eliminating trunk multicuts is equivalent to eliminating normal trunk cuts.

**Theorem 4.12** (Trunk Cut Admissibility)**.** *For positive natural numbers $m$ and $n$, if*

$$\mathcal{R}_{\mathsf{K5}} \vdash \Gamma \Rightarrow \Delta, A^n \mid\mid \mathcal{H}_L \qquad \textit{and} \qquad \mathcal{R}_{\mathsf{K5}} \vdash \Sigma, A^m \Rightarrow \Pi \mid\mid \mathcal{H}_R\ ,$$

*then*

$$\mathcal{R}_{\mathsf{K5}} \vdash \Gamma, \Sigma \Rightarrow \Delta, \Pi \mid\mid \mathcal{H}_L \mid \mathcal{H}_R\ . \tag{5}$$

*In other words,* $\mathsf{MCut}_{\mathsf{t}}$ *is admissible in* $\mathcal{R}_{\mathsf{K5}}$.

*Proof.* Let $\mathcal{D}_L$ and $\mathcal{D}_R$ be derivations of $\Gamma \Rightarrow \Delta, A^n \mid\mid \mathcal{H}_L$ and $\mathcal{R}_{\mathsf{K5}} \vdash \Sigma, A^m \Rightarrow \Pi \mid\mid \mathcal{H}_R$ respectively. Using Proposition 4.4 we may assume that both derivations are normal, i.e., layered in such a way that all the trunk propositional and trunk structural rules occur below the transfer rules, which in turn occur below the crown rules. Since the trunk is not empty in the endsequents of both $\mathcal{D}_L$ and $\mathcal{D}_R$, it follows that the last rule applied in each of them was a trunk rule.

The proof is by double induction on $|A|$ (outer induction) and on the sum of depths of $\mathcal{D}_L$ and $\mathcal{D}_R$ (inner induction). If none of the displayed occurrences of $A$ in the conclusion of $\mathcal{D}_L$ is principal in the last rule $\mathsf{r}_L$ applied in $\mathcal{D}_L$ or if none of the displayed occurrences of $A$ in the conclusion of $\mathcal{D}_R$ is principal in the last rule $\mathsf{r}_R$ applied in $\mathcal{D}_R$, then the induction hypothesis can be applied to the premiss(es) of $\mathcal{D}_L$ and the conclusion of $\mathcal{D}_R$ or to the conclusion of $\mathcal{D}_L$ and the premiss(es) of $\mathcal{D}_R$ respectively, after which an application of $\mathsf{r}_L$ or $\mathsf{r}_R$ respectively yields (5).

If one of $\mathsf{r}_L$ or $\mathsf{r}_R$ is a structural rule with at least one of the displayed occurrences of $A$ being principal, the treatment is standard and left to the reader.

It remains to consider the cases when both $\mathsf{r}_L$ or $\mathsf{r}_R$ are logical rules or initial structures introducing one of the displayed occurrences of $A$. The cases when $A$ is a propositional variable or when the main connective in $A$ is Boolean are standard and left to the reader: we only provide details for the case of $A = \Box B$. In this case, $\mathsf{r}_L$ was an application of $\Box_R$ and $\mathsf{r}_R$ was an application of $\Box_L$. Since $\mathcal{D}_L$ and $\mathcal{D}_R$ are normal, there are only applications of $\Box_L$, $\Box_R$, and crown rules above $\mathsf{r}_L$ in $\mathcal{D}_L$ and there are only applications of $\Box_L$ and crown rules above $\mathsf{r}_R$ in $\mathcal{D}_R$.

If any of the $\Box_R$ applications above $\mathsf{r}_L$ introduce a formula other than one of the displayed occurrences of $A = \Box B$, it can be permuted all the way down towards the endsequent of $\mathcal{D}_L$ as described in the proof of Lemma 4.2 transforming $\mathcal{D}_L$ to $\mathcal{D}'_L$ of the same depth. Since then the last rule of the $\mathcal{D}'_L$ does not introduce any of the displayed $A$'s, we can apply considerations of the previous cases. Similarly, if any of the $\Box_L$ applications above $\mathsf{r}_R$ introduce a formula other than one of the displayed occurrences of $A = \Box B$, it also can be permuted all the way down towards the endsequent of $\mathcal{D}_R$ without affecting the depth of the derivation, and then apply the reasoning of the previous cases. Finally, if any of the $\Box_L$ applications above $\mathsf{r}_L$ does not affect the crown components active in one of the $\Box_R$ rules, it can be permuted all the way down towards the endsequent without changing the depth of the derivation, after which the previous considerations are applied. Since this permutation has not been described in our layering proposition, we present it here: the derivation

$$\begin{array}{cl}
\mathcal{D}_L^1 & \\
\vdots & \\
\Box\Gamma' \Rightarrow || \mathcal{H}'_L \mid \Lambda, C \Rightarrow \Upsilon \mid \Xi_1 \Rightarrow B \mid \cdots \mid \Xi_n \Rightarrow B & \\
\hline
\Box\Gamma', \Box C \Rightarrow || \mathcal{H}'_L \mid \Lambda \Rightarrow \Upsilon \mid \Xi_1 \Rightarrow B \mid \cdots \mid \Xi_n \Rightarrow B & \Box_L \\
\hline\hline
\Box\Gamma', \Box C, \Box\Xi_1, \ldots, \Box\Xi_n \Rightarrow || \mathcal{H}'_L \mid \Lambda \Rightarrow \Upsilon \mid [\Rightarrow B]^n & \Box_L \\
\hline\hline
\Box\Gamma', \Box C, \Box\Xi_1, \ldots, \Box\Xi_n \Rightarrow \Box B^n \; || \mathcal{H}'_L \mid \Lambda \Rightarrow \Upsilon & \Box_R
\end{array}$$

where $[\mathcal{H}]^\ell$ is an abbreviation for $\underbrace{\mathcal{H} \mid \cdots \mid \mathcal{H}}_{\ell \text{ times}}$, is replaced with

$$\begin{array}{cl}
\mathcal{D}_L^1 & \\
\vdots & \\
\Box\Gamma' \Rightarrow || \mathcal{H}'_L \mid \Lambda, C \Rightarrow \Upsilon \mid \Xi_1 \Rightarrow B \mid \cdots \mid \Xi_n \Rightarrow B & \\
\hline\hline
\Box\Gamma', \Box\Xi_1, \ldots, \Box\Xi_n \Rightarrow || \mathcal{H}'_L \mid \Lambda, C \Rightarrow \Upsilon \mid [\Rightarrow B]^n & \Box_L \\
\hline\hline
\Box\Gamma', \Box\Xi_1, \ldots, \Box\Xi_n \Rightarrow \Box B^n \; || \mathcal{H}'_L \mid \Lambda, C \Rightarrow \Upsilon & \Box_R \\
\hline
\Box\Gamma', \Box C, \Box\Xi_1, \ldots, \Box\Xi_n \Rightarrow \Box B^n \; || \mathcal{H}'_L \mid \Lambda \Rightarrow \Upsilon & \Box_L
\end{array}$$

Thus, we consider only the case when all the $\Box_R$ applications above $\mathsf{r}_L$ and all the $\Box_L$ applications above $\mathsf{r}_R$ introduce the displayed $A = \Box B$ and when all the $\Box_L$ applications above $\mathsf{r}_L$ add formulae to the crown components created by the $\Box_R$ rules (when looking upward). This means that $\Delta$, $\Sigma$, and $\Pi$ are empty, whereas $\Gamma = \Box\Xi_1, \ldots, \Box\Xi_n$ contains only formulae introduced by $\Box_L$ rule applications. Tracing the occurrences of the cut formula up to directly above the transfer layer we see that $\mathcal{D}_L$ and $\mathcal{D}_R$ have the forms

$$\begin{array}{cl}
\mathcal{D}_L^1 & \\
\vdots & \\
\Rightarrow || \mathcal{H}_L \mid \Xi_1 \Rightarrow B \mid \cdots \mid \Xi_n \Rightarrow B & \\
\hline\hline
\Box\Xi_1, \ldots, \Box\Xi_n \Rightarrow || \mathcal{H}_L \mid [\Rightarrow B]^n & \Box_L \\
\hline\hline
\Box\Xi_1, \ldots, \Box\Xi_n \Rightarrow \Box B^n \; || \mathcal{H}_L & \Box_R
\end{array}
\qquad
\begin{array}{cl}
\mathcal{D}_R^0 & \\
\vdots & \\
\Rightarrow || \mathcal{H}'_R \mid \Omega_1, B^{m_1} \Rightarrow \Theta_1 \mid \cdots \mid \Omega_k, B^{m_k} \Rightarrow \Theta_k & \\
\hline\hline
\Box B^m \Rightarrow || \mathcal{H}'_R \mid \Omega_1 \Rightarrow \Theta_1 \mid \cdots \mid \Omega_k \Rightarrow \Theta_k & \Box_L
\end{array}$$

respectively, where some of $\Xi_j$'s can be empty, $m = m_1 + \cdots + m_k$, and all $m_i > 0$.

Our goal in this case is to derive (5), which has the form

$$\Box\Xi_1, \ldots, \Box\Xi_n \Rightarrow || \mathcal{H}_L \mid \mathcal{H}'_R \mid \Omega_1 \Rightarrow \Theta_1 \mid \cdots \mid \Omega_k \Rightarrow \Theta_k$$

To achieve this goal we first contract the duplicate displayed occurrences of $B$ in each crown component of the endsequent of $\mathcal{D}_R^0$, obtaining a derivation $\mathcal{D}_R^1$ in $\mathcal{R}_{\mathsf{K5}}$ of the form

$$\begin{array}{cl}
\mathcal{D}_R^0 & \\
\vdots & \\
\Rightarrow || \mathcal{H}'_R \mid \Omega_1, B^{m_1} \Rightarrow \Theta_1 \mid \cdots \mid \Omega_k, B^{m_k} \Rightarrow \Theta_k & \\
\hline\hline
\Rightarrow || \mathcal{H}'_R \mid \Omega_1, B \Rightarrow \Theta_1 \mid \cdots \mid \Omega_k, B \Rightarrow \Theta_k & \mathsf{IC}_L
\end{array}$$

Thus,

$$\mathcal{R}_{\mathsf{K5}} \vdash \; \Rightarrow || \mathcal{H}_L \; || \; \Xi_1 \Rightarrow B \mid \cdots \mid \Xi_n \Rightarrow B \;\; , \tag{6}$$

$$\mathcal{R}_{\mathsf{K5}} \vdash \; \Rightarrow || \mathcal{H}'_R \mid \Omega_1, B \Rightarrow \Theta_1 \mid \cdots \mid \Omega_k, B \Rightarrow \Theta_k \;\; . \tag{7}$$

By Lemma 4.10 applied to (6) and the leftmost displayed occurrence of $B$ in (7)

$$\mathcal{R}_{\mathsf{K5}} \vdash \Rightarrow || \mathcal{H}_L \mid \mathcal{H}'_R \mid \Xi_1, \Omega_1 \Rightarrow \Theta_1 \mid \cdots \mid \Xi_n, \Omega_1 \Rightarrow \Theta_1 \mid \Omega_2, B \Rightarrow \Theta_2 \mid \cdots \mid \Omega_k, B \Rightarrow \Theta_k$$

Applying Lemma 4.10 $k-1$ more times, each time to (6) and the leftmost displayed occurrence of $B$ in the result of the previous application, we obtain

$$\mathcal{R}_{\mathsf{K5}} \vdash \Rightarrow || [\mathcal{H}_L]^k \mid \mathcal{H}'_R \mid \Xi_1, \Omega_1 \Rightarrow \Theta_1 \mid \cdots \mid \Xi_n, \Omega_1 \Rightarrow \Theta_1 \mid \cdots \mid \Xi_1, \Omega_k \Rightarrow \Theta_k \mid \cdots \mid \Xi_n, \Omega_k \Rightarrow \Theta_k$$

It is now easy to derive (5) in $\mathcal{R}_{\mathsf{K5}}$ as follows:

$$\dfrac{\dfrac{\dfrac{\Rightarrow || [\mathcal{H}_L]^k \mid \mathcal{H}'_R \mid \Xi_1, \Omega_1 \Rightarrow \Theta_1 \mid \cdots \mid \Xi_n, \Omega_1 \Rightarrow \Theta_1 \mid \cdots \mid \Xi_1, \Omega_k \Rightarrow \Theta_k \mid \cdots \mid \Xi_n, \Omega_k \Rightarrow \Theta_k}{(\Box\Xi_1, \ldots, \Box\Xi_n)^k \Rightarrow || [\mathcal{H}_L]^k \mid \mathcal{H}'_R \mid [\Omega_1 \Rightarrow \Theta_1 \mid \cdots \mid \Omega_k \Rightarrow \Theta_k]^n}\ \Box_L}{\Box\Xi_1, \ldots, \Box\Xi_n, \Rightarrow || [\mathcal{H}_L]^k \mid \mathcal{H}'_R \mid [\Omega_1 \Rightarrow \Theta_1 \mid \cdots \mid \Omega_k \Rightarrow \Theta_k]^n}\ \mathsf{C}_L}{\Box\Xi_1, \ldots, \Box\Xi_n \Rightarrow || \mathcal{H}_L \mid \mathcal{H}'_R \mid \Omega_1 \Rightarrow \Theta_1 \mid \cdots \mid \Omega_k \Rightarrow \Theta_k}\ \mathsf{EC}$$

where $\Gamma^\ell$ is an abbreviation of $\overbrace{\Gamma, \ldots, \Gamma}^{\ell \text{ times}}$. □

**Theorem 4.13** (Completeness of the Cut-Free System)**.** *The system $\mathcal{R}_{\mathsf{K5}}\mathsf{Cut}$ enjoys full cut elimination. In particular, every* K5 *theorem is derivable in $\mathcal{R}_{\mathsf{K5}}$.*

*Proof.* A derivation $\mathcal{D}$ is turned into a cut-free derivation using an induction on the combined number of applications of $\mathsf{Cut_c}$ and $\mathsf{Cut_t}$. Each topmost application of $\mathsf{Cut_c}$ (with no $\mathsf{Cut_t}$ above it either) can be eliminated by using Theorem 4.8 and each topmost application of $\mathsf{Cut_t}$ (with no $\mathsf{Cut_c}$ above it either) can be eliminated by using Theorem 4.12. □

As usual from the cut elimination theorem we obtain consistency of the logic. We can also use it to show that certain formulae such as the axioms (D), formulated as $\Box\bot \to \bot$ or (T) $\Box A \to A$ are not theorems of the logic.

**Corollary 4.14.** *None of the grafted hypersequents* $\Rightarrow$ , $\Rightarrow \Box\bot \to \bot$, *or* $\Rightarrow \Box p \to p$ *is derivable in* $\mathcal{R}_{\mathsf{K5}}\mathsf{Cut}$.

*Proof.* In Corollary 4.9, we already proved that $\Rightarrow$ is not derivable in $\mathcal{R}_{\mathsf{K5}}\mathsf{Cut_c}$. The statement now follows from the fact that $\mathcal{R}_{\mathsf{K5}} = \mathcal{R}_{\mathsf{K5}}\mathsf{Cut_c} = \mathcal{R}_{\mathsf{K5}}\mathsf{Cut}$.

To show that grafted hypersequents $\Rightarrow \Box A \to A$ are not derivable for $A$ being $\bot$ or a propositional variable, we assume the contrary and consider an arbitrary derivation of such a grafted hypersequent in $\mathcal{R}_{\mathsf{K5}}$, which exists by Theorem 4.13. Its endsequent can be obtained by W from $\Rightarrow$ , which is not derivable, or by $\mathsf{C}_R$ from $\Rightarrow \Box A \to A, \Box A \to A$, or by $\to_R$ from $\Box A \Rightarrow A$. Note that the rule $\Box_L$ is not applicable to $\Box A$ because the crown is empty. Continuing with this line of reasoning, only grafted hypersequents of the form

$$\underbrace{\Box A, \ldots, \Box A}_{n} \Rightarrow \underbrace{\Box A \to A, \ldots, \Box A \to A}_{k}, \underbrace{A, \ldots, A}_{l}$$

can occur in the derivation because the crown remains empty. Given that $A$ is $\bot$ or a propositional variable, no other rule can ever be applied and no initial structure can ever be reached because the unboxed $A$ never occurs in the antecedent of the trunk. □

# 5 Contraction, Decidability, Complexity

Now that we have established cut elimination for the calculus $\mathcal{R}_{\mathsf{K5}}$, our goal is to use this calculus in a decision procedure for the logic K5. The main challenge is to bring the complexity of this decision procedure down to the optimal complexity: While it is known by semantic arguments that the logic K5 is decidable in coNP (and in fact every extension of K5 is as well [HR07]), the (few) existing unlabelled sequent-style calculi used in decidability proofs for this logic either make use of analytic cuts [Ngu01, Tak01] or more complicated structures such as nested sequents [Brü09], resulting in a higher complexity.

The general idea for the decision procedure based on our grafted-sequent calculus for K5 is to employ *backwards proof search*: starting with a grafted hypersequent $\mathcal{G}$, check whether there is a rule which could have been applied last to derive $\mathcal{G}$ and recursively check that all its premisses are derivable. In

Figure 4: The rules $\mathcal{R}^*_{\mathsf{K5}}$ obtained by Kleene'ing the calculus $\mathcal{R}_{\mathsf{K5}}$

Some sample trunk and crown propositional rules:

$$\frac{\Gamma, A \to B \Rightarrow \Delta, A \mid\mid \mathcal{H} \qquad \Gamma, A \to B, B \Rightarrow \Delta \mid\mid \mathcal{H}}{\Gamma, A \to B \Rightarrow \Delta \mid\mid \mathcal{H}} \to^*_L \qquad \frac{\Gamma, A \Rightarrow \Delta, A \to B, B \mid\mid \mathcal{H}}{\Gamma \Rightarrow \Delta, A \to B \mid\mid \mathcal{H}} \to^*_R$$

$$\frac{\Rightarrow \mid\mid \mathcal{H} \mid \Gamma, A \to B \Rightarrow \Delta, A \qquad \Rightarrow \mid\mid \mathcal{H} \mid \Gamma, A \to B, B \Rightarrow \Delta}{\Rightarrow \mid\mid \mathcal{H} \mid \Gamma, A \to B \Rightarrow \Delta} \to^*_L \qquad \frac{\Rightarrow \mid\mid \mathcal{H} \mid \Gamma, A \Rightarrow \Delta, A \to B, B}{\Rightarrow \mid\mid \mathcal{H} \mid \Gamma \Rightarrow \Delta, A \to B} \to^*_R$$

The modal rules:

$$\frac{\Gamma \Rightarrow \Delta, \Box A \mid\mid \mathcal{H} \mid \Rightarrow A}{\Gamma \Rightarrow \Delta, \Box A \mid\mid \mathcal{H}} \Box^*_R \qquad \frac{\Gamma, \Box A \Rightarrow \Delta \mid\mid \mathcal{H} \mid \Sigma, A \Rightarrow \Pi}{\Gamma, \Box A \Rightarrow \Delta \mid\mid \mathcal{H} \mid \Sigma \Rightarrow \Pi} \Box^*_L$$

$$\frac{\Rightarrow \mid\mid \mathcal{H} \mid \Gamma, \Box A \Rightarrow \Delta \mid \Sigma, A \Rightarrow \Pi}{\Rightarrow \mid\mid \mathcal{H} \mid \Gamma, \Box A \Rightarrow \Delta \mid \Sigma \Rightarrow \Pi} \mathsf{5}^* \qquad \frac{\Rightarrow \mid\mid \mathcal{H} \mid \Gamma \Rightarrow \Delta, \Box A \mid \Rightarrow A}{\Rightarrow \mid\mid \mathcal{H} \mid \Gamma \Rightarrow \Delta, \Box A} \mathsf{K}^* \qquad \frac{\Rightarrow \mid\mid \mathcal{H} \mid \Gamma, \Box A, A \Rightarrow \Delta}{\Rightarrow \mid\mid \mathcal{H} \mid \Gamma, \Box A \Rightarrow \Delta} \mathsf{T}^*$$

terms of *alternating Turing machines* [CKS81] the step of checking whether there is a rule which could have been applied to derive $\mathcal{G}$ can be thought of as an *existential guessing* step, while checking that all the premisses of the rule application are derivable amounts to a *universal choosing* step. We show that we can fix the order of applications of rules, thereby eliminating the need for existential guessing steps, which leaves us only with universal choosing steps. This enables us to reduce the complexity from alternating polynomial time (or polynomial space) to the desired $\mathsf{coNP}$. Of course, in order to be able to handle the grafted hypersequents occurring in a derivation efficiently, we also need to show that their size is bounded. For this we need to be able to eliminate applications of the contraction rules, since these rules allow for a potentially unbounded increase in the size of the grafted hypersequent (when seen bottom-up). To show admissibility of the contraction rules we use *Kleene's Trick*, a method first introduced by Kleene in the construction of the G3-type sequent systems for classical and intuitionistic logic [Kle52]. The idea is to copy the relevant parts of the conclusion in a logical rule into the premisses, so that contractions in the conclusion of this rule can be permuted into its premisses. In order to prevent unnecessary blow-up of the structures we omit components of the hypersequent part which can be derived from other components using internal weakenings. Finally, to deal with external contractions involving both principal components of the 5 rule, we add the missing rule. Perhaps not surprisingly this turns out to be the crown version of the standard T rule for reflexive modal logics. This procedure is analogous to adding missing rules to a rule set by internally contracting formulae in the premisses and conclusion so that it satisfies the *closure condition* of [NP01, Neg05] respectively the *contraction closure* condition of [Lel14].

**Definition 5.1** (Modified rules)**.** The modified rules implementing Kleene's trick are given in Figure 4. The rule set including these rules together with initial structures Init and $\bot_L$ both in the trunk and in the crown, as well as the trunk weakening rule W, is called $\mathcal{R}^*_{\mathsf{K5}}$.

Again, in the rules of $\mathcal{R}^*_{\mathsf{K5}}$ we call all the formulae in the $\Gamma, \Delta, \Sigma, \Pi$ and in the components in $\mathcal{H}$ the *contextual formulae*, we call all non-contextual formulae in the conclusion the *principal formulae*, and all non-contextual formulae in the premisses the *active formulae*. In particular the copies of the principal formulae in the premisses are active formulae. The notions of *contextual*, *principal*, and *active* components are as for the system $\mathcal{R}_{\mathsf{K5}}$. Note that the rule set $\mathcal{R}^*_{\mathsf{K5}}$ includes neither the contraction rules EC, $\mathsf{IC}_L$, $\mathsf{IC}_R$, $\mathsf{C}_L$, or $\mathsf{C}_R$ nor the weakening rules EW or IW. The trunk weakening rule W, however, is necessary since the crown rules can only be applied with the empty trunk. It would also be possible to add two new transfer rules analogous to $\Box^*_R$ and $\Box^*_L$ but with empty trunk in the premisses to obtain a system where the trunk weakening rule W is admissible as well. However, in view of the fact that trunk weakening in general is unproblematic, we prefer the system with one structural rule instead of two new logical rules. The resulting crown rules then essentially strike a middle ground between the modal rules of the modified hypersequent system for S5 used in the semantic completeness proof in [Res07] and those given in [Pog10, Chapter 9]: they contain the Kleene'd version of the $\Box_R$ rule as in the former (which is replaced with the un-Kleene'd version in the latter) and use the $\mathsf{T}^*$ rule of the latter (which, in presence of external structural rules, is superfluous in the former). The $\Box_L^*$ rule is present in both calculi. Since we have omitted components in the crown which can be derived from other components using internal weakening $\mathsf{IW}$, we need to show depth-preserving admissibility of $\mathsf{IW}$ before we can show admissibility of contraction.

**Lemma 5.2** (Admissibility of internal and external weakening)**.** *The rules* $\mathsf{IW}$ *and* $\mathsf{EW}$ *of crown internal weakening and external weakening are depth-preserving admissible in* $\mathcal{R}^*_{\mathsf{K5}}$.

*Proof.* Standard by induction on the depth of the derivation. The additional context in the rules $\mathsf{5}^*$ and $\mathsf{K}^*$ ensures that the application of internal weakening can be shifted into the premiss. □

**Lemma 5.3** (Admissibility of contraction)**.** *The rule* $\mathsf{EC}$ *of external contraction, as well as the rules* $\mathsf{IC}_L$, $\mathsf{IC}_R$ $\mathsf{C}_L$, *and* $\mathsf{C}_R$ *of internal trunk and crown contraction, is depth-preserving admissible in* $\mathcal{R}^*_{\mathsf{K5}}$.

*Proof.* The admissibility of internal crown and trunk contraction is shown as usual by an induction on the depth of the derivation. For instance, to contract the principal formula in the conclusion of an application of the rule $\mathsf{5}^*$ shown below left, we apply the the induction hypothesis to the premiss of this application (without increasing the depth) and then use another application of the rule $\mathsf{5}^*$ as shown below right:

$$\dfrac{\begin{array}{c}\vdots\\ \Rightarrow \| \mathcal{H} \mid \Gamma, \Box A, \Box A \Rightarrow \Delta \mid \Sigma, A \Rightarrow \Pi\end{array}}{\Rightarrow \| \mathcal{H} \mid \Gamma, \Box A, \Box A \Rightarrow \Delta \mid \Sigma \Rightarrow \Pi}\ \mathsf{5}^* \qquad \dfrac{\dfrac{\begin{array}{c}\vdots\\ \Rightarrow \| \mathcal{H} \mid \Gamma, \Box A, \Box A \Rightarrow \Delta \mid \Sigma, A \Rightarrow \Pi\end{array}}{\Rightarrow \| \mathcal{H} \mid \Gamma, \Box A \Rightarrow \Delta \mid \Sigma, A \Rightarrow \Pi}\ \mathsf{IH}}{\Rightarrow \| \mathcal{H} \mid \Gamma, \Box A \Rightarrow \Delta \mid \Sigma \Rightarrow \Pi}\ \mathsf{5}^*$$

Similarly, the admissibility of external contraction is shown by induction on the depth of the derivation, where the additional copy of the principal component in the premiss of the rules $\mathsf{5}^*$ and $\mathsf{K}^*$ ensures that a contraction involving this component and a context component can be shifted into the premiss of the rule. If the last applied rule was a crown propositional rule, $\mathsf{5}^*$, $\Box_L^*$, or $\mathsf{T}^*$, to contract its principal component with a context component we need to use Lemma 5.2. For instance, to contract the principal component in the conclusion of an application of $\mathsf{T}^*$

$$\dfrac{\begin{array}{c}\vdots\\ \Rightarrow \| \mathcal{H} \mid \Gamma, \Box A \Rightarrow \Delta \mid \Gamma, \Box A, A \Rightarrow \Delta\end{array}}{\Rightarrow \| \mathcal{H} \mid \Gamma, \Box A \Rightarrow \Delta \mid \Gamma, \Box A \Rightarrow \Delta}\ \mathsf{T}^*$$

by Lemma 5.2 we apply the internal crown weakening (without increasing the depth), then use the induction hypothesis (again without increasing the depth), and then use another application of $\mathsf{T}^*$:

$$\dfrac{\dfrac{\dfrac{\begin{array}{c}\vdots\\ \Rightarrow \| \mathcal{H} \mid \Gamma, \Box A \Rightarrow \Delta \mid \Gamma, \Box A, A \Rightarrow \Delta\end{array}}{\Rightarrow \| \mathcal{H} \mid \Gamma, \Box A, A \Rightarrow \Delta \mid \Gamma, \Box A, A \Rightarrow \Delta}\ \text{adm. } \mathsf{IW}}{\Rightarrow \| \mathcal{H} \mid \Gamma, \Box A, A \Rightarrow \Delta}\ \mathsf{IH}}{\Rightarrow \| \mathcal{H} \mid \Gamma, \Box A \Rightarrow \Delta}\ \mathsf{T}^* \tag{8}$$

The most non-trivial case is when the two principal components of $\mathsf{5}^*$ are contracted. This is treated as follows, illustrating why the addition of the rule $\mathsf{T}^*$ to the system was necessary:

$$\dfrac{\begin{array}{c}\vdots\\ \Rightarrow \| \mathcal{H} \mid \Gamma, \Box A \Rightarrow \Delta \mid \Gamma, \Box A, A \Rightarrow \Delta\end{array}}{\Rightarrow \| \mathcal{H} \mid \Gamma, \Box A \Rightarrow \Delta \mid \Gamma, \Box A \Rightarrow \Delta}\ \mathsf{5}^* \qquad \dfrac{\dfrac{\dfrac{\begin{array}{c}\vdots\\ \Rightarrow \| \mathcal{H} \mid \Gamma, \Box A \Rightarrow \Delta \mid \Gamma, \Box A, A \Rightarrow \Delta\end{array}}{\Rightarrow \| \mathcal{H} \mid \Gamma, \Box A, A \Rightarrow \Delta \mid \Gamma, \Box A, A \Rightarrow \Delta}\ \text{adm. } \mathsf{IW}}{\Rightarrow \| \mathcal{H} \mid \Gamma, \Box A, A \Rightarrow \Delta}\ \mathsf{IH}}{\Rightarrow \| \mathcal{H} \mid \Gamma, \Box A \Rightarrow \Delta}\ \mathsf{T}^*$$

□

Now we can use the previous two lemmata to establish equivalence of the modified calculus with the original one.

**Theorem 5.4** (Equivalence of $\mathcal{R}_{\mathsf{K5}}$ and $\mathcal{R}^*_{\mathsf{K5}}$)**.** *For every grafted hypersequent $\mathcal{G}$ we have that $\mathcal{G}$ is derivable in $\mathcal{R}_{\mathsf{K5}}$ iff it is derivable in $\mathcal{R}^*_{\mathsf{K5}}$.*

*Proof.* First we show that every rule of $\mathcal{R}_{\mathsf{K5}}$ is admissible in $\mathcal{R}^*_{\mathsf{K5}}$. For the missing structural rules this has been shown in Lemma 5.2 and Lemma 5.3. For all the crown and trunk propositional rules, as well as for the transfer rules, of $\mathcal{R}_{\mathsf{K5}}$, to use the corresponding rule of $\mathcal{R}^*_{\mathsf{K5}}$, it is sufficient to use $\mathsf{W}$ or the admissible $\mathsf{IW}$ to add the principal formula to the premiss(es). The crown modal rules $\mathsf{5}$ and $\mathsf{K}$ are translated as

$$\dfrac{\dfrac{\Rightarrow \| \mathcal{H} \mid \Sigma, A \Rightarrow \Pi}{\Rightarrow \| \mathcal{H} \mid \Box A \Rightarrow \mid \Sigma, A \Rightarrow \Pi}\ \text{adm. } \mathsf{EW}}{\Rightarrow \| \mathcal{H} \mid \Box A \Rightarrow \mid \Sigma \Rightarrow \Pi}\ \mathsf{5}^* \qquad \text{and} \qquad \dfrac{\dfrac{\Rightarrow \| \mathcal{H} \mid \Rightarrow A}{\Rightarrow \| \mathcal{H} \mid \Rightarrow \Box A \mid \Rightarrow A}\ \text{adm. } \mathsf{EW}}{\Rightarrow \| \mathcal{H} \mid \Rightarrow \Box A}\ \mathsf{K}^*$$

respectively using the admissible external weakening $\mathsf{EW}$.

Now we show that every rule of $\mathcal{R}^*_{\mathsf{K5}}$ is derivable in $\mathcal{R}_{\mathsf{K5}}$. For all the trunk and crown propositional rules, as well as for the transfer rules, of $\mathcal{R}^*_{\mathsf{K5}}$, first the corresponding rule of $\mathcal{R}_{\mathsf{K5}}$ is used and then the duplicate of the principal formula is contracted by $\mathsf{C}_L$, $\mathsf{C}_R$, $\mathsf{IC}_L$, or $\mathsf{IC}_R$. The crown modal rules $\mathsf{5}^*$ and $\mathsf{K}^*$ are translated as

$$\dfrac{\dfrac{\dfrac{\Rightarrow \| \mathcal{H} \mid \Gamma, \Box A \Rightarrow \Delta \mid \Sigma, A \Rightarrow \Pi}{\Rightarrow \| \mathcal{H} \mid \Gamma, \Box A \Rightarrow \Delta \mid \Box A \Rightarrow \mid \Sigma \Rightarrow \Pi}\ \mathsf{5}}{\Rightarrow \| \mathcal{H} \mid \Gamma, \Box A \Rightarrow \Delta \mid \Gamma, \Box A \Rightarrow \Delta \mid \Sigma \Rightarrow \Pi}\ \mathsf{IW}}{\Rightarrow \| \mathcal{H} \mid \Gamma, \Box A \Rightarrow \Delta \mid \Sigma \Rightarrow \Pi}\ \mathsf{EC} \qquad \text{and} \qquad \dfrac{\dfrac{\dfrac{\Rightarrow \| \mathcal{H} \mid \Gamma \Rightarrow \Delta, \Box A \mid \Rightarrow A}{\Rightarrow \| \mathcal{H} \mid \Gamma \Rightarrow \Delta, \Box A \mid \Rightarrow \Box A}\ \mathsf{K}}{\Rightarrow \| \mathcal{H} \mid \Gamma \Rightarrow \Delta, \Box A \mid \Gamma \Rightarrow \Delta, \Box A}\ \mathsf{IW}}{\Rightarrow \| \mathcal{H} \mid \Gamma \Rightarrow \Delta, \Box A}\ \mathsf{EC}$$

respectively. Finally, the rule $\mathsf{T}^*$ is derived as follows:

$$\dfrac{\dfrac{\dfrac{\Rightarrow \| \mathcal{H} \mid \Gamma, \Box A, A \Rightarrow \Delta}{\Rightarrow \| \mathcal{H} \mid \Box A \Rightarrow \mid \Gamma, \Box A \Rightarrow \Delta}\ \mathsf{5}}{\Rightarrow \| \mathcal{H} \mid \Gamma, \Box A \Rightarrow \Delta \mid \Gamma, \Box A \Rightarrow \Delta}\ \mathsf{IW}}{\Rightarrow \| \mathcal{H} \mid \Gamma, \Box A \Rightarrow \Delta}\ \mathsf{EC}$$

□

The fact that the contraction rules are admissible in $\mathcal{R}^*_{\mathsf{K5}}$ allows us to restrict the grafted hypersequents in the backwards proof search procedure to structures based on sets instead of multisets.

**Definition 5.5** (Set-based structures)**.** A *set-based* sequent is a pair $\Gamma \Rightarrow \Delta$ of sets $\Gamma, \Delta$ of formulae. A *set-based hypersequent* is a set of set-based sequents. A *set-based grafted hypersequent* $\Gamma \Rightarrow \Delta \| \mathcal{H}$ is a set-based sequent $\Gamma \Rightarrow \Delta$ together with a set-based hypersequent $\mathcal{H}$.

The rules of $\mathcal{R}^*_{\mathsf{K5}}$ apply to set-based grafted hypersequents as usual, reading set union $\cup$ for the comma , and for the hypersequent bar $|$.

*Remark* 5.6. In the set-based setting, the $\mathsf{T}^*$ rule becomes an instance of the $\mathsf{5}^*$ rule.

To abbreviate notation we introduce the notion of subsumption of one set-based grafted hypersequent by another. The idea is that one such structure is subsumed by another if each of the components in the crown are components in the crown of the other or derived from such components using weakening, and similarly for the trunk. Formally:

**Definition 5.7** (Subsumption)**.** A set-based grafted hypersequent $\Gamma \Rightarrow \Delta \| \mathcal{H}$ is *subsumed by* another set-based grafted hypersequent $\Sigma \Rightarrow \Pi \| \mathcal{H}'$ if

- $\Gamma \subseteq \Sigma$ and $\Delta \subseteq \Pi$
- for every $\Omega \Rightarrow \Theta \in \mathcal{H}$ there is a $\Omega' \Rightarrow \Theta' \in \mathcal{H}'$ such that $\Omega \subseteq \Omega'$ and $\Theta \subseteq \Theta'$.

We then also write $\Gamma \Rightarrow \Delta \| \mathcal{H} \subseteq \Sigma \Rightarrow \Pi \| \mathcal{H}'$.

Thus if a grafted hypersequent $\mathcal{G}$ is subsumed by another grafted hypersequent $\mathcal{G}'$, then it is possible to derive $\mathcal{G}'$ from $\mathcal{G}$ using only the structural rules, i.e., the different forms of weakening.

**Theorem 5.8** (Decidability and complexity)**.** *The backwards proof search algorithm for* $\mathsf{K5}$ *given as Algorithm 1 decides membership in* $\mathsf{K5}$ *and can be implemented in* $\mathsf{coNP}$.

**Algorithm 1:** Decision procedure for $\mathsf{K5}$

**Input**: a set-based grafted hypersequent $\Gamma \Rightarrow \Delta \mid\mid \mathcal{H}$

**Output**: Is $\iota(\Gamma \Rightarrow \Delta \mid\mid \mathcal{H}) \in \mathsf{K5}$?

set $\Gamma_1 := \Gamma,\ \Delta_1 := \Delta,\ \mathcal{H}_1 := \mathcal{H}$;<br>
**repeat**<br>
    set $\Gamma_2 := \Gamma_1,\ \Delta_2 := \Delta_1,\ \mathcal{H}_2 := \mathcal{H}_1$;<br>
    apply modified propositional trunk rules backwards to each unprocessed trunk formula in $\Gamma_1 \Rightarrow \Delta_1 \mid\mid \mathcal{H}_1$, universally choosing one of the premisses for branching rules, and label these formulae processed ;<br>
**until** $\Gamma_1 \Rightarrow \Delta_1 \mid\mid \mathcal{H}_1 \subseteq \Gamma_2 \Rightarrow \Delta_2 \mid\mid \mathcal{H}_2$;<br>
**if** $\Gamma_1 \Rightarrow \Delta_1 \mid\mid \mathcal{H}_1$ *is a trunk initial structure* **then**<br>
    halt and accept;<br>
**end**<br>
apply $\Box_R^*$ backwards to each formula $\Box A \in \Delta_1$ in $\Gamma_1 \Rightarrow \Delta_1 \mid\mid \mathcal{H}_1$ such that it is not the case that $\Rightarrow \mid\mid \Rightarrow A \subseteq \Gamma_1 \Rightarrow \Delta_1 \mid\mid \mathcal{H}_1$;<br>
apply $\Box_L^*$ backwards to each $\Box A \in \Gamma_1$ and each component of $\mathcal{H}_1$ in $\Gamma_1 \Rightarrow \Delta_1 \mid\mid \mathcal{H}_1$;<br>
apply $\mathsf{W}$ backwards to $\Gamma_1 \Rightarrow \Delta_1 \mid\mid \mathcal{H}_1$ to obtain $\Rightarrow \mid\mid \mathcal{H}_1$;<br>
**repeat**<br>
    set $\mathcal{H}_2 := \mathcal{H}_1$;<br>
    apply modified propositional crown rules backwards to each unprocessed crown formula in $\Rightarrow \mid\mid \mathcal{H}_1$, universally choosing one of the premisses for branching rules, and label these formulae processed ;<br>
    apply $\mathsf{K}^*$ backwards to each consequent formula $\Box A$ in $\Rightarrow \mid\mid \mathcal{H}_1$ such that it is not the case that $\Rightarrow \mid\mid \Rightarrow A \subseteq \Rightarrow \mid\mid \mathcal{H}_1$;<br>
    apply $\mathsf{5}^*$ backwards to each antecedent formula $\Box A$ and each component of $\Rightarrow \mid\mid \mathcal{H}_1$ (including the component with $\Box A$ itself);<br>
    **if** $\Rightarrow \mid\mid \mathcal{H}_1$ *is a crown initial structure* **then**<br>
        halt and accept;<br>
    **end**<br>
**until** $\Rightarrow \mid\mid \mathcal{H}_1 \subseteq \Rightarrow \mid\mid \mathcal{H}_2$;<br>
halt and reject

*Proof.* It is easy to see that all the rules of $\mathcal{R}^*_{\mathsf{K5}}$, except for $\mathsf{W}$, are invertible because they can be read backwards as weakenings, which we proved to be admissible. It is also easy to see that all the possible trunk rules *are* applied before the application of $\mathsf{W}$ in Line 11, that no crown rules *can be* applied before Line 11, and that all the possible crown rules are applied after Line 11. (Here we take into account the fact that no branch of a shortest derivation can visit a grafted hypersequent subsumed by a prior grafted hypersequent from this branch because of depth-preserving admissibility of weakening rules.) Thus, the correctness of the algorithm follows from the completeness of $\mathcal{R}_{\mathsf{K5}}$ (Theorem 4.13) and the equivalence of $\mathcal{R}_{\mathsf{K5}}$ and $\mathcal{R}^*_{\mathsf{K5}}$ (Theorem 5.4).

To see that the complexity of the procedure is indeed $\mathsf{coNP}$, consider a run of the procedure with input $\mathcal{G}$. Given that the algorithm does not use any existential guessing, it is sufficient to demonstrate that each universal-choice branch requires polynomial time. Since all the rules in $\mathcal{R}^*_{\mathsf{K5}}$ have the subformula property, every set-based grafted hypersequent occurring in a derivation of the given grafted hypersequent $\mathcal{G}$ contains only subformulae of formulae occurring in $\mathcal{G}$. If the size of $\mathcal{G}$ is $n$, there are at most $n$ such subformulae and, thus, every set-based sequent containing only subformulae of $\mathcal{G}$ contains at most $2n$ formulae. Since all the trunk propositional rules do not decrease the number of formulae occurring in the trunk, since the **repeat**-loop that starts in Line 2 is terminated after no new formulae are added, and since each formula in the trunk enjoys at most one propositional rule application before being labelled processed, there are at most $2n$ applications of the trunk propositional rules in Line 4 and indeed in the whole **repeat**-loop starting in Line 2. Since the rule $\Box^*_R$ introduces a new crown component from a consequent formula occurring in the trunk, there are at most $n$ applications of $\Box^*_R$ in Line 9 that introduce at most $n$ new components, making the total number of sequents in the crown at most $2n$. Thus, for each of at most $n$ antecedent boxed formula in the trunk the rule $\Box^*_L$ is used at most $2n$ times as described in Line 10, and there are at most $2n^2$ applications of $\Box^*_L$ overall. The only steps that create new crown components within the **repeat**-loop that starts in Line 12 are the applications of $\mathsf{K}^*$ in Line 15. Since for each boxed subformula of a formula occurring in $\mathcal{G}$ at most one crown component would be created by either $\Box^*_R$ or $\mathsf{K}^*$, we can evaluate the total number of applications of $\Box^*_R$ in Line 9 and $\mathsf{K}^*$ in Line 15 as at most $n$ and the total number of components in the crown at any point in the running of the algorithm as at most $2n$. Thus, there are at most $4n^2$ formulas that can occur in the crown in various components: $2n^2$ antecedent formulas and $2n^2$ consequent ones. This means that the propositional crown rules are applied at most $4n^2$ times in Line 14 and that the rule $\mathsf{5}^*$ is applied at most $4n^3$ times: to a combination of each of at most $2n^2$ antecedent formulas with each of at most $2n$ components.

To summarize, the total number of rule instances applied for each branch of universal choices is $\mathcal{O}(n^3)$ steps and it is easy to see that each rule instance can be processed in polynomial time. The whole procedure can, thus, be implemented on a polynomially bounded alternating Turing machine which makes only universal choosing steps. Therefore, the problem of deciding whether a given set-based grafted hypersequent is derivable in $\mathcal{R}^*_{\mathsf{K5}}$ is in the complexity class $A\Pi^p_1 = \mathsf{coNP}$ [CKS81]. □

# 6 Semantic Completeness via Simplified Prefixed Tableaux for K5

In this section, we provide an alternative semantic proof of completeness for our grafted hypersequents. The benefit of such a proof lies in its simplicity and in the alternative representation of grafted hypersequents in the form of simplified prefixed tableaux, which may be of independent interest.

Much like grafted hypersequents for K5 and KD5 are a generalisation of hypersequents for S5, which are known to correspond to the simplified prefixed tableaux for S5 (see, e.g., [FM98, p. 54]), the prefixed tableaux corresponding to grafted hypersequents can be obtained by generalising these simplified prefixed tableaux. For the lack of better terms, we would call the resulting system *grafted tableaux.*

Unlike the rest of the paper, we do not intend this section to be self-contained. It is, after all, an alternative proof of our main result, which can be skipped at will. The reader is expected to know standard prefixed tableaux for modal logics in general and the simplified system for S5 in particular. All the necessary information can be found in the highly readable [FM98], whose terminology and notation we adopt in the following.

However, unlike [FM98], we employ *signed formulae*, i.e., modal formulae signed either T or F. Semantically, formulae signed T are thought of as *true* and those signed F as *false*. Syntactically, T-signed formula correspond to antecedent formulae, whereas F-signed formulae are consequent formulae.

Figure 5: Grafted tableau K5-rules. Here $\ell$ is any prefix, $\mathsf{c}$ and $\mathsf{c}'$ are crown prefixes, $\mathbb{n}$ is a limb prefix, and $n$ is a twig prefix.

$$\frac{\ell : \mathsf{F}\, A \vee B}{\ell : \mathsf{F}\, A \quad \ell : \mathsf{F}\, B} \qquad \frac{\ell : \mathsf{T}\, A \vee B}{\ell : \mathsf{T}\, A \mid \ell : \mathsf{T}\, B} \qquad \frac{\ell : \mathsf{F}\, A \wedge B}{\ell : \mathsf{F}\, A \mid \ell : \mathsf{F}\, B} \qquad \frac{\ell : \mathsf{T}\, A \wedge B}{\ell : \mathsf{T}\, A \quad \ell : \mathsf{T}\, B} \qquad \frac{\ell : \mathsf{F}\, A \to B}{\ell : \mathsf{T}\, A \quad \ell : \mathsf{F}\, B} \qquad \frac{\ell : \mathsf{T}\, A \to B}{\ell : \mathsf{F}\, A \mid \ell : \mathsf{T}\, B}$$

$$\frac{\bullet : \mathsf{F}\, \Box A}{\mathbb{n} : \mathsf{F}\, A}\ \mathbb{n} \text{ new} \qquad \frac{\bullet : \mathsf{T}\, \Box A}{\mathbb{n} : \mathsf{T}\, A}\ \mathbb{n} \text{ occurs} \qquad \frac{\mathsf{c} : \mathsf{F}\, \Box A}{n : \mathsf{F}\, A}\ n \text{ new} \qquad \frac{\mathsf{c} : \mathsf{T}\, \Box A}{\mathsf{c}' : \mathsf{T}\, A}\ \mathsf{c}' \text{ occurs}$$

The natural question of whose antecedent or consequent, is addressed by the prefix. However, to make the tableaux formulation more natural as a stand-alone system, we depart a little bit from the grafted hypersequents as follows. There, we had two types of components, the unique trunk and the components of the crown. The absence of a weakening rule in the prefixed tableaux format suggests to further partition the crown into the lower part, called *limbs* and upper part called *twigs*. Accordingly, we define three types of prefixes:

**Definition 6.1** (Prefixes)**.** Grafted tableaux contain three types of prefixes: the unique *trunk prefix* $\bullet$, *limb prefixes* denoted by thick natural numbers $\mathbb{0}, \mathbb{1}, \mathbb{2}$, etc., and *twig prefixes* denoted by thin natural numbers $0, 1, 2$, etc. To avoid ambiguity, within one tableau proof, each number is used either as a limb or as twig prefix, but not as both, i.e., we never use $\mathbb{n}$ and $n$ for the same number $n$ within one proof. We use $\ell$ to denote an arbitrary prefix and $\mathsf{c}$ or $\mathsf{c}'$ to denote an arbitrary *crown prefix*, i.e., any prefix other than the trunk prefix.

To distinguish boxed formulae with the two signs $\mathsf{F}$ and $\mathsf{T}$, following the standard terminology, we call a signed formula of the form $\ell : \mathsf{F}\,\Box A$ a *possibility formula*, one of the form $\ell : \mathsf{T}\,\Box A$ a *necessity formula*.

**Definition 6.2** (Grafted tableau system for K5)**.** The rules of the *grafted tableau calculus for* K5 can be found in Figure 5. The *conjunctive* and *disjunctive rules* in the first row consist of the standard propositional rules applied to an arbitrary prefix. The last rule in the second row is exactly the S5 necessity rule from [FM98, Def. 2.3.7] (it is restricted to crown prefixes only because the trunk prefix is not used for S5) and is called the *crown necessity rule*. The S5 possibility rule from [FM98, Def. 2.3.6] is modified in that the new prefix created from a crown prefix must be a twig prefix; it is the third rule in the second row and is called the *crown possibility rule*. Finally, two more rules are added: the *trunk necessity rule* that propagates the unboxed part of a true boxed formula in the trunk to the limbs (but not to the twigs) and the *trunk possibility rule* that creates a new limb prefix from a false boxed formula in the trunk.

A *grafted* K5*-tableau* for $S$ is any object constructed from an initial finite set $S$ of prefixed signed formulae not containing twig prefixes by means of grafted tableau K5-rules. (The restriction to limb crown prefixes in initiating sets ensures that twig prefixes can only appear in the presence of at least one limb prefix.)

The *closure conditions for a branch* are standard: for some label $\ell$ the branch must either contain both $\ell : \mathsf{T}\, A$ and $\ell : \mathsf{F}\, A$ or contain $\ell : \mathsf{T}\,\bot$. A tableau is *closed* when all its branches are closed. A *grafted tableau proof* of a formula $A$ is a closed tableau for $\{\bullet : \mathsf{F}\, A\}$. As usual, a branch is *open* if it is not closed, and analogously a tableau is *open* if it is not closed, i.e., if it contains an open branch.

**Example 6.3.** As an illustration, Example *a* below shows a grafted K5-tableau proof of the shifted transitivity axiom $\Box(\Box B \to \Box\Box B)$. The open grafted K5-tableau in Example *b* below shows that the transitivity axiom $\mathsf{4}$ itself, $\Box B \to \Box\Box B$, is not generally valid in $\mathsf{K5}$.[3]

| Example $a$ | | Example $b$ | |
|---|---|---|---|
| $\bullet : \mathsf{F}\,\Box(\Box B \to \Box\Box B)$ | $a.1.$ | | |
| $\mathbb{1} : \mathsf{F}\,\Box B \to \Box\Box B$ | $a.2.$ | $\bullet : \mathsf{F}\,\Box p \to \Box\Box p$ | $b.1.$ |
| $\mathbb{1} : \mathsf{T}\,\Box B$ | $a.3.$ | $\bullet : \mathsf{T}\,\Box p$ | $b.2.$ |
| $\mathbb{1} : \mathsf{F}\,\Box\Box B$ | $a.4.$ | $\bullet : \mathsf{F}\,\Box\Box p$ | $b.3.$ |
| $2 : \mathsf{F}\,\Box B$ | $a.5.$ | $\mathbb{2} : \mathsf{F}\,\Box p$ | $b.4.$ |
| $3 : \mathsf{F}\,B$ | $a.6.$ | $3 : \mathsf{F}\,p$ | $b.5.$ |
| $2 : \mathsf{T}\,B$ | $a.7.$ | $\mathbb{2} : \mathsf{T}\,p$ | $b.6.$ |
| $3 : \mathsf{T}\,B$ | $a.8.$ | | |

In Example $a$, formula $a.2.$ is from $a.1.$ by the trunk possibility rule, creating the new limb prefix $\mathbb{1}$. Formulae $a.3.$ and $a.4.$ are from $a.2.$ by a propositional rule. Formula $a.5.$ is from $a.4.$ by the crown possibility rule, creating the new twig prefix 2. Similarly, formula $a.6.$ is from $a.5.$ by the same rule, creating the twig prefix 3. Formulae $a.7.$ and $a.8.$ are from $a.3.$ by the crown necessity rule. The only branch in the tableau is closed by $a.6.$ and $a.8.$

Example $b$ is essentially the same, but now the step corresponding to the crucial inference yielding line $a.8.$ is not possible anymore, since the twig prefix 3, which could be accessed from the limb prefix $\mathbb{1}$ in Example $a$, cannot be accessed from the trunk prefix $\bullet$ here.

Before showing how our hypersequent crown from the previous sections separates into the lower limbs and upper twigs, we prove completeness of this grafted tableaux system (and, thereby, cut admissibility for it) with respect to $\mathsf{K5}$. The simplicity of the argument is the main point of this section. We start with soundness.

**Definition 6.4** ($\mathsf{K5}$-Satisfiability)**.** A set $S$ of prefixed signed formulae is $\mathsf{K5}$-*satisfiable* if there exists a Euclidean model $(W, R, \sigma)$ and a mapping $\theta$ from the set of prefixes of $S \cup \{\bullet\}$ to $W$ such that:

1. for each limb prefix $\mathbb{n}$ occurring in $S$, we have $\theta(\bullet)R\theta(\mathbb{n})$;
2. for arbitrary crown prefixes $\mathsf{c}$ and $\mathsf{c}'$ occurring in $S$ (possibly identical), we have $\theta(\mathsf{c})R\theta(\mathsf{c}')$;
3. for each $\ell : \mathsf{T}\,A \in S$, we have $(W, R, \sigma), \theta(\ell) \Vdash A$;
4. for each $\ell : \mathsf{F}\,A \in S$, we have $(W, R, \sigma), \theta(\ell) \nVdash A$.

As always, a grafted tableau branch is $\mathsf{K5}$-*satisfiable* if the set of prefixed signed formulae occurring on this branch is satisfiable. A grafted tableau is $\mathsf{K5}$-*satisfiable* if one of its branches is.

It is quite clear that a closed grafted tableau is not satisfiable. The proof of this fact is trivial rather than standard. The main lemma necessary for soundness is

**Lemma 6.5** (Soundness lemma)**.** *Any* $\mathsf{K5}$ *rule applied to a* $\mathsf{K5}$*-satisfiable grafted* $\mathsf{K5}$*-tableau yields a* $\mathsf{K5}$*-satisfiable grafted* $\mathsf{K5}$*-tableau.*

*Proof.* We only consider the non-propositional rules from the second row of Figure 5. Let $\mathcal{B}$ be a $\mathsf{K5}$-satisfiable branch of a given grafted $\mathsf{K5}$-tableau.

For the trunk necessity rule, assume that $\bullet : \mathsf{T}\,\Box A$ is present on $\mathcal{B}$, that $\mathbb{n}$ is a limb prefix occurring on $\mathcal{B}$, and that $\theta$ witnesses the satisfiability of $\mathcal{B}$. In particular, $\theta$ must be defined on both $\bullet$ and $\mathbb{n}$. Then $(W, R, \sigma), \theta(\bullet) \Vdash \Box A$ and $\theta(\bullet)R\theta(\mathbb{n})$. Thus, $(W, R, \sigma), \theta(\mathbb{n}) \Vdash A$, meaning that adding $\mathbb{n} : \mathsf{T}\,A$ to $\mathcal{B}$ does not spoil its satisfiability. The case of the crown necessitation rule is analogous.

For the trunk possibility rule, assume that $\bullet : \mathsf{F}\,\Box A$ is present on $\mathcal{B}$, that $\mathbb{n}$ is a new limb prefix for $\mathcal{B}$, and that $\theta$ witnesses the satisfiability of $\mathcal{B}$. In particular, $\theta$ must be defined on $\bullet$ but not on $\mathbb{n}$. Then $(W, R, \sigma), \theta(\bullet) \nVdash \Box A$. Thus, there must exist a world $w \in W$ such that $(W, R, \sigma), w \nVdash A$ and $\theta(\bullet)Rw$. It is easy to show using Euclideanity of $R$ that $\theta(\mathsf{c})Rw$ and $wR\theta(\mathsf{c})$ for each crown prefix $\mathsf{c}$ occurring on $\mathcal{B}$ and that $wRw$ (one should remember that twig prefixes cannot appear on a branch that has no limb prefixes). Thus, $\theta' := \theta \cup \{(\mathbb{n}, w)\}$ witnesses the satisfiability of $\mathcal{B}$ extended with $\mathbb{n} : \mathsf{F}\,A$. The case of the crown possibility rule is even simpler since the requirement that the new world be accessible from $\theta(\bullet)$ is not applicable anymore. $\Box$

[3] Since there exist formulae $B$ making $\Box B \to \Box\Box B$ valid, we disprove it for the case of an atomic $B = p$.

**Corollary 6.6** (Tableau soundness)**.** *If $A$ is not* K5*-valid, no grafted* K5*-tableau for* $\bullet : \mathsf{F}\, A$ *can close.*

*Proof.* It is sufficient to show that the tableau initiated with the single formula $\bullet : \mathsf{F}\, A$ is satisfiable. Since $A$ is not valid, there must exist a Euclidean model $(W, R, \sigma)$ and a world $w \in W$ such that $(W, R, \sigma), w \not\Vdash A$. It remains to note that $\theta := \{(\bullet, w)\}$ witnesses the satisfiability of this tableau because no conditions on the accessibility are imposed in the absence of crown prefixes. □

As is often the case with modal tableaux, it is generally impossible to truly complete a tableau. For instance, in Example 6.3*b*, it is still possible to create new limb prefixes out of *b*.3. and new twig prefixes out of *b*.4. Hence, we use the weaker notion of saturation:

**Definition 6.7** (Saturation)**.** A grafted K5-tableau is called K5-*saturated* if, for every branch that is not closed, the following conditions are fulfilled:

1. if a prefixed signed formula other than a possibility or necessity formula occurs on the branch, the applicable rule has been applied to it on the branch;
2. if a trunk necessity formula occurs on the branch, the trunk necessity rule has been applied to it on the branch for each limb prefix $\mathbb{n}$ that occurs on the branch;
3. if a crown necessity formula occurs on the branch, the crown necessity rule has been applied to it on the branch for each crown prefix $\mathsf{c}$ that occurs on the branch;
4. if a trunk possibility formula $\bullet : \mathsf{F}\, \Box A$ occurs on the branch, $\mathbb{n} : \mathsf{F}\, A$ occurs on the branch for some limb prefix $\mathbb{n}$;
5. if a crown possibility formula $\mathsf{c} : \mathsf{F}\, \Box A$ occurs on the branch, $n : \mathsf{F}\, A$ occurs on the branch for some twig prefix $n$;

It is easy to see that the grafted K5-tableau in Example 6.3*b* is K5-saturated.

**Lemma 6.8** (Saturation termination)**.** *For any grafted* K5*-tableau, any run of the following non-deterministic systematic procedure terminates and leads to a* K5*-saturated grafted* K5*-tableau.*

**Systematic procedure.** Non-deterministically pick an open branch and a formula from this branch that violates the K5-saturation conditions on the branch and apply the rule (or, in case of $\ell : \mathsf{T}\, \Box A$ formulae, several necessity rules) to ensure that this formula does not violate the K5-saturation conditions on this branch anymore.

*Proof.* The proof is quite standard and is omitted here. The argument used on the sequent side in our Algorithm 1 for grafted hypersequents from the previous section is essentially a variant of this systematic procedure with a more restrictive order of rule applications. □

The key lemma, from which completeness is a simple corollary, is the following:

**Lemma 6.9** (Completeness lemma)**.** *In any open* K5*-saturated grafted* K5*-tableau, at least one branch is* K5*-satisfiable.*

*Proof.* An open K5-saturated grafted K5-tableau must contain an open branch, which must satisfy all the K5-saturation conditions. The model and the witnessing function are constructed out of this branch as follows. The set of possible worlds $W$ is the set of prefixes occurring on this branch, so the witnessing function is simply the identity function on $W$. The accessibility relation is defined as follows:

$$R := \{(\bullet, \mathbb{n}) \mid \mathbb{n} \text{ occurs on the branch}\} \cup \{(\mathsf{c}, \mathsf{c}') \mid \mathsf{c} \text{ and } \mathsf{c}' \text{ occur on the branch}\} \ .$$

Finally, we define the valuation $\sigma$ by

$$\ell \in \sigma(p) \qquad \Longleftrightarrow \qquad \ell : \mathsf{T}\, p \text{ is present on the branch}$$

for each propositional variable $p \in \mathcal{V}$ and any prefix $\ell$ occurring on the branch. It is a simple exercise to check that this model is Euclidean and that the identity function on $W$ witnesses the satisfiability of the branch. □

**Corollary 6.10** (Tableau completeness)**.** *If $A$ is* $\mathsf{K5}$*-valid, it has a grafted tableau proof in the system for* $\mathsf{K5}$*.*

*Proof.* By Lemma 6.8, the single-node tableau $\bullet : \mathsf{F}\, A$ can be saturated by using the systematic procedure. Were the resulting tableau open, by Lemma 6.9 it would have contained a $\mathsf{K5}$-satisfiable branch. Since every branch begins at the root $\bullet : \mathsf{F}\, A$, in particular, we would have $(W, R, \sigma), \theta(\bullet) \nVdash A$ for some Euclidean model $(W, R, \sigma)$, contradicting the $\mathsf{K5}$-validity of $A$. Thus, the resulting saturated tableau is closed and is a grafted tableau proof of $A$. □

As usual, this semantic completeness proof for the simplified prefixed tableaux calculus is characteristically simpler than the syntactic completeness proof for the grafted hypersequent calculus by cut-elimination. For the reader who is not interested in the intricacies of the cut elimination proof, we now provide bidirectional translations between grafted hypersequents and grafted tableaux to make this simpler proof a shortcut for proving cut admissibility for the grafted hypersequent system. More precisely, we give the translation between grafted tableaux and the Kleene'd version of the grafted hypersequent calculus $\mathcal{R}^*_{\mathsf{K5}}$ from Figure 4, shown to be equivalent to $\mathcal{R}_{\mathsf{K5}}$ in Theorem 5.4, whereby all sequent components are treated as set based.

Before we delve into the technical details of the bidirectional translation, it might make sense to explain how grafted tableaux differ from grafted hypersequents and talk about the pitfall they both are trying to avoid, each by means more natural to their area. The pitfall is transitivity. We have already observed in Remark 3.1 that the rule $\mathsf{K}$ is not generally sound if the trunk is allowed to contain formulae. Allowing that would amount to postulating transitivity of the modal logic. However, unlike destructive tableaux, prefixed tableaux do not have tools for removing trunk formulae from the branch the way the weakening rule $\mathsf{W}$ does on the sequent side (seen bottom-up).

Not willing to mix the two types of tableaux, we opted for introducing the third type of sequent components/prefixes instead. This is again supported by the semantic intuition since all the non-root worlds in a connected rooted $\mathsf{K5}$ model are divided into worlds directly accessible from it, and worlds accessible from it in no less than two steps. We call the former limbs as they are closer to the trunk and the latter twigs because they are growing out of limbs or other twigs but not out of the trunk. Note that all the non-root worlds in this connected model, limbs and twigs alike, still form a totally-connected cluster, within which transitivity is satisfied. Thus, the only world that may violate transitivity is the root. In other words, the only situation we need to prevent is when a necessity formula from the trunk directly affects a twig, i.e., a world created out of another crown world rather than out of the trunk.

On the sequent side, this is achieved by forcing an order on the rules: no crown rule is applied and, hence, no twig is created, while there are formulae, in particular, necessity formulae, in the trunk. The benefit of this approach is in minimising the number of structural connectives used to distinguish the types of sequent components. On the tableau side, we do this explicitly, allowing trunk necessity formulas to affect limbs but not twigs. This approach enables us to dispense with the (somewhat ad hoc) ordering requirements by letting the different types of prefixes take care of the structural distinctions.

**Theorem 6.11** (Equivalence)**.** *A grafted hypersequent*

$$\Gamma \Rightarrow \Delta \mid\mid \Sigma_1 \Rightarrow \Pi_1 \mid \ldots \mid \Sigma_n \Rightarrow \Pi_n$$

*is derivable in $\mathcal{R}^*_{\mathsf{K5}}$ (with set-based sequent components) iff there is a closed grafted* $\mathsf{K5}$*-tableau for*

$$\{\bullet : \mathsf{T}\, A \mid A \in \Gamma\} \cup \{\bullet : \mathsf{F}\, B \mid B \in \Delta\} \cup \bigcup_{j=1}^{n} \Big( \{\mathring{\mathbb{1}}_j : \mathsf{T}\, C \mid C \in \Sigma_i\} \cup \{\mathring{\mathbb{1}}_j : \mathsf{F}\, D \mid D \in \Pi_i\} \Big) \ , \qquad (9)$$

*where $\mathring{\mathbb{1}}_1 < \cdots < \mathring{\mathbb{1}}_n$ for $n \geq 0$.*

*Proof (sketch):* In this sketch, we rely on the standard view of tableaux as upside-down refutations of sequents, which we do no flesh out in detail.

**From sequents to tableaux.** From a derivation of the grafted hypersequent $\Gamma \Rightarrow \Delta \mid\mid \Sigma_1 \Rightarrow \Pi_1 \mid \ldots \mid \Sigma_n \Rightarrow \Pi_n$ in $\mathcal{R}^*_{\mathsf{K5}}$ we construct a closed grafted $\mathsf{K5}$-tableau for (9) by mimicking the applied grafted hypersequent rules (from conclusion to initial sequents) by grafted tableaux rules. It is clear from the comparison of rules in Figure 4 with those in Figure 5 that each application of one of these grafted hypersequent rules can be mirrored by the corresponding grafted tableaux rule, with branching rules corresponding to branching rules. The only rule of $\mathcal{R}^*_{\mathsf{K5}}$ without a tableau analog is $\mathsf{W}$ (note that it is depicted in Figure 1 rather than Figure 4). We provide the table of rule correspondences in Figure 6.

Figure 6: The correspondence between Kleene'd grafted non-propositional hypersequent rules and grafted tableau rules

| $\mathcal{R}^*_{\mathsf{K5}}$ | $\mathsf{K5}$ tableau |
|---|---|
| $\Box^*_L$ | trunk necessity |
| $\Box^*_R$ | trunk possibility |
| $\mathsf{W}$ | — |
| $\mathsf{5}^*$ | crown necessity, $\mathsf{c} \neq \mathsf{c}'$ |
| $\mathsf{T}^*$ | crown necessity, $\mathsf{c} = \mathsf{c}'$ |
| $\mathsf{K}^*$ | crown possibility |

By the same argument as in the proof of Lemma 4.1, it is clear that in any derivation in $\mathcal{R}^*_{\mathsf{K5}}$, the order of rule applications along each branch is as follows: trunk rules, then $\mathsf{W}$, then crown rules. Given the absence of twig prefixes in (9) and the fact that new twig prefixes are only created by the crown possibility rule, i.e., after all $\Box^*_L$ rules along the branch have already been translated, there is no mismatch between the scopes of the $\mathsf{5}^*$ rule applicable to any crown component and the trunk necessity rule applicable only to limb prefixes. Indeed, when $\Box^*_L$ rules are translated into tableau rules, limb prefixes are the only crown prefixes present.

Since each branch of a grafted hypersequent proof ends with an initial structure, each branch in the corresponding grafted tableau is closed, making it a grafted tableau proof.

Note that in the process of translating the grafted hypersequent derivation into a grafted tableau proof in general also new twig labels will be introduced, while they are not allowed to occur in (9). This is not a problem, however, since (9) corresponds only to the conclusion of a grafted hypersequent derivation, while the twig labels introduced in the tableau correspond to additional crown components introduced in the grafted hypersequent derivation.

**From tableaux to sequents.** Due to the layering of sequent derivations, not every tableau proof can be translated directly. We first need to permute trunk rules to precede crown rules in the construction of the tableau. Assume that there is a closed grafted $\mathsf{K5}$-tableau $\mathcal{T}$ for a set $S$ of prefixed signed formulae that contains no twig prefixes. We use Lemma 6.8 to construct a $\mathsf{K5}$-saturated tableau $\mathcal{T}'$ for the same set $S$ using the systematic procedure in such a way that first all formulae with the prefix $\bullet$ that violate the saturation conditions are processed, and only when no such formulae remain, do we start to process formulae with crown prefixes. Since any systematic procedure terminates, there must be a moment in the construction of $\mathcal{T}'$ when all $\bullet$-formulae are saturated. While trunk propositional formulae and trunk possibility formulae are saturated once and for all, trunk necessity formulae could potentially become unsaturated again if a new limb prefix was introduced. For that to happen, however, one would need a new trunk possibility formula, which could only be produced by a trunk propositional rule, and all such rules have already been applied previously. Thus, the tableau $\mathcal{T}'$ indeed is a saturated tableau for $S$ with trunk rules preceding crown rules along each branch. Let us call such tableaux *normal*.

We now argue that this saturated tableau $\mathcal{T}'$ must also be closed. Indeed, were it open, it would have been satisfiable by Lemma 6.9. In particular, the set $S$ would have been satisfiable. But a satisfiable initial tableau for $S$ then could not have lead to the non-satisfiable closed tableau $\mathcal{T}$ for $S$ we started with. This contradiction with Lemma 6.5 shows that our saturated normal tableau $\mathcal{T}'$ is only saturated because it is closed. This is not a problem, however, since it is a closed normal tableau we needed to be able to translate into sequents.

The translation then works in the reverse direction to the sequent-to-tableaux translation, with the same correspondence between rules, only used in the opposite direction. Along each branch, at the boundary between the trunk and crown tableau rules, the corresponding grafted hypersequent derivation has a $\mathsf{W}$ rule removing all trunk formulae, thus, giving green light to use the crown hypersequent rules corresponding to crown tableau rules now that the trunk is empty. □

Given the (usual) close relationship between the sequent-style and tableau-style systems, we strongly believe that the decidability proof and complexity estimates could have been equally well established using the grafted tableau system. Some would argue that the tableau environment is, in fact, more suitable for the task. However, we make it a point to show that optimal complexity can be achieved by means of sequent-like calculi, which, admittedly, are suboptimal in terms of compactness of notation.

# 7 Extensions and Modifications

So far we restricted our investigations to a grafted hypersequent calculus for the logic $\mathsf{K5}$. However, the framework of grafted hypersequents is more general than that and allows to capture other logics as well. The idea of the calculus for $\mathsf{K5}$ was to have rules from a nested sequent setting governing the behaviour of the trunk of a grafted hypersequent and rules from a hypersequent setting governing the behaviour of the crown. In the case of $\mathsf{K5}$ the trunk rules corresponded to the standard modal logic $\mathsf{K}$, while the crown rules were those of $\mathsf{S5}$, in close analogy with the semantic intuition that $\mathsf{K5}$ is the logic of frames where all successor states of a state are part of an $\mathsf{S5}$-subframe, i.e., are part of a clique. Thus, to vary the calculus we have two main options: varying its trunk rules or its crown rules. As an example for the first option, we consider the logic $\mathsf{KD5}$, followed by an example for the second option in the form of a calculus for the logic of shift-reflexive frames.

## 7.1 A Calculus for KD5

The logic $\mathsf{KD5}$ extends the logic $\mathsf{K5}$ with the additional axiom $(\mathsf{D}) = \Diamond\top \equiv_{\mathsf{K5}} \Box\bot \to \bot = \neg\Box\bot$ and is the logic of Euclidean and *serial* frames, i.e., of Euclidean frames additionally satisfying the frame property $\forall x\,\exists y\; xRy$. As mentioned in the introduction, constructing a grafted hypersequent calculus for this logic is of independent interest, since together with $\mathsf{K5}$ it is one of the logics captured in the nested sequent framework (yielding a decision procedure of suboptimal complexity) for which so far no calculus in a simpler framework, such as that of hypersequents, could be found. In order to construct a grafted hypersequent calculus for this logic we need to add rules to the calculus $\mathcal{R}_{\mathsf{K5}}$ to ensure that $(\mathsf{D})$ is derivable both in the trunk and in the crown (the latter is necessary in the completeness proof for future use of $\mathsf{nec}$, see the proof of Theorem 3.6). However, the following derivation shows, that without adding any extra rules we already have $\mathcal{R}_{\mathsf{K5}} \vdash\; \Rightarrow \,||\, \Rightarrow \neg\Box\bot$:

$$\dfrac{\dfrac{\dfrac{\dfrac{\dfrac{}{\Rightarrow \,||\, \bot \Rightarrow}\;\bot_L}{\Rightarrow \,||\, \Box\bot \Rightarrow \,|\, \Rightarrow}\;\mathsf{5}}{\Rightarrow \,||\, \Box\bot \Rightarrow \,|\, \Box\bot \Rightarrow}\;\mathsf{IW}}{\Rightarrow \,||\, \Box\bot \Rightarrow}\;\mathsf{EC}}{\Rightarrow \,||\, \Rightarrow \neg\Box\bot}\;\neg_R$$

Thus, it suffices to add another transfer rule that allows to derive the axiom $(\mathsf{D})$ in the trunk. For this we borrow the following rule $\Box^{\mathsf{D}}_L$ from the nested sequent setting [Pog09, Pog10]:

$$\dfrac{\Gamma \Rightarrow \Delta \,||\, \mathcal{H} \,|\, A \Rightarrow}{\Gamma, \Box A \Rightarrow \Delta \,||\, \mathcal{H}}\;\Box^{\mathsf{D}}_L$$

The calculi for the logic $\mathsf{KD5}$ are defined as

$$\mathcal{R}_{\mathsf{KD5}} := \mathcal{R}_{\mathsf{K5}} \cup \{\Box^{\mathsf{D}}_L\} \qquad \text{and} \qquad \mathcal{R}_{\mathsf{KD5}}\mathsf{Cut} := \mathcal{R}_{\mathsf{K5}}\mathsf{Cut} \cup \{\Box^{\mathsf{D}}_L\}\;.$$

Then by extending the soundness and completeness arguments of Section 3 in a straightforward way we obtain:

**Theorem 7.1** (Soundness and Completeness)**.** *For every formula $A$ we have:*

$$\mathcal{R}_{\mathsf{KD5}} \vdash\; \Rightarrow A \qquad\Longrightarrow\qquad A \in \mathsf{KD5} \qquad\Longrightarrow\qquad \mathcal{R}_{\mathsf{KD5}}\mathsf{Cut} \vdash\; \Rightarrow A$$

*Proof.* Soundness of all the rules apart from $\Box^{\mathsf{D}}_L$ is shown as in the proof of Proposition 3.3. For the rule $\Box^{\mathsf{D}}_L$, assume that the negation of the formula interpretation of its conclusion is satisfiable, i.e., that writing $H$ for $\iota(\;\Rightarrow \,||\, \mathcal{H})$ the formula

$$\neg\iota(\Gamma, \Box A \Rightarrow \Delta \,||\, \mathcal{H}) \quad \equiv_{\mathsf{KD5}} \quad \bigwedge \Gamma \wedge \Box A \wedge \neg \bigvee \Delta \wedge \neg H$$

holds in a serial and Euclidean model $(W, R, \sigma)$ at world $w$, where $A \equiv_{\mathsf{KD5}} B$ naturally means that $(A \to B) \wedge (B \to A) \in \mathsf{KD5}$. Since $R$ is serial there is a world $v \in W$ with $wRv$. Further, since $\Box A$ holds at $w$, we have that $A$ holds at $v$ and $\Diamond A$ holds at $w$. Therefore, the formula

$$\neg\iota(\Gamma \Rightarrow \Delta \mid\mid \mathcal{H} \mid A \Rightarrow\ ) \quad \equiv_{\mathsf{KD5}} \quad \bigwedge \Gamma \wedge \neg \bigvee \Delta \wedge \neg H \wedge \Diamond A$$

holds at world $w$.

For completeness, as we have seen, the axiom $(\mathsf{D})$ is derivable on the hypersequent level. It is also derivable in the trunk as follows:

$$\dfrac{\dfrac{\dfrac{}{\Rightarrow \mid\mid \bot \Rightarrow}\ \bot_L}{\Box\bot \Rightarrow}\ \Box_L^{\mathsf{D}}}{\Rightarrow \neg\Box\bot}\ \neg_R$$

Since all the remaining axioms are derivable and $\mathsf{MP}$ and $\mathsf{nec}$ preserve derivability as shown in the proof of Theorem 3.6, completeness follows. □

The next step is to show cut elimination for $\mathcal{R}_{\mathsf{KD5}}$ via an extension of the arguments for $\mathcal{R}_{\mathsf{K5}}$.

**Theorem 7.2** (Cut elimination for $\mathcal{R}_{\mathsf{KD5}}$)**.** *The rules* $\mathsf{Cut_t}$ *and* $\mathsf{Cut_c}$ *are admissible in* $\mathcal{R}_{\mathsf{KD5}}$.

*Proof.* It is easy to show that the new rule $\Box_L^{\mathsf{D}}$ can be permuted over all other trunk rules except for the rule $\Box_L$, and over other instances of $\Box_L^{\mathsf{D}}$ rules. It is also easy to show that both $\Box_R$ and $\Box_L$ can be permuted over the new rule $\Box_L^{\mathsf{D}}$. This entails the following layering of the derivations in $\mathcal{R}_{\mathsf{KD5}} \cup \{\mathsf{Cut_c}\}$:

the crown layer (possibly with applications of $\mathsf{Cut_c}$),

applications of $\Box_L$,

applications of $\Box_R$ and $\Box_L^{\mathsf{D}}$,

the trunk layer.

The analogue of Definition 4.3 for *normal* derivations, thus, instead of Clause 3 has the clause

3′. $\Box_R$ and $\Box_L^{\mathsf{D}}$ do not occur below any trunk rules other than $\Box_L$, $\Box_R$, and $\Box_L^{\mathsf{D}}$

and the analogue of Proposition 4.4 is shown straightforwardly. Since the crown rules of $\mathcal{R}_{\mathsf{KD5}}$ are exactly those of $\mathcal{R}_{\mathsf{K5}}$, it is also clear that the proofs of Shift Right Lemma 4.6 and of Shift Left Lemma 4.7, as well as of cut elimination for $\mathsf{Cut_c}$ (Theorem 4.8), of the non-derivability of $\Rightarrow \mid\mid \Rightarrow$ and $\Rightarrow$ (Corollary 4.9), and of the generalized $\mathsf{Cut_c}$-elimination (Lemma 4.10) go through as before.

To show the admissibility of the trunk cut rule $\mathsf{Cut_t}$ we need to extend the argument in the proof of Theorem 4.13. Using the depth-preserving permutability of $\Box_R$ and $\Box_L^{\mathsf{D}}$ over each other, we assume that $\mathcal{D}_L$ and $\mathcal{D}_R$ are structured from top to bottom in different ways:

$\mathcal{D}_L$: the crown layer, then $\Box_L$, then $\Box_R$, then $\Box_L^{\mathsf{D}}$, then the trunk layer;

$\mathcal{D}_R$: the crown layer, then $\Box_L$, then $\Box_L^{\mathsf{D}}$, then $\Box_R$, then the trunk layer.

The case of the last applied rule being a propositional or structural trunk rule works as before. The case when $\mathsf{r}_L$ or $\mathsf{r}_R$ is an application of $\Box_L^{\mathsf{D}}$ with none of the the displayed occurrences of $A$ being principal is processed in the similar way. The main differences lie in the case when $A = \Box B$ and both $\mathsf{r}_L$ and $\mathsf{r}_R$ introduce one of the displayed occurrences of $A$ because now $\mathsf{r}_R$ can be an application of either $\Box_L$ (as before) or $\Box_L^{\mathsf{D}}$. The case when $\mathsf{r}_R$ is an application of $\Box_L$ is, in fact, exactly the same as in the cut elimination proof for $\mathcal{R}_{\mathsf{K5}}$ because in this case, $\Box_L^{\mathsf{D}}$ rules do not occur above either $\mathsf{r}_L$ or $\mathsf{r}_R$ due to our choice of layering for the corresponding derivations. Thus, we only consider in detail the case when $\mathsf{r}_R$ is an application of $\Box_L^{\mathsf{D}}$. As before, if one of the $\Box_L^{\mathsf{D}}$ rules above $\mathsf{r}_R$ introduces a formula other than a displayed occurrence of $A$, it can be permuted downwards and dealt with by the induction hypothesis on the combined depth of the derivations. So we assume that all $\Box_L^{\mathsf{D}}$ rules introduce displayed occurrences of $A$. Further, if one of the $\Box_L$ rules above $\mathsf{r}_R$ does not affect any of the crown components created by the $\Box_L^{\mathsf{D}}$ rules, it can be permuted downwards in a depth-preserving way as follows: the derivation

$$\begin{array}{cl}
\mathcal{D}_R^1 & \\
\vdots & \\
\Box\Gamma', \Box B^{m''} \Rightarrow \,||\, \mathcal{H}'_R \mid \Omega_1, C, B^{i_1} \Rightarrow \Theta_1 \mid \Omega_2, B^{i_2} \Rightarrow \Theta_2 \mid \cdots \mid \Omega_z, B^{i_z} \Rightarrow \Theta_z \mid \Psi_1, B^{j_1} \Rightarrow \;\mid \cdots \mid \Psi_k, B^{j_k} \Rightarrow & \\
\hline
\Box\Gamma', \Box C, \Box B^{m''} \Rightarrow \,||\, \mathcal{H}'_R \mid \Omega_1, B^{i_1} \Rightarrow \Theta_1 \mid \cdots \mid \Omega_z, B^{i_z} \Rightarrow \Theta_z \mid \Psi_1, B^{j_1} \Rightarrow \;\mid \cdots \mid \Psi_k, B^{j_k} \Rightarrow & \Box_L \\
\hline\hline
\Box\Gamma', \Box C, \Box\Psi_1, \ldots, \Box\Psi_k, \Box B^{m'} \Rightarrow \,||\, \mathcal{H}'_R \mid \Omega_1 \Rightarrow \Theta_1 \mid \cdots \mid \Omega_z \Rightarrow \Theta_z \mid [B \Rightarrow\,]^k & \Box_L \\
\hline\hline
\Box\Gamma', \Box C, \Box\Psi_1, \ldots, \Box\Psi_k, \Box B^{m} \Rightarrow \,||\, \mathcal{H}'_R \mid \Omega_1 \Rightarrow \Theta_1 \mid \cdots \mid \Omega_z \Rightarrow \Theta_z & \Box_L^{\mathsf{D}}
\end{array}$$

where $z > 0$, $k > 0$, $m = m' + k = m'' + i_1 + \cdots + i_z + j_1 + \cdots + j_k$, all $j_x$'s and $i_2, \ldots, i_z$ are positive, $i_1 \geq 0$, and some of the $\Psi_t$ can be empty, is replaced with

$$\begin{array}{cl}
\mathcal{D}_R^1 & \\
\vdots & \\
\Box\Gamma', \Box B^{m''} \Rightarrow \,||\, \mathcal{H}'_R \mid \Omega_1, C, B^{i_1} \Rightarrow \Theta_1 \mid \Omega_2, B^{i_2} \Rightarrow \Theta_2 \mid \cdots \mid \Omega_z, B^{i_z} \Rightarrow \Theta_z \mid \Psi_1, B^{j_1} \Rightarrow \;\mid \cdots \mid \Psi_k, B^{j_k} \Rightarrow & \\
\hline\hline
\Box\Gamma', \Box\Psi_1, \ldots, \Box\Psi_k, \Box B^{m'} \Rightarrow \,||\, \mathcal{H}'_R \mid \Omega_1, C \Rightarrow \Theta_1 \mid \Omega_2 \Rightarrow \Theta_2 \mid \cdots \mid \Omega_z \Rightarrow \Theta_z \mid [B \Rightarrow\,]^k & \Box_L \\
\hline\hline
\Box\Gamma', \Box\Psi_1, \ldots, \Box\Psi_k, \Box B^{m} \Rightarrow \,||\, \mathcal{H}'_R \mid \Omega_1, C \Rightarrow \Theta_1 \mid \Omega_2 \Rightarrow \Theta_2 \mid \cdots \mid \Omega_z \Rightarrow \Theta_z & \Box_L^{\mathsf{D}} \\
\hline
\Box\Gamma', \Box C, \Box\Psi_1, \ldots, \Box\Psi_k, \Box B^{m} \Rightarrow \,||\, \mathcal{H}'_R \mid \Omega_1 \Rightarrow \Theta_1 \mid \cdots \mid \Omega_z \Rightarrow \Theta_z & \Box_L
\end{array}$$

So once again, the induction hypothesis on the combined depth of the derivations suffices. As follows from all these considerations, it is sufficient to consider the case when $\mathcal{D}_L$ has the form

$$\begin{array}{cl}
\mathcal{D}_L^1 & \\
\vdots & \\
\Rightarrow \,||\, \mathcal{H}_L \mid \Xi_1 \Rightarrow B \mid \cdots \mid \Xi_n \Rightarrow B & \\
\hline\hline
\Box\Xi_1, \ldots, \Box\Xi_n \Rightarrow \,||\, \mathcal{H}_L \mid [\,\Rightarrow B]^n & \Box_L \\
\hline\hline
\Box\Xi_1, \ldots, \Box\Xi_n \Rightarrow \Box B^n \,||\, \mathcal{H}_L & \Box_R
\end{array}$$

and $\mathcal{D}_R$ has the form

$$\begin{array}{cl}
\mathcal{D}_R^0 & \\
\vdots & \\
\Rightarrow \,||\, \mathcal{H}'_R \mid \Omega_1, B^{i_1} \Rightarrow \Theta_1 \mid \cdots \mid \Omega_z, B^{i_z} \Rightarrow \Theta_z \mid \Psi_1, B^{j_1} \Rightarrow \;\mid \cdots \mid \Psi_k, B^{j_k} \Rightarrow & \\
\hline\hline
\Box\Psi_1, \ldots, \Box\Psi_k, \Box B^{m'} \Rightarrow \,||\, \mathcal{H}'_R \mid \Omega_1 \Rightarrow \Theta_1 \,||\, \cdots \mid \Omega_z \Rightarrow \Theta_z \mid [B \Rightarrow\,]^k & \Box_L \\
\hline\hline
\Box\Psi_1, \ldots, \Box\Psi_k, \Box B^{m} \Rightarrow \,||\, \mathcal{H}'_R \mid \Omega_1 \Rightarrow \Theta_1 \,||\, \cdots \mid \Omega_z \Rightarrow \Theta_z & \Box_L^{\mathsf{D}}
\end{array}$$

where $z \geq 0$, $k > 0$, $m = m' + k = i_1 + \cdots + i_z + j_1 + \cdots + j_k$, all $i_x$'s and $j_y$'s are positive, and some of the $\Psi_t$ can be empty.

Our goal is to derive

$$\Box\Xi_1, \ldots, \Box\Xi_n, \Box\Psi_1, \ldots, \Box\Psi_k \Rightarrow \,||\, \mathcal{H}_L \mid \mathcal{H}'_R \mid \Omega_1 \Rightarrow \Theta_1 \mid \cdots \mid \Omega_z \Rightarrow \Theta_z \tag{10}$$

To achieve this goal we first construct a derivation $\mathcal{D}_R^1$ by contracting the duplicate $B$'s:

$$\begin{array}{cl}
\mathcal{D}_R^0 & \\
\vdots & \\
\Rightarrow \,||\, \mathcal{H}'_R \mid \Omega_1, B^{i_1} \Rightarrow \Theta_1 \mid \cdots \mid \Omega_z, B^{i_z} \Rightarrow \Theta_z \mid \Psi_1, B^{j_1} \Rightarrow \;\mid \cdots \mid \Psi_k, B^{j_k} \Rightarrow & \\
\hline\hline
\Rightarrow \,||\, \mathcal{H}'_R \mid \Omega_1, B \Rightarrow \Theta_1 \mid \cdots \mid \Omega_z, B \Rightarrow \Theta_z \mid \Psi_1, B \Rightarrow \;\mid \cdots \mid \Psi_k, B \Rightarrow & \mathsf{IC}_L
\end{array}$$

and then apply the analogue of Lemma 4.10 $k + z$ times: the first time to the endsequent of $\mathcal{D}_L$ and the leftmost displayed occurrence of $B$ in the endsequent of $\mathcal{D}_R^1$, each consecutive time to the endsequent of $\mathcal{D}_L$ and to the leftmost displayed occurrence of $B$ in the result of the previous application. As the result we obtain the derivability in $\mathcal{R}_{\mathsf{KD5}}$ of

$$\Rightarrow \,||\, [\mathcal{H}_L]^{k+z} \mid \mathcal{H}'_R \mid \Xi_1, \Omega_1 \Rightarrow \Theta_1 \mid \cdots \mid \Xi_n, \Omega_1 \Rightarrow \Theta_1 \mid \cdots \mid \Xi_1, \Omega_z \Rightarrow \Theta_z \mid \cdots \mid \Xi_n, \Omega_z \Rightarrow \Theta_z \mid \\ \mid \Xi_1, \Psi_1 \Rightarrow \mid \cdots \mid \Xi_n, \Psi_1 \Rightarrow \mid \cdots \mid \Xi_1, \Psi_k \Rightarrow \mid \cdots \mid \Xi_n, \Psi_k \Rightarrow$$

Since all the formulas from the $\Xi_x$ and from the $\Psi_y$ can be transferred to the trunk by $\Box_L$ rules with $\Box_L^{\mathsf{D}}$ used for the last formula in each $\Xi_x, \Psi_y \Rightarrow$ to remove the component itself, removing duplicates by means of contraction rules would suffice in order to derive (10), except for the cases when one of the $\Xi_x, \Psi_y \Rightarrow$ has the form $\Rightarrow$. If at least one such crown component is present it is to be weakened to any other existing non-empty crown component and contracted with it, after which the derivation proceeds as just described. The only problem, therefore, might occur when the only crown components are the empty ones, i.e., of the form $\Rightarrow$. Even in this case, if $z > 0$ or if at least one of $\mathcal{H}_L$ and $\mathcal{H}'_R$ is not empty, then the extra empty components can be contracted with those present in (10). Thus, the only remaining case could be when $\mathcal{D}_L$ and $\mathcal{D}_R$ had the forms

$$\dfrac{\begin{matrix}\mathcal{D}_L^1 \\ \vdots \\ \Rightarrow \,||\, [\Rightarrow B]^n\end{matrix}}{\Rightarrow \Box B^n}\;\Box_R \qquad \text{and} \qquad \dfrac{\begin{matrix}\mathcal{D}_R^0 \\ \vdots \\ \Rightarrow \,||\, [B \Rightarrow]^m\end{matrix}}{\Box B^m \Rightarrow}\;\Box_L^{\mathsf{D}}$$

respectively. However, this would mean that we can derive $\Rightarrow \,||\, \Rightarrow$, which we have shown to be impossible. ☐

Having established cut elimination, again we would like to use the calculus in a decision procedure for the logic KD5. The strategy is the same as in the case of K5: we first modify the (cut-free) calculus so as to make contraction admissible. This allows us to consider set-based grafted hypersequents, and a slight modification in the backwards proof search algorithm for K5 will give us the decidability and complexity result. The *modified rules* of the calculus $\mathcal{R}^*_{\mathsf{KD5}}$ are those of the calculus $\mathcal{R}^*_{\mathsf{K5}}$ together with the modified version $\Box_L^{\mathsf{D}^*}$ of the $\Box_L^{\mathsf{D}}$ rule given by:

$$\dfrac{\Gamma, \Box A \Rightarrow \Delta \,||\, \mathcal{H} \mid A \Rightarrow}{\Gamma, \Box A \Rightarrow \Delta \,||\, \mathcal{H}}\;\Box_L^{\mathsf{D}^*}$$

As in the case of $\mathcal{R}^*_{\mathsf{K5}}$ we then show admissibility of internal and external weakening and the contraction rules and equivalence of the two calculi.

**Lemma 7.3** (Admissibility of internal and external weakening and contraction)**.** *The rules* IW, EW, *and* EC, *as well as* $\mathsf{IC}_L, \mathsf{IC}_R$ *and* $\mathsf{C}_L, \mathsf{C}_R$ *are depth-preserving admissible in* $\mathcal{R}^*_{\mathsf{KD5}}$.

*Proof.* As before. The additional rule $\Box_L^{\mathsf{D}^*}$ is treated like the rule $\Box_L^*$. ☐

**Theorem 7.4** (Equivalence of $\mathcal{R}_{\mathsf{KD5}}$ and $\mathcal{R}^*_{\mathsf{KD5}}$.)**.** *A grafted hypersequent $\mathcal{G}$ is derivable in $\mathcal{R}_{\mathsf{KD5}}$ iff it is derivable in $\mathcal{R}^*_{\mathsf{KD5}}$.*

*Proof.* As for Theorem 5.4. ☐

The decision procedure for the logic KD5 is the straightforward adaption of the one for K5 from Algorithm 1. The only difference is the additional Line 9.D inserted between Line 9 and Line 10 that corresponds to the additional rule $\Box_L^{\mathsf{D}^*}$.

9.D apply $\Box_L^{\mathsf{D}^*}$ backwards to the each formula $\Box A \in \Gamma_1$ in $\Gamma_1 \Rightarrow \Delta_1 \,||\, \mathcal{H}_1$ such that it is not the case that $\Rightarrow \,||\, A \Rightarrow \subseteq \Gamma_1 \Rightarrow \Delta_1 \,||\, \mathcal{H}_1$

**Theorem 7.5.** *The backwards proof search procedure given by the modified algorithm decides validity in* KD5 *and can be implemented in* coNP.

Figure 7: Additional grafted tableau rule for $\mathsf{KD5}$. As before, $\mathbb{n}$ is a limb prefix.

$$\frac{\bullet : \mathsf{T}\, \Box A}{\mathbb{n} : \mathsf{T}\, A}\ \mathbb{n} \text{ new}$$

Figure 8: The crown modal rules of the calculi for shift reflexivity

$$\frac{\Rightarrow \,||\, \mathcal{H} \mid A_1, \ldots, A_n \Rightarrow B}{\Rightarrow \,||\, \mathcal{H} \mid \Box A_1, \ldots, \Box A_n, \Gamma \Rightarrow \Box B, \Delta}\ \mathsf{K}_n \qquad \frac{\Rightarrow \,||\, \mathcal{H} \mid \Gamma, A \Rightarrow \Delta}{\Rightarrow \,||\, \mathcal{H} \mid \Gamma, \Box A \Rightarrow \Delta}\ \mathsf{T}$$

*Proof.* A straightforward adaption of the proof of Theorem 5.8. Concerning the complexity (and again writing $n$ for the number of subformulae of the input grafted hypersequent), the additional step of applying the rule $\Box_L^{\mathsf{D}*}$ backwards in Line 9.D introduces at most $n$ new crown components in at most $n$ applications of $\Box_L^{\mathsf{D}}$. Thus, the total number of crown components at any moments becomes at most $3n$ instead of $2n$ for $\mathsf{K5}$. Thus, the rule $\Box_L$ is used at most $3n^2$ times instead of $2n^2$, there are at most $6n^2$ applications of crown propositional rules, and each rule $5^*$ is applied at most $9n^3$ times. It is clear that the asymptotic behavior of the algorithm is, however, unaffected. ☐

As in the case of $\mathsf{K5}$ our grafted hypersequent calculus can be recast into the tableau framework by adding to the grafted tableau system for $\mathsf{K5}$ the special necessity rule from Figure 7. Soundness and completeness, as well as translations between this grafted tableaux system and $\mathcal{R}^*_{\mathsf{KD5}}$, are essentially the same as for the case of $\mathsf{K5}$, with the additional new case left to the reader as an easy exercise.

*Remark* 7.6. Note that it is more common in tableau systems to postulate seriality by allowing to add $\bullet : \mathsf{T}\, \Diamond A$ to a branch containing $\bullet : \mathsf{T}\, \Box A$ (see, e.g., [FM98, Def. 2.3.1]). For our language, however, after the defined connectives are unfolded, this would mean adding $\bullet : \mathsf{T}\, \Box(A \to \bot) \to \bot$. Given that the latter yields exactly the same result but with an overhead, including an additional closed branch, we saw no reason to violate analyticity just to conform to the historical consensus.

## 7.2 Calculi for Shift Reflexivity

As an example for the second possibility of varying the calculus we construct grafted hypersequent calculi for logics of *shift reflexive* frames, which are also called *secondary reflexive* or *almost reflexive*. This property is given axiomatically by the axiom

$$(\mathsf{T}_\Box) \quad \Box(\Box A \to A)$$

and semantically by the frame condition $\forall w \forall v (wRv \to vRv)$, i.e., the property that every successor is reflexive. We write $\mathsf{KT}_\Box$ for the logic given by the normal modal logic $\mathsf{K}$ together with the axiom $(\mathsf{T}_\Box)$, and $\mathsf{KDT}_\Box$ for the extension of this logic with the seriality axiom $(\mathsf{D})$, which, predictably, is the logic of shift reflexive and serial frames. Under a deontic interpretation of the $\Box$ as the modality $\mathcal{O}$ (read: "It is obligatory that…"), the logic $\mathsf{KDT}_\Box$ is also known as the logic $\mathsf{SDL}^+$, a very natural extension of the standard deontic logic $\mathsf{SDL}$ ($= \mathsf{KD}$) [McN14]. In order to construct grafted hypersequent calculi for these logics, we now graft a hypersequent calculus for the logic $\mathsf{KT}$ onto a nested sequent calculus for the logic $\mathsf{K}$ or $\mathsf{KD}$ respectively.

**Definition 7.7** ($\mathcal{R}_{\mathsf{KT}_\Box}$ and $\mathcal{R}_{\mathsf{KDT}_\Box}$)**.** The calculus $\mathcal{R}_{\mathsf{KT}_\Box}$ contains the trunk rules of $\mathcal{R}_{\mathsf{K5}}$ as given in Figure 1 and the crown rules of $\mathcal{R}_{\mathsf{K5}}$ as given in Figure 2, except that the rules $\mathsf{K}$ and $\mathsf{5}$ from Figure 2 are replaced with new crown modal rules $\mathsf{K}_n$ for every $n \geq 0$ and $\mathsf{T}$ as given in Figure 8. The calculus $\mathcal{R}_{\mathsf{KDT}_\Box}$ contains the rules of $\mathcal{R}_{\mathsf{KT}_\Box}$ together with the additional rule $\Box_L^{\mathsf{D}}$ from Section 7.1.

The notions of contextual, principal and active formulae and components are defined as expected, with the exception that we call the formulae occurring in $\Gamma, \Delta$ in the rules $\mathsf{K}_n$ the *weakening context*. Note that the rules $\mathsf{K}_n$ and $\mathsf{T}$ from Figure 8 do not make use of the hypersequent mechanism. In particular, only one hypersequent component is principal in their conclusion: they are sequent-style rules in a hypersequent setting. Thus, by standard arguments it can be shown that a grafted hypersequent $\Rightarrow \,||\, \mathcal{H}$ is derivable in $\mathcal{R}_{\mathsf{KT}_\Box}$ or $\mathcal{R}_{\mathsf{KDT}_\Box}$, iff $\Rightarrow || \Gamma \Rightarrow \Delta$ is derivable in the same system for some $\Gamma \Rightarrow \Delta \in \mathcal{H}$ iff $\bigwedge \Gamma \to \bigvee \Delta$ is derivable in the modal logic $\mathsf{KT}$ for some $\Gamma \Rightarrow \Delta \in \mathcal{H}$.[4] This would seem to suggest that it suffices to graft a *sequent* calculus for $\mathsf{KT}$ instead of a hypersequent calculus onto the nested calculus for $\mathsf{K}$ or $\mathsf{KD}$ respectively. However, this would necessitate a modification of the transfer rule $\Box_R$, which introduces a new crown component after each application. Since we would like to emphasise the uniform character of our approach, we are unwilling to make this modification.

As usual, we define $\mathcal{R}_{\mathsf{KT}_\Box}\mathsf{Cut}$ and $\mathcal{R}_{\mathsf{KDT}_\Box}\mathsf{Cut}$ to be $\mathcal{R}_{\mathsf{KT}_\Box}$ and $\mathcal{R}_{\mathsf{KDT}_\Box}$ respectively extended with both cut rules $\mathsf{Cut}_\mathrm{t}$ and $\mathsf{Cut}_\mathrm{c}$.

Soundness and completeness of the calculi are established readily.

**Theorem 7.8** (Soundness and completeness with cut)**.** *For every formula $D$ we have:*

$$\begin{array}{ccccc} \mathcal{R}_{\mathsf{KT}_\Box} \vdash \Rightarrow D & \Longrightarrow & D \in \mathsf{KT}_\Box & \Longrightarrow & \mathcal{R}_{\mathsf{KT}_\Box}\mathsf{Cut} \vdash \Rightarrow D \\ \mathcal{R}_{\mathsf{KDT}_\Box} \vdash \Rightarrow D & \Longrightarrow & D \in \mathsf{KDT}_\Box & \Longrightarrow & \mathcal{R}_{\mathsf{KDT}_\Box}\mathsf{Cut} \vdash \Rightarrow D \end{array}$$

*Proof.* For soundness, again we show that all the rules preserve validity under the formula interpretation, which was proved for the propositional rules and most of the transfer rules in Proposition 3.3 and for the rule $\Box_L^{\mathsf{D}}$ for serial frames in Theorem 7.1. For the rule $\mathsf{T}$, a straightforward argument shows that if the negation of a conclusion of an instance of $\mathsf{T}$ is satisfied on a shift reflexive frame at world $w$, then so is the negation of the premiss. For the rules $\mathsf{K}_n$, suppose that the negation of the conclusion of an instance of the rule $\mathsf{K}_n$ is satisfied in a shift reflexive (and serial in the case of $\mathsf{KDT}_\Box$) model, i.e., there is a model $\mathfrak{M} = (W, R, \sigma)$ with shift reflexive (and serial) $R$ and a world $w \in W$ such that the formula

$$\neg\iota(\Rightarrow || \mathcal{H} \mid \Box A_1, \dots, \Box A_n, \Gamma \Rightarrow \Box B, \Delta) \equiv_{\mathsf{KT}_\Box} \neg H \wedge \Diamond \left( \bigwedge_{j=1}^{n} \Box A_j \wedge \bigwedge \Gamma \wedge \neg \Box B \wedge \neg \bigvee \Delta \right)$$

holds at $w$ in $\mathfrak{M}$, where $A \equiv_{\mathsf{KT}_\Box} B$ means that $(A \to B) \wedge (B \to A) \in \mathsf{KT}_\Box$ and $H := \iota(\Rightarrow || \mathcal{H})$. Note that $\neg H \equiv_{\mathsf{KT}_\Box} \bigwedge \Diamond \Upsilon$ for some finite (possibly empty) set $\Upsilon$ of formulas. Then for every $B \in \Upsilon$, there is a $v_B \in W$ such that $wRv_B$ and $\mathfrak{M}, v_B \Vdash B$. In addition, there is a $x \in W$ such that $wRx$ and $\mathfrak{M}, x \Vdash \bigwedge_{j=1}^{n} \Box A_j \wedge \bigwedge \Gamma \wedge \neg \Box B \wedge \neg \bigvee \Delta$. Further, there must exist a $y \in W$ such that $xRy$ and $\mathfrak{M}, y \Vdash \bigwedge_{j=1}^{n} A_j \wedge \neg B$. Now for a fresh world $z \notin W$ let $\mathfrak{M}^*$ be the model $(W \cup \{z\},\ R^*,\ \sigma^*)$ where $R^* := R \cup \{(z, v_B) : B \in \Upsilon\} \cup \{(z, y)\}$ and the valuation $\sigma^*$ is the same as $\sigma$ on worlds of $W$ and is arbitrary on $z$. Since the frame underlying $\mathfrak{M}$ is shift reflexive, the frame underlying $\mathfrak{M}^*$ is shift reflexive as well, and if the former frame is serial, then so is the latter. Moreover, since the new world $z$ is not accessible from any world, for any world $w \in W$ and any formula $C$, we have $\mathfrak{M}, w \Vdash C$ iff $\mathfrak{M}^*, w \Vdash C$. In particular, $\mathfrak{M}^*, v_B \Vdash B$ for each $B \in \Upsilon$. Since $\mathfrak{M}^*, z \Vdash \Diamond B$ for every $B \in \Upsilon$, we have $\mathfrak{M}^*, z \Vdash \neg H$. Similarly, $\mathfrak{M}^*, y \Vdash \bigwedge_{j=1}^{n} A_j \wedge \neg B$ and, hence, $\mathfrak{M}^*, z \Vdash \Diamond \left( \bigwedge_{j=1}^{n} A_j \wedge \neg B \right)$. Summarizing,

$$\neg\iota(\Rightarrow || \mathcal{H} \mid A_1, \dots, A_n \Rightarrow B) \quad \equiv_{\mathsf{KT}_\Box} \quad \neg H \wedge \Diamond \left( \bigwedge_{j=1}^{n} A_j \wedge \neg B \right)$$

holds at world $z$ in $\mathfrak{M}^*$, and, thus, the negation of the premiss of this instance of $\mathsf{K}_n$ is satisfied in a shift reflexive (and serial) model.

For completeness, again we first show by induction on the complexity of the formula $A$ that generalised axioms, i.e., the grafted hypersequents

$$\Gamma, A \Rightarrow \Delta, A \,||\, \mathcal{H} \qquad \text{and} \qquad \Rightarrow || \mathcal{H} \mid \Sigma, A \Rightarrow \Pi, A$$

are derivable in both calculi. The only necessary modification to the proof of Lemma 3.5 is that the case of boxed formulae in the crown is handled by a single application of $\mathsf{K}_1$ instead of the derivation (2). Then again we derive all the axioms of $\mathsf{KT}_\Box$ or of $\mathsf{KDT}_\Box$ respectively both in the crown and in the trunk and use the two cut rules to simulate $\mathsf{MP}$. The rule $\mathsf{nec}$ in the trunk is still simulated by using the induction hypothesis for the crown and the rule $\Box_R$, whereas $\mathsf{nec}$ in the crown now has to be processed by $\mathsf{K}_0$ instead of $\mathsf{K}$. To derive the new axiom ($\mathsf{T}_\Box$) in the trunk and crown, we append the derivation

$$\dfrac{\dfrac{\dfrac{}{\Rightarrow \,||\, A \Rightarrow A}\text{ gen. }\mathsf{Init}}{\Rightarrow \,||\, \Box A \Rightarrow A}\mathsf{T}}{\Rightarrow \,||\, \Rightarrow \Box A \to A}\to_R$$

by the rule $\Box_R$ or $\mathsf{K}_0$ respectively. In addition, to derive the axiom ($\mathsf{K}$) in the crown we have to modify the middle of the derivation (3) from the proof of Theorem 3.6: the sequence of rules $\mathsf{5}$, $\mathsf{K}$, and $\mathsf{merge}$ is replaced with a single application of $\mathsf{K}_2$. $\Box$

[4]Here $\mathsf{KT}$ is the logic of reflexive frames obtained by adding the axiom $(\mathsf{T})$ to $\mathsf{K}$, see [BdRV01].

Cut elimination is then shown by the obvious adaption of the Shift Left Lemma and the Shift Right Lemma.

**Theorem 7.9** (Cut elimination for $\mathcal{R}_{\mathsf{KT}_\Box}$ and $\mathcal{R}_{\mathsf{KDT}_\Box}$)**.** *The rules* $\mathsf{Cut}_\mathsf{t}$ *and* $\mathsf{Cut}_\mathsf{c}$ *are admissible in* $\mathcal{R}_{\mathsf{KT}_\Box}$ *and* $\mathcal{R}_{\mathsf{KDT}_\Box}$.

*Proof.* The layering of the derivations is established as for $\mathcal{R}_{\mathsf{K5}}$ (see Proposition 4.4). In the proof of the analogue of the Shift Right Lemma 4.6, the case when the last rule applied in $\mathcal{D}_R$ was $\mathsf{T}$ and it did not introduce any of the formulae to be cut presents no difficulties and is handled again by the induction hypothesis. The same situation for a rule $\mathsf{K}_k$ can only mean that all occurrences of the formulae to be cut in the principal sequent of $\mathsf{K}_k$ were part of the weakening context, meaning that the same result can be achieved by weakenings with a prior cross cut if necessary. For $A = \Box B$ and the main case of one of the occurrences to be cut being a principal formula in $\mathcal{D}_R$ we now have some $\mathsf{K}_k$ as the last rule applied in $\mathcal{D}_L$ and either $\mathsf{K}_\ell$ or $\mathsf{T}$ as the last rule applied in $\mathcal{D}_R$. For the case of $\mathsf{T}$ we have

$$\dfrac{\begin{matrix}\mathcal{D}'_L\\ \vdots\\ \Rightarrow \,||\, \mathcal{H}_L \mid C_1, \ldots, C_k \Rightarrow B\end{matrix}}{\Rightarrow \,||\, \mathcal{H}_L \mid \Box C_1, \ldots, \Box C_k, \Gamma' \Rightarrow \Box B, \Delta}\mathsf{K}_k$$

and

$$\dfrac{\begin{matrix}\mathcal{D}'_R\\ \vdots\\ \Rightarrow \,||\, \mathcal{H}_R \mid \Sigma_1, \Box B^{m_1} \Rightarrow \Pi_1 \mid \cdots \mid \Sigma_{n-1}, \Box B^{m_{n-1}} \Rightarrow \Pi_{n-1} \mid \Sigma_n, \Box B^{m_n - 1}, B \Rightarrow \Pi_n\end{matrix}}{\Rightarrow \,||\, \mathcal{H}_R \mid \Sigma_1, \Box B^{m_1} \Rightarrow \Pi_1 \mid \cdots \mid \Sigma_{n-1}, \Box B^{m_{n-1}} \Rightarrow \Pi_{n-1} \mid \Sigma_n, \Box B^{m_n} \Rightarrow \Pi_n}\mathsf{T}$$

Using the induction hypothesis on the depth of $\mathcal{D}_R$ we obtain a derivation of

$$\begin{aligned}\Rightarrow \,||\, \mathcal{H}_L \mid \mathcal{H}_R \mid \Box C_1, \ldots, \Box C_k, \Gamma', \Sigma_1 \Rightarrow \Delta, \Pi_1 \mid \cdots \mid \Box C_1, \ldots, \Box C_k, \Gamma', \Sigma_{n-1} \Rightarrow \Delta, \Pi_{n-1} \mid \\ \mid \underbrace{\Box C_1, \ldots, \Box C_k, \Gamma',}_{\text{if } m_n > 1} \Sigma_n, B \Rightarrow \underbrace{\Delta,}_{\text{if } m_n > 1} \Pi_n\end{aligned}$$

Further, using $\mathsf{Cut}_\mathsf{c}$ for $B$ of a smaller size with the premiss of the $\mathsf{K}_k$ rule we obtain a derivation of

$$\begin{aligned}\Rightarrow \,||\, \mathcal{H}_L \mid \mathcal{H}_L \mid \mathcal{H}_R \mid \Box C_1, \ldots, \Box C_k, \Gamma', \Sigma_1 \Rightarrow \Delta, \Pi_1 \mid \cdots \mid \Box C_1, \ldots, \Box C_k, \Gamma', \Sigma_{n-1} \Rightarrow \Delta, \Pi_{n-1} \mid \\ \mid C_1, \ldots, C_k, \underbrace{\Box C_1, \ldots, \Box C_k, \Gamma',}_{\text{if } m_n > 1} \Sigma_n \Rightarrow \underbrace{\Delta,}_{\text{if } m_n > 1} \Pi_n\end{aligned}$$

From this grafted hypersequents, it is sufficient to first use $k$ instances of $\mathsf{T}$ to box all $C'_i s$ and then use contractions (and weakenings if $m_n = 1$) to obtain a derivation of the desired result:

$$\Rightarrow \,||\, \mathcal{H}_L \mid \mathcal{H}_R \mid \Box C_1, \ldots, \Box C_k, \Gamma', \Sigma_1 \Rightarrow \Delta, \Pi_1 \mid \cdots \mid \Box C_1, \ldots, \Box C_k, \Gamma', \Sigma_n \Rightarrow \Delta, \Pi_n$$

For the case of the last applied rule in $\mathcal{D}_R$ being $\mathsf{K}_\ell$ (note that $\ell > 0$) we have

$$\dfrac{\begin{matrix}\mathcal{D}'_R\\ \vdots\\ \Rightarrow \,||\, \mathcal{H}_R \mid \Sigma_1, \Box B^{m_1} \Rightarrow \Pi_1 \mid \cdots \mid \Sigma_{n-1}, \Box B^{m_{n-1}} \Rightarrow \Pi_{n-1} \mid F_1, \ldots, F_j, B^{\ell - j} \Rightarrow D\end{matrix}}{\Rightarrow \,||\, \mathcal{H}_R \mid \Sigma_1, \Box B^{m_1} \Rightarrow \Pi_1 \mid \cdots \mid \Sigma_{n-1}, \Box B^{m_{n-1}} \Rightarrow \Pi_{n-1} \mid \Sigma'_n, \Box F_1, \ldots, \Box F_j, \Box B^{m_n} \Rightarrow \Box D, \Pi'_n}\mathsf{K}_\ell$$

where $j < \ell$ and $\ell - j \leq m_n$. Using the induction hypothesis on the depth of $\mathcal{D}_R$ we obtain a derivation of

$$\Rightarrow || \mathcal{H}_L \mid \mathcal{H}_R \mid \Box C_1, \ldots, \Box C_k, \Gamma', \Sigma_1 \Rightarrow \Delta, \Pi_1 \mid \cdots \mid \Box C_1, \ldots, \Box C_k, \Gamma', \Sigma_{n-1} \Rightarrow \Delta, \Pi_{n-1} \mid \\ \mid F_1, \ldots, F_j, B^{l-j} \Rightarrow D$$

Further, using $\mathsf{Cut_c}$ for $B$ of a smaller size with the premiss of the $\mathsf{K}_k$ rule we obtain a derivation of

$$\Rightarrow || \mathcal{H}_L \mid \mathcal{H}_L \mid \mathcal{H}_R \mid \Box C_1, \ldots, \Box C_k, \Gamma', \Sigma_1 \Rightarrow \Delta, \Pi_1 \mid \cdots \mid \Box C_1, \ldots, \Box C_k, \Gamma', \Sigma_{n-1} \Rightarrow \Delta, \Pi_{n-1} \mid \\ \mid C_1, \ldots, C_k, F_1, \ldots, F_j \Rightarrow D$$

From this grafted hypersequents, it is sufficient to first use $\mathsf{K}_{k+j}$ for the last crown component with antecedent weakened by $\Gamma', \Sigma'_n \Rightarrow \Delta, \Pi'_n$ and and then use contractions to obtain a derivation of the desired result:

$$\Rightarrow || \mathcal{H}_L \mid \mathcal{H}_R \mid \Box C_1, \ldots, \Box C_k, \Gamma', \Sigma_1 \Rightarrow \Delta, \Pi_1 \mid \cdots \mid \Box C_1, \ldots, \Box C_k, \Gamma', \Sigma_{n-1} \Rightarrow \Delta, \Pi_{n-1} \mid \\ \mid \Box C_1, \ldots, \Box C_k, \Gamma', \Sigma'_n, \Box F_1, \ldots, \Box F_j \Rightarrow \Delta, \Box D, \Pi'_n$$

In the proof of the analogue of the Shift Left Lemma 4.7, the case where $A$ is the formula $\Box B$ and one occurrence of it is principal in the last applied rule in $\mathcal{D}_L$, which must be $\mathsf{K}_k$ is as follows. In this case the derivation $\mathcal{D}_L$ ends with

$$\dfrac{\begin{array}{c}\mathcal{D}'_L \\ \vdots \\ \Rightarrow || \mathcal{H}_L \mid \Gamma_1 \Rightarrow \Delta_1, \Box B^{m_1} \mid \cdots \mid \Gamma_{n-1} \Rightarrow \Delta_{n-1}, \Box B^{m_{n-1}} \mid C_1, \ldots, C_k \Rightarrow B\end{array}}{\Rightarrow || \mathcal{H}_L \mid \Gamma_1 \Rightarrow \Delta_1, \Box B^{m_1} \mid \cdots \mid \Gamma_{n-1} \Rightarrow \Delta_{n-1}, \Box B^{m_{n-1}} \mid \Box C_1, \ldots, \Box C_k, \Omega \Rightarrow \Box B, \Lambda} \ \mathsf{K}_k$$

If $n = 1$, then we can directly apply the analogue of Shift Right Lemma 4.6. Otherwise, we first perform a cross cut by applying the induction hypothesis to the premiss of this rule $\mathsf{K}_k$, and then apply $\mathsf{K}_k$ to the result:

$$\dfrac{\dfrac{\begin{array}{c}\mathcal{D}'_L \\ \vdots \\ \Rightarrow || \mathcal{H}_L \mid \Gamma_1 \Rightarrow \Delta_1, \Box B^{m_1} \mid \cdots \mid \Gamma_{n-1} \Rightarrow \Delta_{n-1}, \Box B^{m_{n-1}} \mid C_1, \ldots, C_k \Rightarrow B\end{array} \qquad \begin{array}{c}\mathcal{D}_R \\ \vdots \\ \Rightarrow || \mathcal{H}_R \mid \Sigma, \Box B \Rightarrow \Pi\end{array}}{\Rightarrow || \mathcal{H}_L \mid \mathcal{H}_R \mid \Gamma_1, \Sigma \Rightarrow \Delta_1, \Pi \mid \cdots \mid \Gamma_{n-1}, \Sigma \Rightarrow \Delta_{n-1}, \Pi \mid C_1, \ldots, C_k \Rightarrow B} \ \mathsf{IH}}{\Rightarrow || \mathcal{H}_L \mid \mathcal{H}_R \mid \Gamma_1, \Sigma \Rightarrow \Delta_1, \Pi \mid \cdots \mid \Gamma_{n-1}, \Sigma \Rightarrow \Delta_{n-1}, \Pi \mid \Box C_1, \ldots, \Box C_k, \Omega \Rightarrow \Box B, \Lambda} \ \mathsf{K}_k$$

Since now there is only one displayed occurrence of the formula $\Box B$ which is moreover principal in the last applied rule, we can use the analogue of the Shift Right Lemma 4.6 for the cut formula $\Box B$, applied to the conclusions of this derivation and of $\mathcal{D}_R$, obtaining a derivation of

$$\Rightarrow || \mathcal{H}_L \mid \mathcal{H}_R \mid \Gamma_1, \Sigma \Rightarrow \Delta_1, \Pi \mid \cdots \mid \Gamma_{n-1}, \Sigma \Rightarrow \Delta_{n-1}, \Pi \mid \mathcal{H}_R \mid \Box C_1, \ldots, \Box C_k, \Omega, \Sigma \Rightarrow \Lambda, \Pi \ .$$

It now only remains to remove duplicate sequents using $\mathsf{EC}$.

Finally, root level cuts are eliminated as in the proof of Theorem 4.13 or Theorem 7.2 for the serial case. ☐

The strategy for obtaining a decision procedure using the calculi $\mathcal{R}_{\mathsf{KT}_\Box}$ and $\mathcal{R}_{\mathsf{KDT}_\Box}$ is the same as for the calculi $\mathcal{R}_{\mathsf{K5}}$ and $\mathcal{R}_{\mathsf{KD5}}$: first modify the rules using Kleene's Trick to ensure admissibility of the structural rules and thus equivalence to set-based grafted hypersequents, then perform backwards proof search on these structures. The modified versions of the rules $\mathsf{K}_n$ and $\mathsf{T}$ are given in Figure 9 (the second rule is identical to the rule we had to add in Figure 4 in order to make the Kleene'd systems complete for $\mathsf{K5}$ and $\mathsf{KD5}$). Again, the principal component in the rule $\mathsf{T}$ is not copied into the premiss, since it is subsumed by the active component of the premiss. Then the *modified rule sets* $\mathcal{R}^*_{\mathsf{KT}_\Box}$ and $\mathcal{R}^*_{\mathsf{KDT}_\Box}$ contain the rules of the calculi $\mathcal{R}^*_{\mathsf{K5}}$ and $\mathcal{R}^*_{\mathsf{KD5}}$ respectively, with the rules $\mathsf{K}^*_n$ and $\mathsf{T}^*$ instead of $\mathsf{K}^*$ and $\mathsf{5}^*$. Unlike these previous cases, we also add the rule of external weakening $\mathsf{EW}$. This is not, strictly speaking, necessary because this rule would have been admissible otherwise. We include it primarily to be used in the algorithm for the backward proof search. Then as above we obtain:

Figure 9: The Kleene'd rules for the calculi $\mathcal{R}^*_{\mathsf{KT}_\Box}$ and $\mathcal{R}^*_{\mathsf{KDT}_\Box}$

$$\frac{\Rightarrow \mid\mid \mathcal{H} \mid \Box A_1, \dots, \Box A_n, \Gamma \Rightarrow \Box B, \Delta \mid A_1, \dots, A_n \Rightarrow B}{\Rightarrow \mid\mid \mathcal{H} \mid \Box A_1, \dots, \Box A_n, \Gamma \Rightarrow \Box B, \Delta}\ \mathsf{K}^*_n \qquad \frac{\Rightarrow \mid\mid \mathcal{H} \mid \Gamma, \Box A, A \Rightarrow \Delta}{\Rightarrow \mid\mid \mathcal{H} \mid \Gamma, \Box A \Rightarrow \Delta}\ \mathsf{T}^*$$

**Lemma 7.10** (Admissibility of internal weakening and contraction)**.** *The rules* $\mathsf{IW}$ *and* $\mathsf{EC}$*, as well as* $\mathsf{IC}_L$*,* $\mathsf{IC}_R$*,* $\mathsf{C}_L$*, and* $\mathsf{C}_R$ *are depth-preserving admissible in* $\mathcal{R}^*_{\mathsf{KT}_\Box}$ *and* $\mathcal{R}^*_{\mathsf{KDT}_\Box}$*.*

*Proof.* Again we first show depth-preserving admissibility of $\mathsf{IW}$ by induction on the depth of the derivation. This is then used to show admissibility of the contraction rules. In particular, the admissibility of $\mathsf{EC}$ with last applied rule $\mathsf{T}^*$ is shown exactly the same way as in (8) in the proof of Lemma 5.3. The remaining cases are standard, except that one may need to use the induction hypothesis twice for the admissibility of $\mathsf{IC}_L$, if the last applied rule was $\mathsf{K}^*_n$. $\Box$

**Theorem 7.11** (Equivalence of the calculi)**.** *Let* $\mathcal{G}$ *be a grafted hypersequent. Then*

1. $\mathcal{G}$ *is derivable in* $\mathcal{R}_{\mathsf{KT}_\Box}$ *iff it is derivable in* $\mathcal{R}^*_{\mathsf{KT}_\Box}$
2. $\mathcal{G}$ *is derivable in* $\mathcal{R}_{\mathsf{KDT}_\Box}$ *iff it is derivable in* $\mathcal{R}^*_{\mathsf{KDT}_\Box}$*.*

*Proof.* As above. $\Box$

For the decision procedure we need to modify the algorithm slightly. This is due to the fact that in contrast to the hypersequent calculus for the logic $\mathsf{S5}$ which we used at the crown level of the calculi $\mathcal{R}_{\mathsf{K5}}$ and $\mathcal{R}_{\mathsf{KD5}}$, in the calculus for $\mathsf{KT}$ used at the crown level of the calculi $\mathcal{R}_{\mathsf{KT}_\Box}$ and $\mathcal{R}_{\mathsf{KDT}_\Box}$ we cannot fix the order of the rule applications. Thus we need to existentially guess the last applied modal rule, as captured in Line 21 of the decision procedure for $\mathsf{KT}_\Box$ given as Algorithm 2. For the logic $\mathsf{KDT}_\Box$ we add Line 9.D from p. 31 between lines 9 and 10 as before.

*Remark* 7.12. Algorithm 2 could also be modified to a slightly more efficient version: In Line 9 it would be sufficient to existentially guess only one consequent formula $\Box A$ from the trunk and apply rule $\Box^*_R$ backwards to it. Then the existential guessing step of Line 12 becomes superfluous, and instead of first creating many crown components and then deleting all but one of these we would only create one in the first place. While this would slightly increase efficiency, for the sake of greater transparency in the correctness proof we chose the current formulation.

**Theorem 7.13** (Decidability and complexity)**.** *The backwards proof search procedure for the calculi* $\mathcal{R}^*_{\mathsf{KT}_\Box}$ *and* $\mathcal{R}^*_{\mathsf{KDT}_\Box}$ *given in Algorithm 2 decides the validity problem for the logics* $\mathsf{KT}_\Box$ *and* $\mathsf{KDT}_\Box$ *respectively and can be implemented in* $\mathsf{PSPACE}$*.*

*Proof.* As before, all the rules of $\mathcal{R}^*_{\mathsf{K5}}$, except for $\mathsf{W}$ and $\mathsf{K}^*_\ell$, are invertible and it can be seen that all the possible trunk rules are applied before the application of $\mathsf{W}$ in Line 11, that no crown rules can be applied before Line 11, and that all the possible crown rules are applied after Line 11. The ability to choose one crown component in Line 12 follows from our earlier observation that hypersequents are not necessary to deal with the crown. Every time after the `repeat`-loop of Line 14 terminates, the only remaining applicable rules are instances of $\mathsf{K}^*_\ell$. Completeness is guaranteed by existentially choosing among sufficiently many possibilities to subsume all other possible applications of these rules, i.e., those when some of the boxed formulae in $\Sigma_1$ are not taken to be principal. Finally, the ability to use $\mathsf{EW}$ to remove the old component after an application of $\mathsf{K}^*_\ell$ follows from the same observation about hypersequents not being necessary to deal with the crown and the fact that deleting the newly created component would simply cancel the preceding application of $\mathsf{K}^*_\ell$. Thus, the correctness of the algorithm follows from the completeness of $\mathcal{R}_{\mathsf{KT}_\Box}$ and $\mathcal{R}_{\mathsf{KDT}_\Box}$ and the equivalence of $\mathcal{R}_{\mathsf{KT}_\Box}$ and $\mathcal{R}_{\mathsf{KDT}_\Box}$ to $\mathcal{R}^*_{\mathsf{KT}_\Box}$ and $\mathcal{R}^*_{\mathsf{KDT}_\Box}$ respectively (Theorem 7.11).

For the complexity, again we write $n$ for the size of the input. Since the part of the procedure before Line 12 is the same as in Algorithm 1 or in the algorithm for $\mathcal{R}^*_{\mathsf{KD5}}$, the number of rule applications up to this point in Algorithm 2 or in its modification for $\mathsf{KDT}_\Box$ again is $\mathcal{O}(n^2)$ and the resulting grafted hypersequent has at most $2n$ crown components. Thus there are at most $2n$ possibilities for the existential

**Algorithm 2:** Decision procedure for $\mathsf{KT}_\Box$

**Input**: a set-based grafted hypersequent $\Gamma \Rightarrow \Delta \mid\mid \mathcal{H}$

**Output**: Is $\iota(\Gamma \Rightarrow \Delta \mid\mid \mathcal{H}) \in \mathsf{KT}_\Box$?

set $\Gamma_1 := \Gamma,\ \Delta_1 := \Delta,\ \mathcal{H}_1 := \mathcal{H}$;

**repeat**

- set $\Gamma_2 := \Gamma_1,\ \Delta_2 := \Delta_1,\ \mathcal{H}_2 := \mathcal{H}_1$;
- apply modified propositional trunk rules backwards to each unprocessed trunk formula in $\Gamma_1 \Rightarrow \Delta_1 \mid\mid \mathcal{H}_1$, universally choosing one of the premisses for branching rules, and label these formulae processed;

**until** $\Gamma_1 \Rightarrow \Delta_1 \mid\mid \mathcal{H}_1 \ \subseteq\ \Gamma_2 \Rightarrow \Delta_2 \mid\mid \mathcal{H}_2$;

**if** $\Gamma_1 \Rightarrow \Delta_1 \mid\mid \mathcal{H}_1$ *is a trunk initial structure* **then**

- halt and accept;

**end**

apply $\Box^*_R$ backwards to each formula $\Box A \in \Delta_1$ in $\Gamma_1 \Rightarrow \Delta_1 \mid\mid \mathcal{H}_1$ such that it is not the case that $\Rightarrow \mid\mid \Rightarrow A \ \subseteq\ \Gamma_1 \Rightarrow \Delta_1 \mid\mid \mathcal{H}_1$;

apply $\Box^*_L$ backwards to each $\Box A \in \Gamma_1$ and each component of $\mathcal{H}_1$ in $\Gamma_1 \Rightarrow \Delta_1 \mid\mid \mathcal{H}_1$;

apply $\mathsf{W}$ backwards to $\Gamma_1 \Rightarrow \Delta_1 \mid\mid \mathcal{H}_1$ to obtain $\Rightarrow \mid\mid \mathcal{H}_1$;

existentially guess a crown component $\Sigma_1 \Rightarrow \Pi_1 \in \mathcal{H}_1$ and apply $\mathsf{EW}$ backwards to $\Rightarrow \mid\mid \mathcal{H}_1$ several times to obtain $\Rightarrow \mid\mid \Sigma_1 \Rightarrow \Pi_1$;

**repeat**

- **repeat**
  - set $\Sigma_2 := \Sigma_1$, $\Pi_2 := \Pi_1$;
  - apply modified propositional crown rules and $\mathsf{T}^*$ backwards to each unprocessed crown formula in $\Rightarrow \mid\mid \Sigma_1 \Rightarrow \Pi_1$, universally choosing one of the premisses for branching rules, and label these formulae processed ;
- **until** $\Rightarrow \mid\mid \Sigma_1 \Rightarrow \Pi_1 \ \subseteq\ \Rightarrow \mid\mid \Sigma_2 \Rightarrow \Pi_2$;
- **if** $\Rightarrow \mid\mid \Sigma_1 \Rightarrow \Pi_1$ *is a crown initial structure* **then**
  - halt and accept;
- **end**
- existentially guess a formula $\Box B \in \Pi_1$ and apply the rule $\mathsf{K}^*_l$ backwards to $\Rightarrow \mid\mid \Sigma_1 \Rightarrow \Pi_1$ with this $\Box B$ and all the boxed formulae from $\Sigma_1$ as principal formulae, where $l$ is the number of such formulae, to obtain $\Rightarrow \mid\mid \Sigma_1 \Rightarrow \Pi_1 \mid \Phi \Rightarrow B$;
- apply $\mathsf{EW}$ backwards to $\Rightarrow \mid\mid \Sigma_1 \Rightarrow \Pi_1 \mid \Phi \Rightarrow B$ to obtain $\Rightarrow \mid\mid \Phi \Rightarrow B$;
- set $\Sigma_1 := \Phi$, $\Pi_1 := \{B\}$;

**until** $0 = 1$;

guessing step in Line 12 and the rule W is applied no more than $2n - 1$ times. Note that in Line 12 the algorithm halts and rejects if the crown is empty. Immediately after Line 12 there is only one crown component left, containing at most $2n$ formulae in total. The rest of the algorithm is essentially backwards proof search in a sequent calculus for the logic KT. The repeat loop of Line 14 applies no more than $2n$ rule instances because it processes each formula in the only crown component no more than once. There are no more than $n$ possibilities to choose a formula from the consequent in Line 21. Note that the algorithm halts and rejects if the consequent contains no boxed formulae. Finally, the repeat loop of Line 13 terminates after at most $n$ cycles because the maximal modal nesting depth of the new component created in Line 21 is strictly smaller than that of the other component deleted in the next line. Thus the total running time is polynomial in the size of the input, and since we alternate between universal choices and existential guesses, the algorithm runs in alternating polynomial time, i.e., the problem is in PSPACE [CKS81]. □

Since a modal formula $A$ is a theorem of the modal logic KT iff the formula $\Box A$ is a theorem of modal logic $\mathsf{KT}_\Box$ (and analogously for $\mathsf{KDT}_\Box$), and since the decision problem for KT is known to be PSPACE-complete [Lad77], it is clear that the complexity bound witnessed by the algorithm is in fact optimal.

*Remark* 7.14. The closest prefixed tableaux systems for the logics of this section seem to be Massacci's calculi from [Mas94], which correspond to nested sequents rather than to our grafted hypersequents. To mimic the calculi of this section better, the shape of the rules $\mathsf{K}_n$ (with or without $*$) suggests the use of destructive tableaux. However, this path leads, in our opinion to only slight modifications of the standard destructive tableaux for T, whose modal rules consist of the general destructive rule for K and the non-destructive T rule for capturing reflexivity (see, e.g., [Fit83, Ch. 2, Sect. 1]):

- for $\mathsf{KT}_\Box$, the use of the non-destructive T rule should be restricted until after the first application of a destructive rule;
- for $\mathsf{SDL}^+$, in addition, the use of the destructive rule for D should be (optionally) permitted before the destructive rule for K is used for the first time.

Such destructive systems are closer to the proof search Algorithm 2 than to the actual grafted hypersequent calculi and do not seem to be too novel. For this reason, as well as to keep the size of the paper reasonable, we refrain from providing their detailed descriptions.

# 8 Conclusion

In this article we have presented a novel proof-theoretic framework based on grafting a hypersequent calculus on top of a bounded-depth nested sequent calculus. In this framework we obtained natural cut-free calculi for the modal logics K5 and KD5 as well as calculi for extensions of K or KD with the axiom for shift reflexivity. The latter extension constitutes a calculus for the important deontic logic $\mathsf{SDL}^+$. For all the calculi we established syntactical cut elimination, admissibility of the structural rules in a slightly modified version of the calculi, and decidability of the derivability problem via backwards proof search. Notably, all the decision procedures are of optimal complexity, in particular, those for the logics K5 and KD5 are in coNP. To the best of our knowledge our calculi for these logics are the first analytic sequent-style formulations that give rise to decision procedures of optimal complexity. Further, we developed simplified prefixed tableaux calculi corresponding to the grafted hypersequent calculi, resulting in alternative semantic proofs of cut-free completeness.

**Future work.** We plan to extend and generalise these particular results in several different directions. For the first direction, we would like to investigate whether the methods developed here on the basis of specific examples can be used to handle logics of a bigger family uniformly. As one approach, it should be possible to plug in the generic cut elimination proof for hypersequent calculi from [Lel14] for the crown level part of the cut elimination proof for grafted hypersequents. As long as the crown level versions of the standard modal rules stay sound, this should give rise to analytic grafted hypersequent calculi for all extensions of K or KD with axioms of the form $\bigvee_{i=1}^{n} \Box \varphi_i$ where the $\varphi_i$ have modal nesting depth at most one and only contain negative occurrences of boxes. As another approach, we plan to investigate strengthenings and modifications of the nested sequent part of the calculi, e.g., to handle transitive logics or calculi where the nested sequent part has depth greater than one. E.g., we conjecture that the framework of grafted hypersequents should be able to capture the modal logic of *shift euclideanity* axiomatised by the boxed version $\Box(\Diamond\Box p \to \Box p)$ of axiom (5) by considering the system for K5 given in this paper adapted to grafted hypersequents where the trunk has depth 1 instead of depth 0. For the second direction, we would like to see whether the methods for proving limitative results about which logics cannot be captured by rules of a certain format from [LP13, Lel14] can be transfered to the grafted hypersequent framework. A very interesting question as raised by one of the reviewers then would be to investigate whether the depth of the trunk necessary to capture a modal logic gives rise to a natural hierarchy of logics. Finally, it would be interesting to see whether the grafted hypersequent framework can be adapted to capture multi-modal or first-order modal logics.

**Acknowledgements.** We would like to thank Agata Ciabattoni for always being ready to extend a helping hand, Christian Fermüller for his support, as well as Revantha Ramanayake and Lutz Straßburger for pointing out positive and negative facets of our ideas respectively. We would also like to thank the reviewers for the constructive comments, in particular for the suggestion to investigate semantic completeness proofs and the tableaux calculi corresponding to the calculi in our framework, which led to the results of Section 6.